\documentclass[nobibnotes,preprintnumbers,aps,prl,superscriptaddress,twocolumn,twoside,sort&compress]{revtex4-2}				 
\usepackage{graphicx}																												 
\usepackage{titlesec}	
\usepackage{multirow}	
\usepackage{gensymb}		
\usepackage{balance}
						
\usepackage{blindtext} 
\usepackage[colorlinks=true,linkcolor=blue,citecolor=red,urlcolor=blue]{hyperref}													 

\begin{document}																													 
%

\newcommand{\kvec}{\mbox{{\scriptsize {\bf k}}}}
\newcommand{\lvec}{\mbox{{\scriptsize {\bf l}}}}
\newcommand{\qvec}{\mbox{{\scriptsize {\bf q}}}}

\newcommand{\mycomment}[1]{}
\def\eq#1{(\ref{#1})}
\def\fig#1{\hspace{0mm}Fig. \ref{#1}}
\def\figur#1{\hspace{1mm}Figure \ref{#1}}
\def\tab#1{\hspace{1mm}Table \ref{#1}}
\title{
A recipe for an effective selection of promising candidates for high-temperature superconductors among binary hydrides}

\author{Izabela A. Wrona}
\affiliation{Institute of Physics, Czestochowa University of Technology, Ave. Armii Krajowej 19, 42-200 Czestochowa, Poland}
\author{Pawe{\l} Niegodajew}
\affiliation{Department of Thermal Machinery, Czestochowa University of Technology, Ave. Armii Krajowej 21, 42-200 Czestochowa, Poland}
\author{Artur P. Durajski} \email{artur.durajski@pcz.pl}
\affiliation{Institute of Physics, Czestochowa University of Technology, Ave. Armii Krajowej 19, 42-200 Czestochowa, Poland}

\begin{abstract}

Recent research on compressed binary hydrides have unveiled the potential for achieving superconductivity at near-room-temperature. Nevertheless, the available decision-making procedures standing behind the selection of constituent elements that may potentially exhibit high values of critical temperature ($T_c$) are far from optimal. In other words, a lot of experimental and numerical effort is wasted on exploring unpromising compounds. By conducting a deep study of a database containing over $580$ binary hydride superconductors, we were able to observe some interesting relationships between $T_c$ and selected physico-chemical properties of examined compounds. Among studied parameters, the ratio of the sum of the molecular weight of heavier atoms to the total mass of all hydrogen atoms in the chemical formula of hydride ($M_X/M_H$) was found to be the most valuable indicator that can help to screen for new promising superconductor candidates. This is because the highest $T_c$ requires the lowest $M_X/M_H$ ratio. Statistical analysis indicates a $28\%$ chance of finding $T_c > 200$ K within $0 < M_X/M_H < 15$. It is expected that these findings will not only allow for more efficient use of resources by improving future superconductor candidates selection but also they will accelerate ongoing experimental and numerical research that should bring new exciting discoveries in a much shorter time.
\end{abstract}
\maketitle
%
\section{I. Introduction}

Discovered by Onnes (1911) the effect of zero electrical resistance accompanied by strong diamagnetism in mercury initiated a global research effort aimed at searching for new materials characterized by ever-increasing superconducting critical temperatures $T_c$. 
Although this trend has been going on for more than a century, the enthusiasm associated with it has remained undiminished since a progressively growing number of novel compounds exhibiting superconducting properties are experimentally synthesized or numerically predicted year by year.

The hypothesis put forth by \citet{Ashcroft1968A}, suggesting that under very high pressure, hydrogen could potentially exhibit high-temperature superconductivity, although unconfirmed, due to experimental challenges associated with hydrogen metallization, drawn widespread attention among the scientific community. While more recently \citet{Ashcroft2004A} proposed that hydrides could be metalized using available experimental techniques, by leveraging chemical ``pre-compression'' from relatively heavier elements, increased attention has been paid to hydrogen-based compounds.
Although some of the initial findings of superconducting properties in hydrides were not very encouraging, due to low critical temperatures \cite{Skoskiewicz1972, Satterthwaite1970, goncharenko2008, eremets2008}, the breakthrough came with some recent experiments. Among them, one can distinguish H$_3$S with $T_c = 203$ K \cite{drozdov2015}, LaH$_{10}$ with $T_c = 250$ K \cite{drozdov2019}, YH$_{9}$ with $T_c =243$ \cite{kong2021}, YH$_{6}$ with $T_c = 224$ K \cite{troyan2021} and CaH$_{6}$ with $T_c = 215$ K \cite{ma2022}. These findings confirmed that the superconducting state at room temperature is possible to reach and made hydrogen-based materials the most serious candidates for that purpose.

To accelerate the superconducting materials exploration process, a thorough understanding of the physico-chemical mechanisms of the superconductivity induction process in specific groups of compounds is yet to be discovered.
So far, criteria guiding the selection of materials with potentially high values of $T_c$ have lacked robust empirical analysis of existing data. It is paramount to prevent the choice of unpromising compounds from being research objects at the initial stages since experimental approaches are costly while computational methods are time-consuming.

Several theories have already been formulated to support searching for high $T_c$ superconductors. A simple assumption that a high-temperature superconductor should have high-energy phonons was found to be insufficient \cite{bi2019, Flores-Livas2020A, Pickard2020}. The activities aimed at classifying hydrogen-rich superconductors based on structural chemical and electronic properties ended with the conclusion that only materials with highly symmetrical structures and a high density of states (DOS) at the Fermi level ensure high $T_c$ \cite{Flores-Livas2020A, semenok2020A, Ishikawa2019A}. However, these conditions were later found to be necessary but insufficient for such a purpose. Recently, \citet{Belli2021} explored 178 hydrogen-based superconductors, intending to explain the origin of the high $T_c$. The authors found that $T_c$ is well correlated with the networking value multiplied by both the hydrogen fraction and the cubic root of the fraction of the DOS at Fermi energy coming from the hydrogen atoms.
The last finding, although interesting, is not of much use at the stage when exploring new compounds that could potentially exhibit high $T_c$ since the information about the networking value is accessible when completing ab initio crystal structure prediction via numerical simulation or after experimental measurements, for example, using X-ray diffraction. Therefore, further research is needed to search for correlations that could ensure a fast initial assessment of any compound and whether it possibly possesses promising superconductive properties with high $T_c$.

In this study, we investigated over $250$ binary hydride superconductors, which were discovered through experimental or ab initio methods utilizing density functional theory (DFT). We aimed to discern potential correlations between $T_c$ and the physical or chemical properties that characterize these compounds. As a result, we found that $T_c$ is inversely proportional to the ratio of the sum of the molecular weight of heavier atoms to the sum of the molecular weight of light hydrogen atoms. This finding allows experimentalists to indicate a potential chemical configuration for which the probability of obtaining high $T_c$ is the highest. On the other hand, our results deliver information on where the search efforts should not be directed. We expect that the outcomes presented herein will significantly expedite the exploration of room-temperature superhydrides.

\section{II. Explored databases}

Experimental research, despite its considerable expense, remains irreplaceable in advancing the quest for high-temperature superconductors. The experimental efforts involve synthesizing hydrides under immense pressure using diamond anvil cells or other high-pressure devices \cite{Shen2016A, Bassett2009, Jayaraman1983, sakata2020} in most cases to confirm superconducting properties, critical temperatures, and other key features already predicted by computational methods like DFT. Also, various methods such as X-ray diffraction, Raman spectroscopy, electrical resistivity measurements, and magnetic susceptibility tests are employed to determine structural properties, phonon spectra, and superconducting transitions of investigated compounds \cite{bhattacharyya2024, minkov2023}. Very high costs and complex fabrication processes are the key reasons for so little experimental work in the field. This may suggest that the so far progress in superconductivity is driven mostly by theoretical predictions.

Not less important are theoretical calculations since they allow screening materials with diversified structures at relatively low costs compared to experimental approaches. Among available theoretical methods, DFT found the broadest popularity since it enables predicting the electronic structure, phonon properties, and superconducting behavior of hydrogen-rich compounds under varying conditions including high pressures \cite{gebreyohannes2022, PhysRevMaterials.7.L101801}. 
\begin{figure}[b]
\begin{center}
\includegraphics[width=0.9\columnwidth]{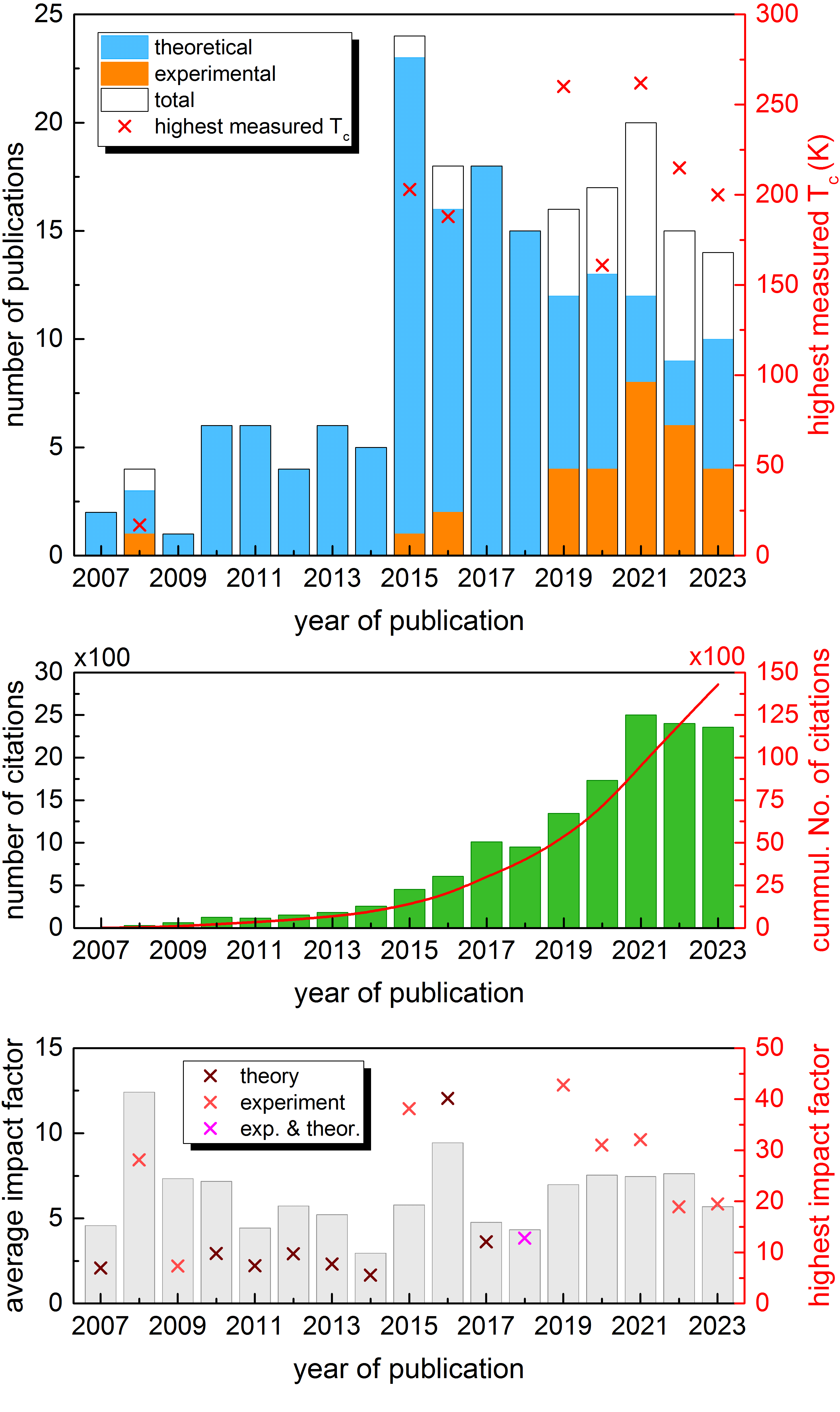}
\caption{(upper panel) Bar chart of the annual distribution of papers focused on binary hydrogen-rich superconductors between 2007 and 2023. The highest $T_c$ values discovered experimentally in a given year are marked with red crosses.
(central panel) Bar chart of the number of citations of the above papers. The cumulative number of citations is marked with a red line.
(bottom panel) Bar chart of the average impact factor of journals published above papers. The highest IF values of journals published above papers in a given year are marked with crosses.
}
\label{f01}
\end{center}
\end{figure}

In this work, we explored a database containing results on superconducting binary hydrides (X$_n$H$_m$) so far obtained within theoretical and experimental studies. The database includes $296$ compounds with confirmed superconducting properties (database available in the ZENODO \cite{zenodo}).
In some studies, a single compound was exposed to different pressures, and in turn various structures with different values of $T_c$ were obtained. This allowed extending the database up to over $580$ unique compounds \cite{
abe2011,abe2013,abe2015,abe2017,abe2018,abe2018A,abe2019,abe2021,akashi2015,
bi2017,bi2019,bi2021,
chang2020,chen2008,chen2014,chen2015,chen2021,chen2021A,chen2021B,chen2021C,chen2022,cheng2015,cui2015,
davariesfahani2016,davariesfahani2017,drozdov2015,drozdov2019,du2021,duan2010,duan2014,duan2015,duda2018,durajski2018,
einaga2016,eremets2008,errea2013,errea2016,
feng2015,flores-livas2012,flores-livas2016,flores-livas2016A,fu2016,
gao2008,gao2010,gao2011,gao2013,gao2021A,gebreyohannes2022,gebreyohannes2022A,gu2017,
hai2021,han2017,he2023,heil2019,hong2022,hooper2013,hou2015,hou2015A,hu2013,huo2022,hutcheon2020,
ishikawa2016,
jeon2022,jin2010,
kim2010,kim2011,kong2021,kruglov2018,kruglov2020,kuzovnikov2019,kvashnin2018,kvashnin2018A,
li2010,li2014,li2015,li2016,li2017,li2017A,li2019,li2019A,li2020,li2022,li2023,liao2020,liu2015,liu2015A,liu2015B,liu2015C,liu2016,liu2016A,liu2017,liu2017A,liu2018,liu2023,lonie2013,lu2015,
ma2022,majumdar2017,martinez-canales2009,matsuoka2019,mishra2021,
ning2017,
papaconstantopoulos2018,papaconstantopoulos2020,pena-alvarez2020,peng2017,
qian2017,
saha2023,salke2019,sano2016,scheler2011,semenok2018,semenok2020,semenok2020A,shamp2015,shamp2016,shanavas2016,shao2018,shao2019,shao2021,shao2021A,shipley2021,snider2021,somayazulu2019,song2020,song2021,sun2020,szczesniak2015,szczesniak2016,
tanaka2017,tikhonov2023,troyan2016,troyan2021,tse2007,tsuppayakorn-aek2020,tsuppayakorn-aek2021,tsuppayakorn-aek2023,
wang2012,wang2017,wang2019,wang2021,wang2021A,wang2022,wang2023,wang2023A,wei2016,wu2018,
xiao2019,xiao2019A,xie2014,xie2020,xie2020A,
yan2015,yan2022,yan2023,yang2019,yang2023,yao2007,yao2023,ye2018,yong2023,yu2014,yu2015,yuan2019,
zeng2017,zhang2010,zhang2011,zhang2015,zhang2015A,zhang2015B,zhang2015C,zhang2016,zhang2020,zhang2022,zhang2022A,zhang2022B,zheng2018,zhong2012,zhong2016,zhong2022,zhou2011,zhou2012,zhou2020,zhou2020A,zhuang2017,zhuang2017A,zhuang2018, minkov2023, gavriliuk2023}.

A total of 191 publications on binary hydrogen-rich superconductors from the period of 2007 to 2023 were included in our sample. As depicted in \fig{f01}, the total number of publications suddenly increased by leaps and bounds in 2015. This is strongly correlated with the experimental observation of superconductivity in H$_3$S \cite{drozdov2015}. Since then, the annual number of publications has been gradually decreasing among the years. However, in 2019, a slight shift in this trend occurred due to a new series of inspiring experimental discoveries involving high $T_c$ superconductors. 
The number of citations and the impact factor (IF) of academic publications are key indicators for measuring development trends in a particular research field. Over the past seventeen years, the number of citations for publications in the field of superconductivity in binary hydrides has gradually increased over the years. This trend suggests that the research community is increasingly recognizing the significance of this area, highlighting its potential for future scientific breakthroughs and technological applications.
The average IF of journals publishing papers included in our database oscillates around a value of $6$. Notably, over the past five years, papers published in journals with the highest IF have predominantly featured experimental results. In contrast, purely theoretical papers have tended to be published in journals with lower IF. This trend highlights a growing emphasis on experimental validation and practical application in high-impact research, while theoretical work, despite its importance, is increasingly directed towards specialized journals with lower impact factors. This shift may reflect the scientific community's prioritization of experimentally verified advancements over purely theoretical explorations.

As suggested by Flores-Livas \textit{et al.} \cite{Flores-Livas2020A}, there is an upper limit of $300$ binary hydrides that exhibit superconducting properties and are thermodynamically stable. Therefore, one can be sure that in the present study, the number of compounds that were taken into consideration can be regarded as representative in the search for general conclusions and trends when exploring the considered group of binary hydrogen-rich superconductors. In particular, using the collected results we focused on studying correlations between $T_c$ and fundamental physical or chemical properties that characterize these compounds to provide a recipe for upcoming studies on near room-temperature experimental.

\begin{figure}[!h]
\begin{center}
\includegraphics[width=0.9\columnwidth]{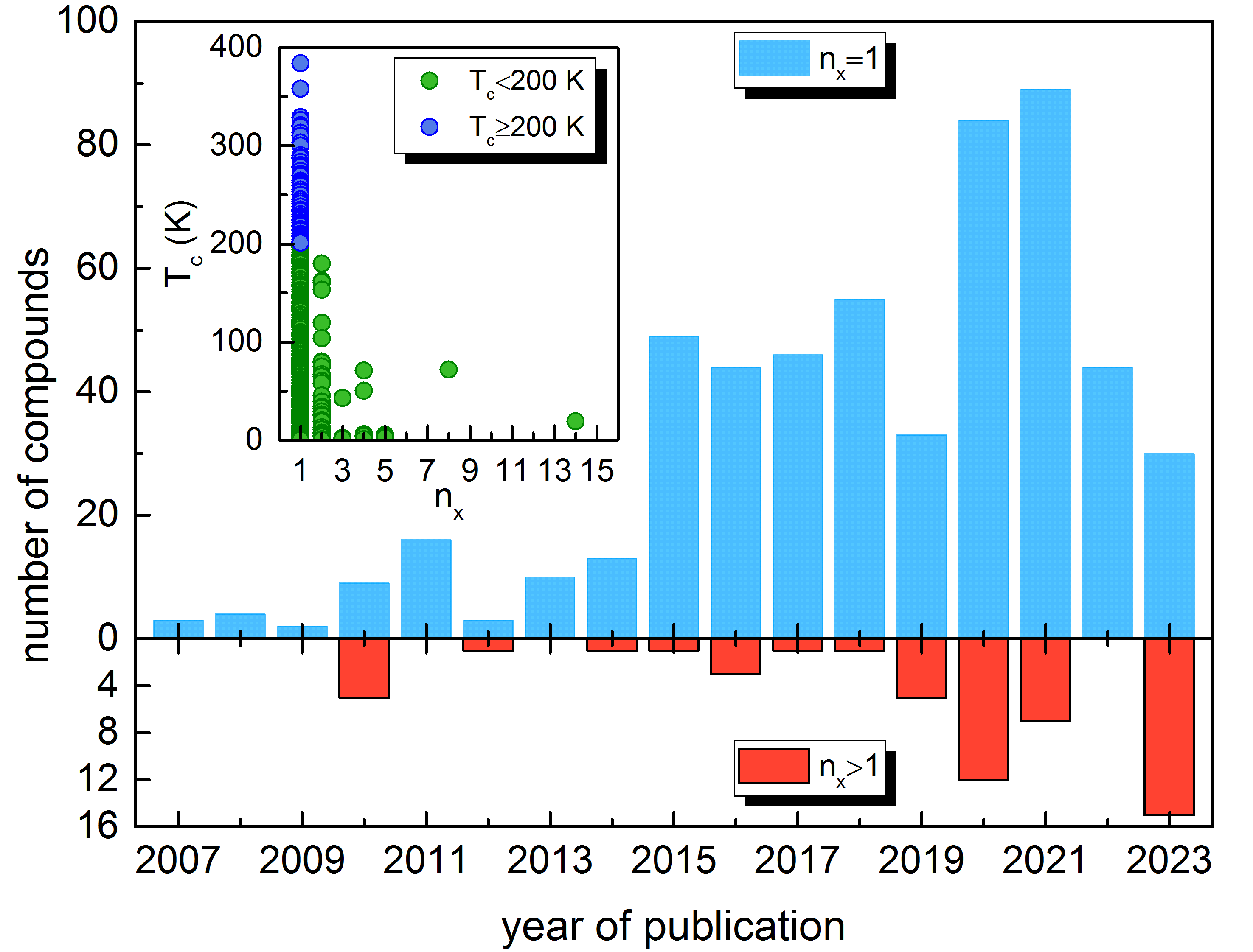}
\caption{The annual distribution of binary hydrogen-rich superconductors between $2007$ and $2023$ taking into account the division into compounds consisting of one or more heavy elements.}
\label{f02}
\end{center}
\end{figure}

Upon comprehensive analysis of our database, a general observation emerges that most of the hydrogenated systems (over $90$\%) contain only one heavy atom. The yearly distribution of binary hydrogen-rich superconductors, categorized based on compounds containing varying number of heavy elements, is presented in \fig{f02}. In particular, the inset in \fig{f02} highlights the pronounced interest in compounds comprising solely one heavy element, stemming from their propensity to achieve critical temperatures exceeding $200$ K. 
Within the group of X$_n$H$_m$ systems under investigation, where $n=1$, more than $14.5\%$ compounds exhibited a critical temperature surpassing $200$ K, accounting for $13.2\%$ of the whole dataset.

\section{III. Results and discussion}

\begin{figure}[!h]
\begin{center}
\includegraphics[width=0.9\columnwidth]{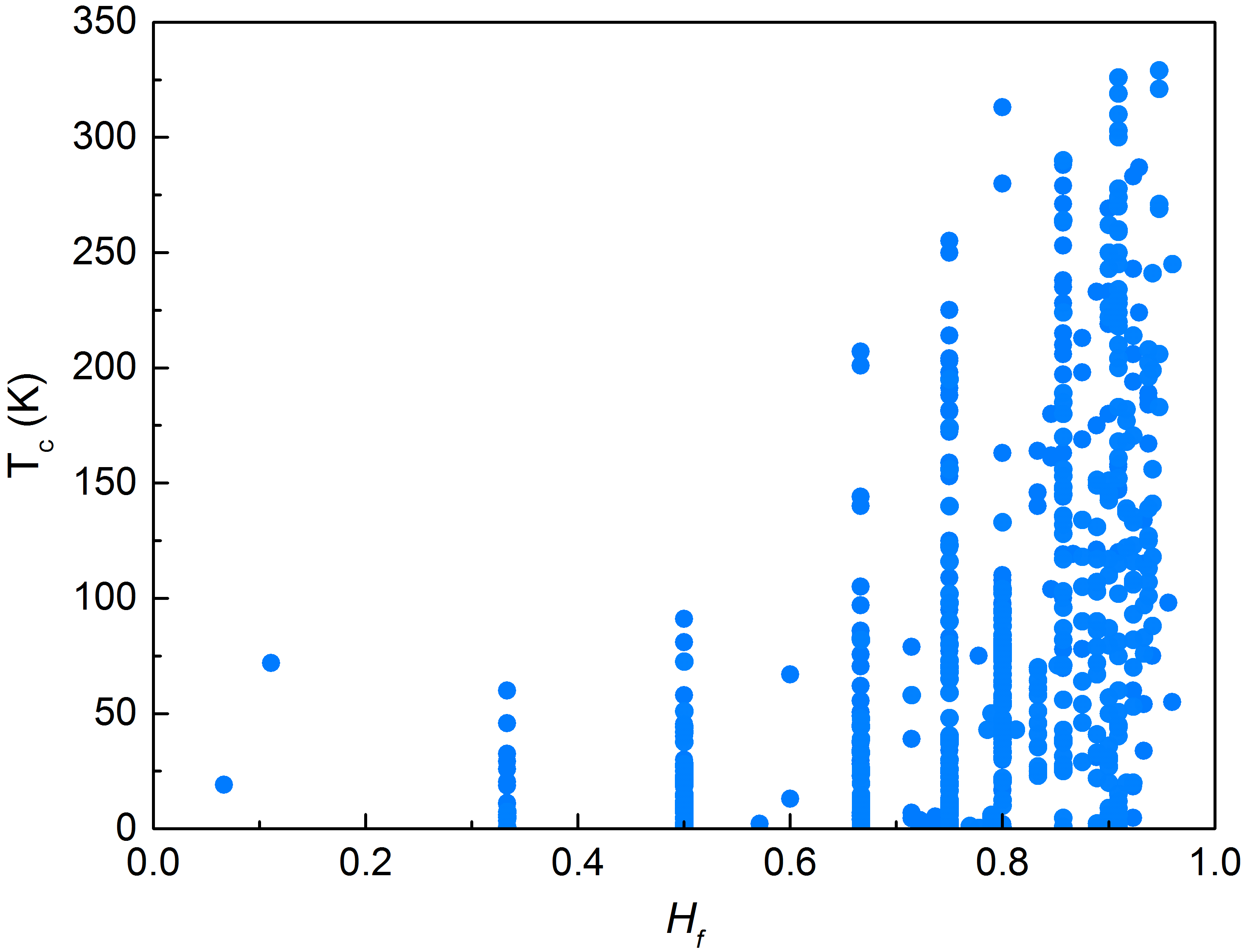}
\caption{Critical temperature as a function of the hydrogen fraction in the compound. }\vspace{0.5cm}
\label{f03}
\includegraphics[width=0.9\columnwidth]{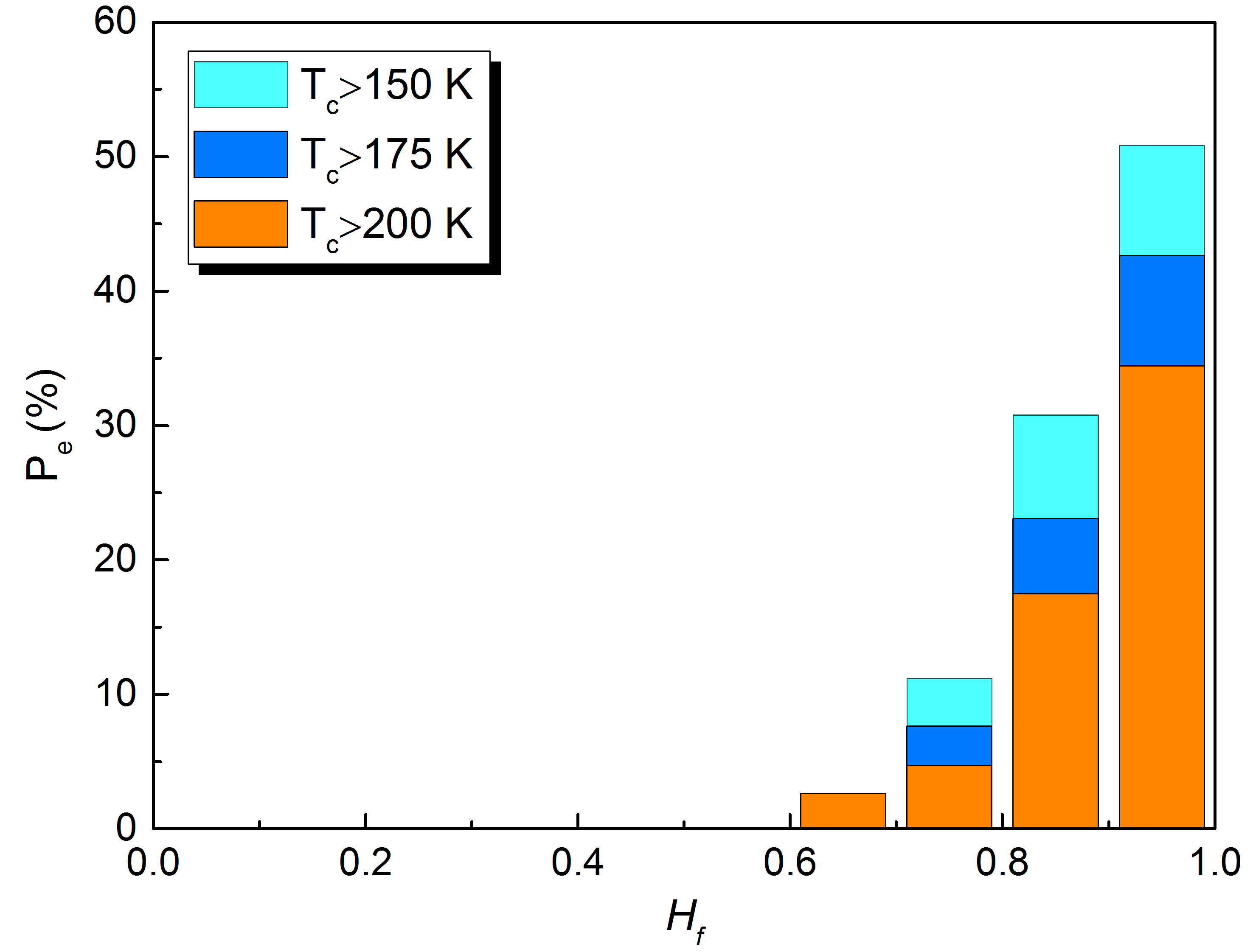}
\caption{Probability of exceeding a critical temperature threshold as a function of hydrogen fraction.}\vspace{0.5cm}
\label{f04}
\includegraphics[width=0.9\columnwidth]{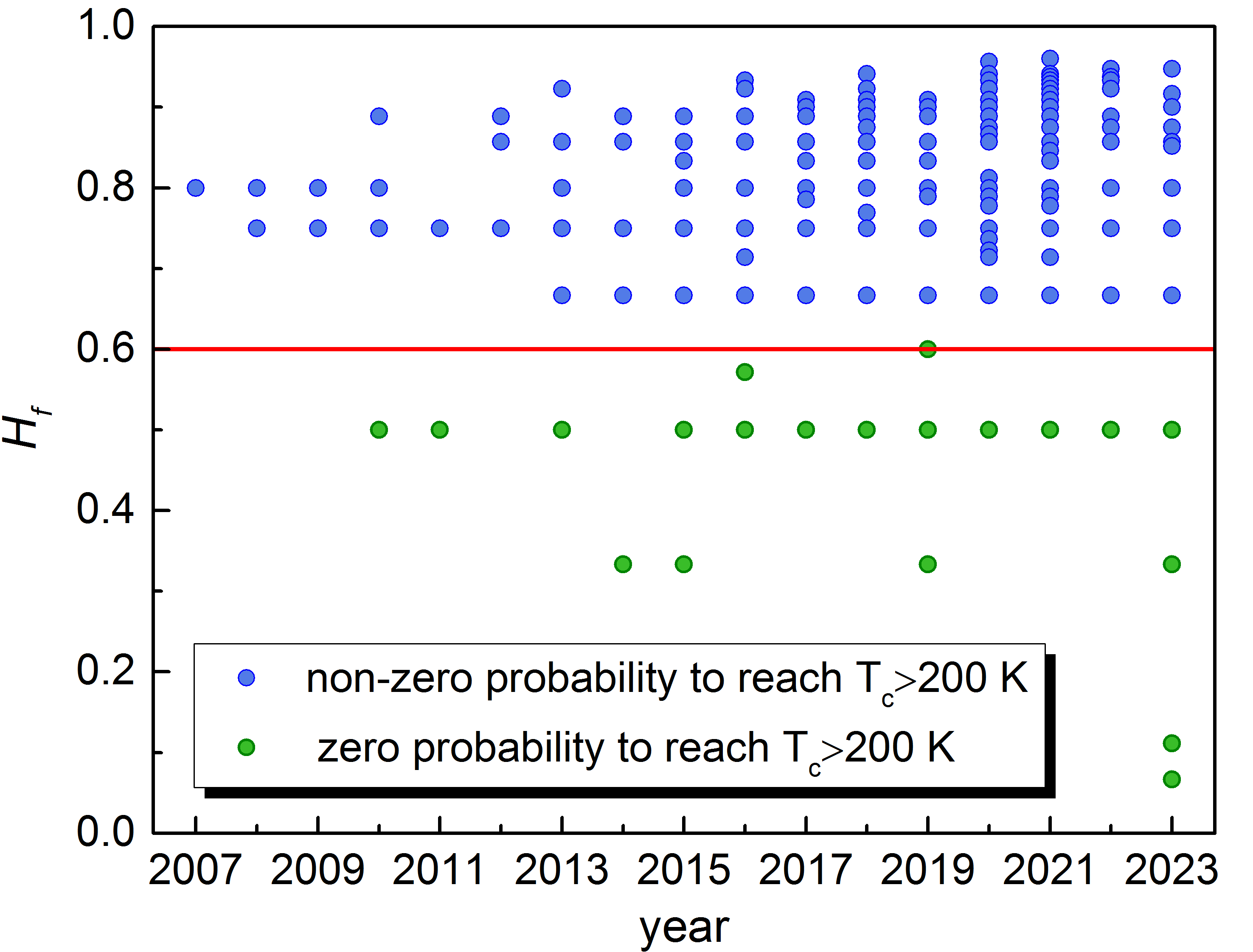}
\caption{The annual distribution of the hydrogen fraction of the studied superconducting compounds. The red line indicates the threshold below which a critical temperature higher than $200$ K cannot be achieved.}
\label{f05}
\end{center}
\end{figure}

When embarking on the exploration of novel superconducting binary hydride compounds, researchers initially did not possess a strategic development pathway that could instruct them on selecting a promising candidate. In particular, they did not know what physical parameters the materials should be characterized by to ensure a high probability of achieving a near room value of $T_c$. These parameters include fundamental features, such as the count of constituent atoms and their respective molecular weights. Molecular weights offer insights into the compound's mass distribution, while the number and types of atoms present shed light on its elemental composition, influencing its chemical and physical properties.
Nonetheless, one should bear in mind that searching for superconductors exhibiting high $T_c$ values together with unique thermodynamic properties is a very difficult, very time-consuming and complex process.  
Herein, we aimed to discern potential correlations between $T_c$ and molecular weight of heavy atoms and the number of heavy atoms or number of hydrogen atoms in binary hydrides X$_n$H$_m$. Let us first start from presenting the critical temperature as a function of hydrogen fraction in a compound (\fig{f03}). The hydrogen fraction $H_{f}$ takes the following form:
\begin{equation}
H_{f}=\frac{n_{\rm{H}}}{n_{\rm{X}}+n_{\rm{H}}},
\label{EQ_T_c}
\end{equation}
%
\begin{figure}[!h]
\begin{center}
\includegraphics[width=0.9\columnwidth]{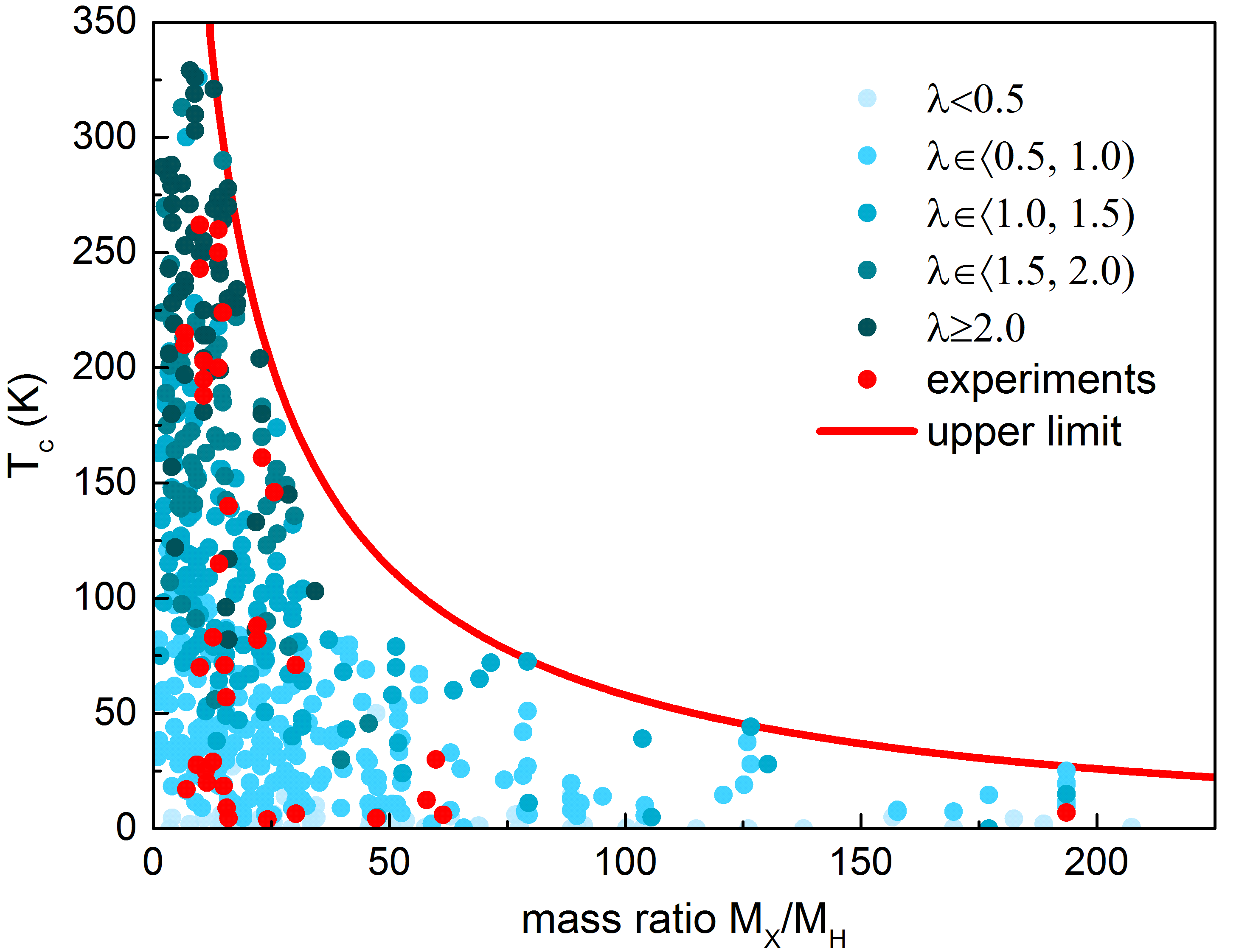}
\caption{Superconducting temperature as a function of $M_x/M_H$. Points are colored according to different $\lambda$.}\vspace{0.5cm}
\label{f06}
\includegraphics[width=0.9\columnwidth]{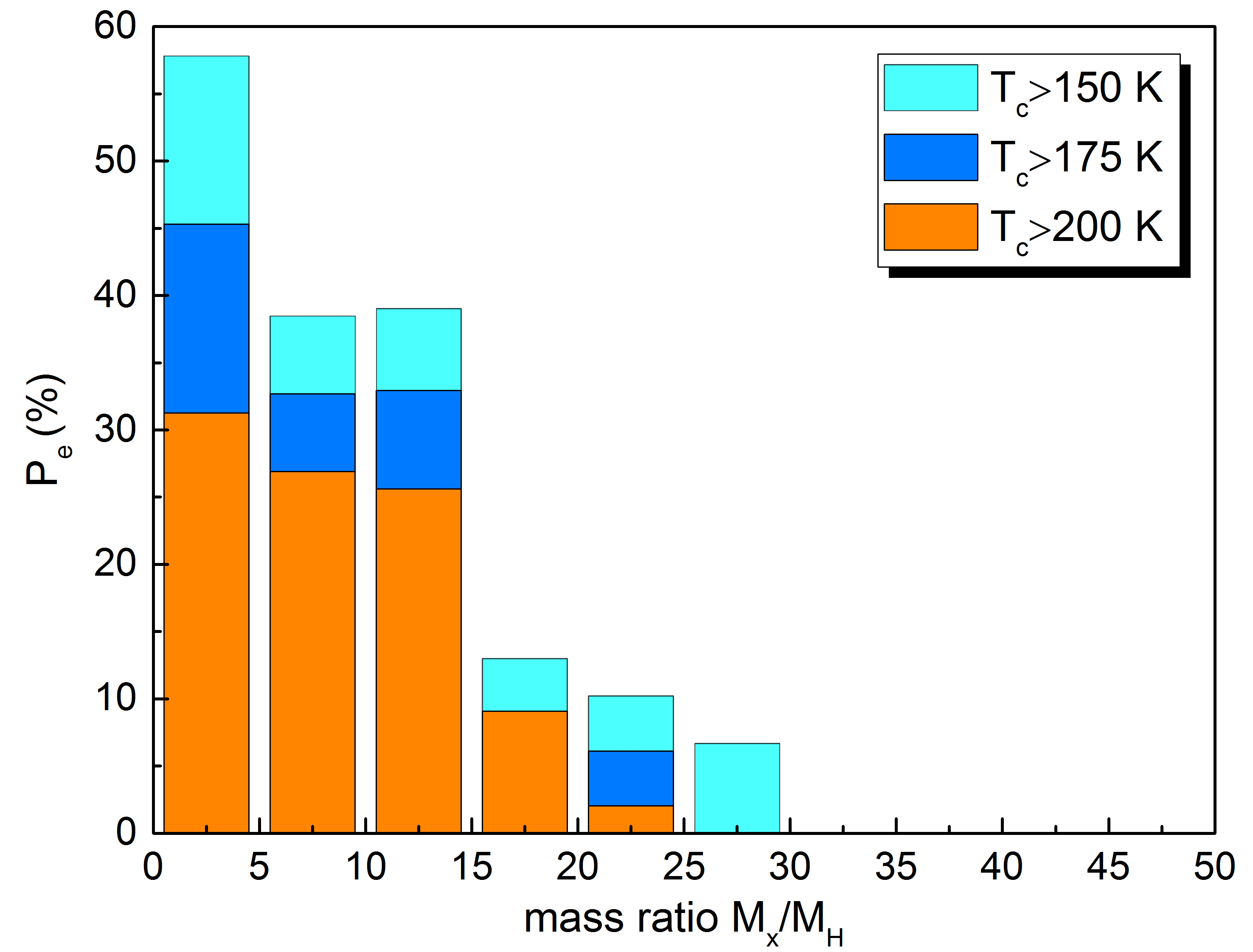}
\caption{Probability of exceeding a critical temperature threshold as a function of mass ratio.}\vspace{0.5cm}
\label{f07}
\includegraphics[width=0.9\columnwidth]{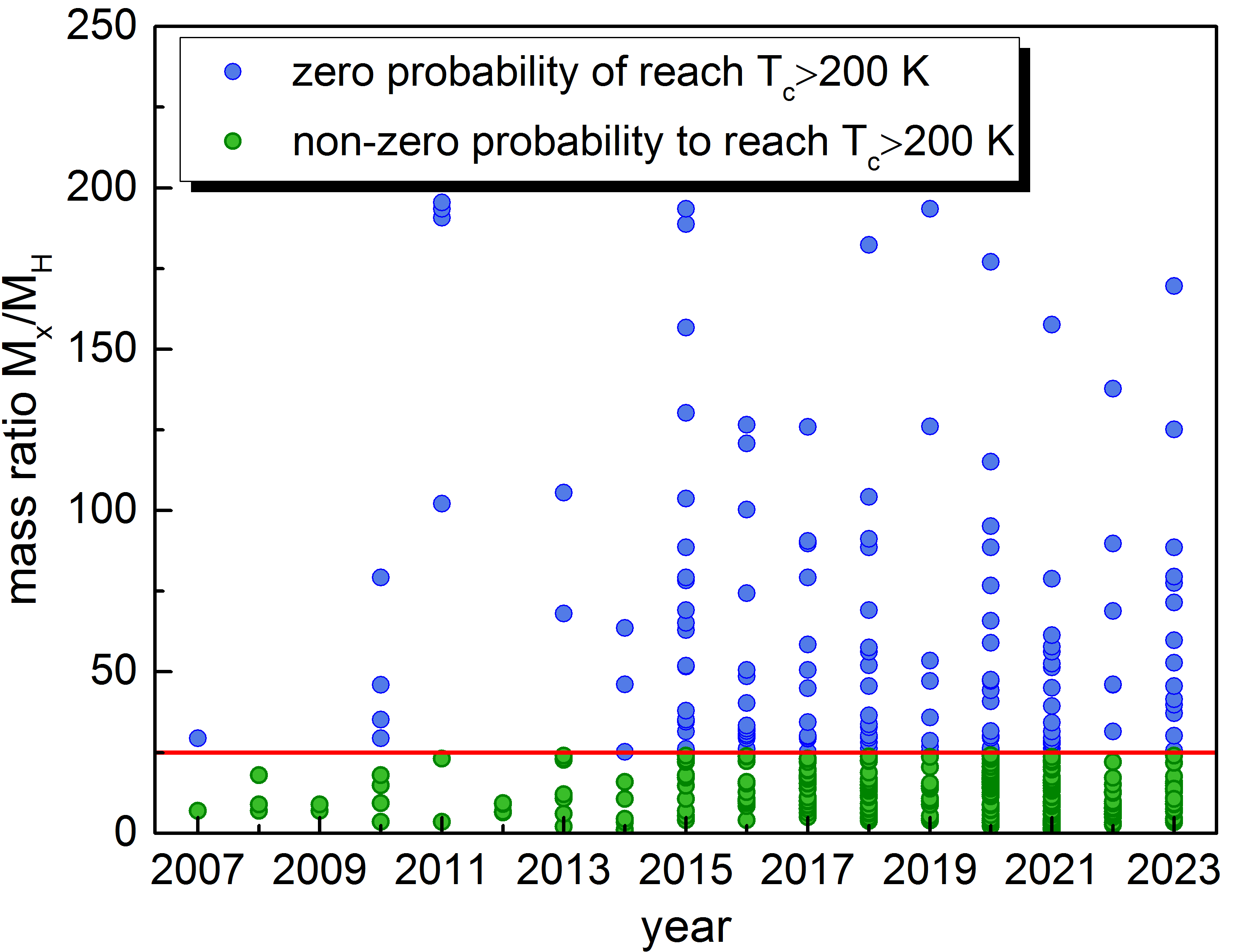}
\caption{The annual distribution of the mass ratio of the studied superconducting compounds. The red line indicates the threshold above which a critical temperature higher than $200$ K cannot be achieved.}
\label{f08}
\end{center}
\end{figure}
%
\clearpage
where $n_{\rm{H}}$ and $n_{\rm{X}}$ are the number of hydrogen and heavier element atoms in the compound, respectively. 
One can note that the highest critical temperatures for the investigated materials are always seen when values of $H_{f}$ are close to unity. This is the case when a compound contains several H atoms per heavier atom.  

A statistical summary of the findings presented in \fig{f03} in the form of the exceedance probability $P_e$ \cite{Kunreuther_2002,TAMMINEN20134577} for three different threshold levels of $T_c$ ($150$ K, $175$ K and $200$ K) is presented in \fig{f04}. One should note that further in the paper we mostly focus on the highest threshold value whereas the smaller ones are included just for demonstrative purposes. 
Selection of the cut-off temperature of $200$ K was inspired by the relatively recent breakthrough discovery of superconduction properties of H$_3$S at just slightly more than $200$ K \cite{drozdov2015}. 
As can be seen, for the given database, $H_{f}$ must exceed $0.6$ for the first value of $T_c>200$ to emerge. As $H_f$ increases, a gradual growth in $P_e$ is observed with its maximum of $34\%$ for $H_{f}\in(0.9,1.0)$. 
It is an important observation since it directly eliminates compounds with zero probability of exceeding $T_c=200$ K. In the given database, about $12.5\%$ of all the compounds have $H_{f} < 0.6$ (points below the red line in \fig{f05}) and so the effort associated with the calculations or measurements could be avoided if having this information in the past. Not less interesting is that the share of compounds with $H_{f}\in(0.9,1.0)$ in the database is at the level of $21\%$.

When presenting $T_c$ as a function of the ratio of the total mass of all heavier atoms to the total mass of all light hydrogen atoms $M_X/M_H$ (\fig{f06}), one can see that all points fall below the decaying exponential line: 
\begin{equation}
T_c^{\rm limit}=6T_0exp{\left(-\frac{\sqrt{M_X/M_H}}{10}\right)} /\sqrt{M_X/M_H},
\label{EQ_T_c}
\end{equation}
where $T_c^{\rm limit}$ is a upper limit of $T_c$ and $T_0$ is the freezing point of water ($273$ K). It clearly indicates that the necessary condition that must be met to achieve a high $T_c$ is to have a compound with the lowest possible $M_X/M_H$ ratio. On the other hand, \fig{f06} also suggests that materials characterized by high $M_X/M_H$ values cannot be considered as good candidates for high-temperature superconductors.
One should also note that for more detailed insight the points in \fig{f06} are in addition categorized into five groups diversified by different ranges of the electron-phonon coupling constant $\lambda$. 
At first glance, groups of points exhibit layered arrangement which manifests that a high value of $T_c$ is guaranteed when $\lambda$ takes a high value as well.

It is important to note that the upper $T_c$ limit (the upper limit imposed by the Eq. \ref{EQ_T_c}) can be regarded as pressure independent. It is because \fig{f06} includes also the data for a number of the same components exposed to different values of pressure. 
A good example is the compound SiH$_4$, for which we collected seven sets of results from five theoretical works: $T_c = 55$ K (125 GPa, $C$2/$c$ space group) \cite{yao2007}, $T_c = 20$ K (150 GPa, $Cmca$ space group) \cite{chen2008}, $T_c = 16.5$ K (190 GPa, $Pbcn$ space group) \cite{martinez-canales2009}, $T_c = 35.1$ K (300 GPa, $P$-3 space group) \cite{cui2015}, $T_c = 35$ K (400 GPa, $P2_1$/$c$ space group) and $T_c = 110$ K (610 GPa, $C$2/$m$ space group) \cite{zhang2015B} and one experimental work: $T_c = 17$ K ($96$ GPa) \cite{eremets2008}.


Let us now look at the statistics, namely, what is the probability $P_e$ of exceeding a certain $T_c$ threshold value within the given mass ratio $M_X/M_H$ range (\fig{f07}). As is seen, $P_e$ takes non-zero value (for $T_c>200$ K) when $M_X/M_H$ is less than $25$. Yet, this probability is at a very small level of $\sim$$2\%$. When $M_X/M_H$ takes the value from the range between $15$ and $20$, an elevation of $P_e$ up to $\sim9\%$ is seen. 
A further reduction of $M_X/M_H$ to the values from $0$ to $15$ ensures a substantial growth in $P_e$ up to the averaged share of $28\%$. Such a remarkable growth in probability can be regarded as exceptionally high. 
One should note that the share of the compounds with $0<M_X/M_H<15$ in the considered database is equaled to $43\%$.
Similarly, as it was in the case of statistical accounting of $H_f$, the analysis of $M_X/M_H$ ratio indicates that $35\%$ of the total tested compounds could have been initially dismissed from consideration in the quest for achieving a high-temperature superconducting state (see \fig{f08}). 

\begin{figure}[!h]
\begin{center}
\includegraphics[width=1\columnwidth]{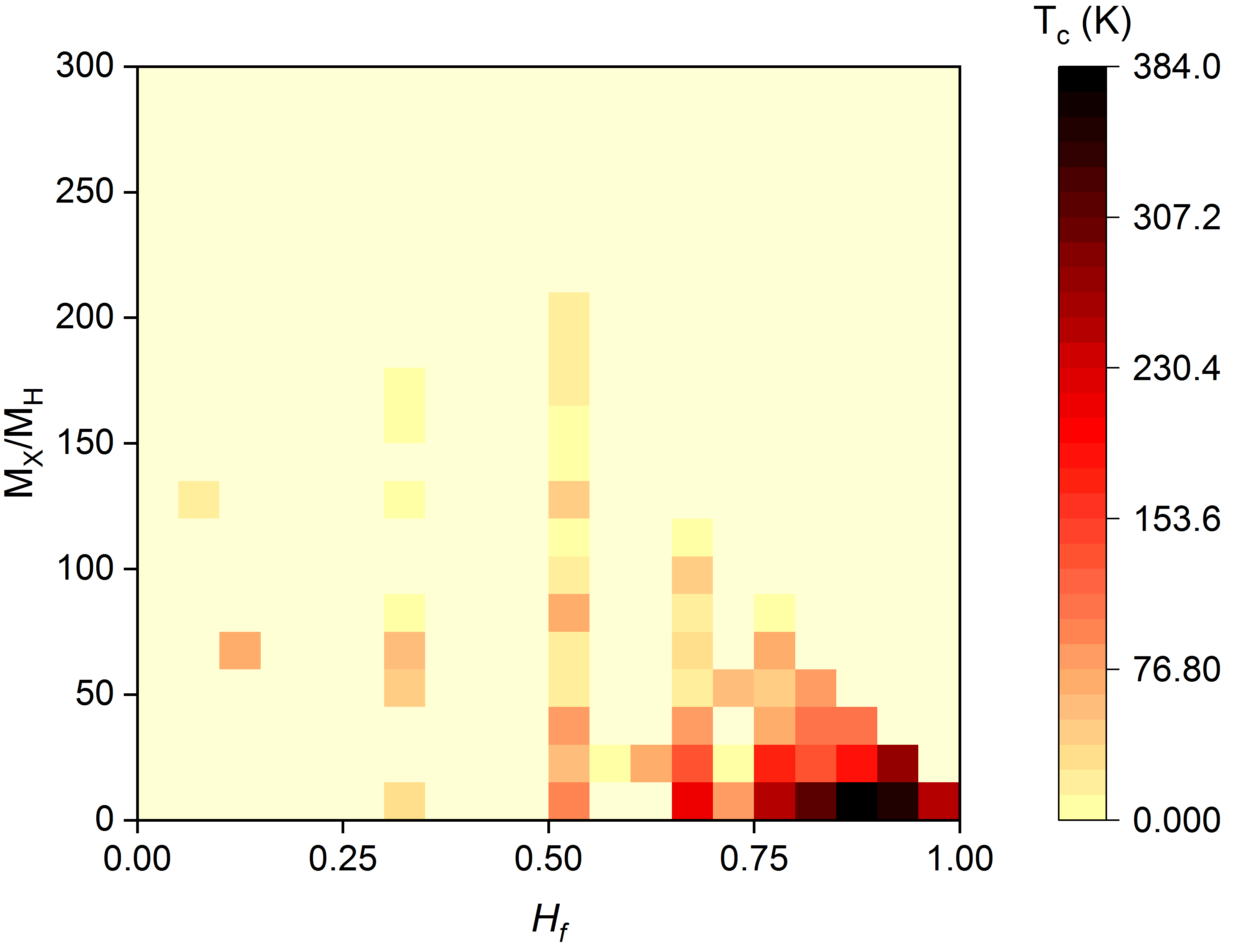}
\caption{The contour map of the maximum $T_c(M_X/M_H, H_f)$.}
\label{f09}
\end{center}
\end{figure}

Figure \ref{f09} summarizes the so far presented results by demonstrating the contour map of the maximum value of $T_c$ plotted vs $H_f$ and $M_X/M_H$. It directly allows identifying what are the best combinations of these two physico-chemical properties to ensure achieving a superconducting state at high $T_c$. A clear local maximum surrounded by the region with elevated values of $T_c$ limited by $0.75<H_f<1$ and $0<M_X/M_H<50$ indicates the area that should be of immense interest in the future.

\section{Conclusions and summary}

In this paper, we explored a database containing over $580$ binary hydrides exhibiting superconductive properties. For all these compounds, values of $T_c$ were predicted using numerical simulation or measuring techniques.
Some interesting observations were made by studying the relationships between $T_c$ and selected physico-chemical properties of the compounds such as $n_X$, $H_f$, and $M_X/M_H$ ratio. 
The key finding was that when plotting $T_c$ as a function of $M_X/M_H$ all the data points fall below the decaying exponential curve indicating an evident upper limit of $T_c$. Therefore, the necessary condition that must be met to achieve a high $T_c$ is to have a compound with the lowest possible $M_X/M_H$ ratio. On the other hand, it is also clear now that materials characterized by high $M_X/M_H$ values cannot be considered as good candidates for high-temperature superconductors. This finding seems to be pressure independent which can be supported by the fact that a number of the same components (from the database) were exposed to different values of pressure and so exhibited different values of $T_c$. 
Moreover, the statistical analysis revealed that when exploring compounds with $0<M_X/M_H<15$, there is about $28\%$ chances to discover a superconductor with $T_c>200$ K.
The results also indicated that high $T_c$ binary hydride superconductors should be characterized by the number of heavier atoms $n_X=1$ and as high as possible value of the hydrogen fraction $H_f$. 

It is expected that in the near future, these findings will improve the decision-making process regarding the selection of constituent elements within tested binary superconducting compounds. In particular, the given pathway should enable the scientific community to thoroughly and in-depthly explore a broader group of much more promising potential superconductor candidates without wasting time and resources on investigating unpromising systems.

\section{Additional Data}
Following the reviewers' suggestions, we have reviewed recent experimental results published in 2024 \cite{guo2024, cross2024, he2024}. While these new data support our hypothesis and strengthen the conclusions of our study, they were not included in our analysis. This paper focuses on examining the results from the full years 2007-2023, and incorporating partial data from 2024 could introduce unnecessary confusion.

\section{Data availability}
The database containing information on over $580$ binary hydride superconductors is available on the Zenodo public database under accession code XXXX \cite{zenodo}.

\section{Declaration of competing interest}
The authors declare no competing financial interest.

\section{Supplementary information} The online version contains supplementary material available at https://doi.org/XXXX

\section{Acknowledgements}
A.P.D. acknowledges financial support from the National Science Centre (Poland) under Project No. 2022/47/B/ST3/00622. 


\bibliography{bibliography}

\begin{thebibliography}{210}%
\makeatletter
\providecommand \@ifxundefined [1]{%
 \@ifx{#1\undefined}
}%
\providecommand \@ifnum [1]{%
 \ifnum #1\expandafter \@firstoftwo
 \else \expandafter \@secondoftwo
 \fi
}%
\providecommand \@ifx [1]{%
 \ifx #1\expandafter \@firstoftwo
 \else \expandafter \@secondoftwo
 \fi
}%
\providecommand \natexlab [1]{#1}%
\providecommand \enquote  [1]{``#1''}%
\providecommand \bibnamefont  [1]{#1}%
\providecommand \bibfnamefont [1]{#1}%
\providecommand \citenamefont [1]{#1}%
\providecommand \href@noop [0]{\@secondoftwo}%
\providecommand \href [0]{\begingroup \@sanitize@url \@href}%
\providecommand \@href[1]{\@@startlink{#1}\@@href}%
\providecommand \@@href[1]{\endgroup#1\@@endlink}%
\providecommand \@sanitize@url [0]{\catcode `\\12\catcode `\$12\catcode
  `\&12\catcode `\#12\catcode `\^12\catcode `\_12\catcode `\%12\relax}%
\providecommand \@@startlink[1]{}%
\providecommand \@@endlink[0]{}%
\providecommand \url  [0]{\begingroup\@sanitize@url \@url }%
\providecommand \@url [1]{\endgroup\@href {#1}{\urlprefix }}%
\providecommand \urlprefix  [0]{URL }%
\providecommand \Eprint [0]{\href }%
\providecommand \doibase [0]{http://dx.doi.org/}%
\providecommand \selectlanguage [0]{\@gobble}%
\providecommand \bibinfo  [0]{\@secondoftwo}%
\providecommand \bibfield  [0]{\@secondoftwo}%
\providecommand \translation [1]{[#1]}%
\providecommand \BibitemOpen [0]{}%
\providecommand \bibitemStop [0]{}%
\providecommand \bibitemNoStop [0]{.\EOS\space}%
\providecommand \EOS [0]{\spacefactor3000\relax}%
\providecommand \BibitemShut  [1]{\csname bibitem#1\endcsname}%
\let\auto@bib@innerbib\@empty
\bibitem [{\citenamefont {Ashcroft}(1968)}]{Ashcroft1968A}%
  \BibitemOpen
  \bibfield  {author} {\bibinfo {author} {\bibfnamefont {N.~W.}\ \bibnamefont
  {Ashcroft}},\ }\bibfield  {title} {\enquote {\bibinfo {title} {Metallic
  hydrogen: a high-temperature superconductor?}}\ }\href {\doibase
  https://doi.org/10.1103/PhysRevLett.21.1748} {\bibfield  {journal} {\bibinfo
  {journal} {Phys. Rev. Lett.}\ }\textbf {\bibinfo {volume} {21}},\ \bibinfo
  {pages} {1748} (\bibinfo {year} {1968})}\BibitemShut {NoStop}%
\bibitem [{\citenamefont {Ashcroft}(2004)}]{Ashcroft2004A}%
  \BibitemOpen
  \bibfield  {author} {\bibinfo {author} {\bibfnamefont {N.~W.}\ \bibnamefont
  {Ashcroft}},\ }\bibfield  {title} {\enquote {\bibinfo {title} {Hydrogen
  dominant metallic alloys: high temperature superconductors?}}\ }\href
  {\doibase https://doi.org/10.1103/PhysRevLett.92.187002} {\bibfield
  {journal} {\bibinfo  {journal} {Phys. Rev. Lett.}\ }\textbf {\bibinfo
  {volume} {92}},\ \bibinfo {pages} {187002} (\bibinfo {year}
  {2004})}\BibitemShut {NoStop}%
\bibitem [{\citenamefont {Skoskiewicz}(1972)}]{Skoskiewicz1972}%
  \BibitemOpen
  \bibfield  {author} {\bibinfo {author} {\bibfnamefont {T.}~\bibnamefont
  {Skoskiewicz}},\ }\bibfield  {title} {\enquote {\bibinfo {title}
  {Superconductivity in the palladium-hydrogen and palladium-nickel-hydrogen
  systems},}\ }\href {\doibase https://doi.org/10.1002/pssa.2210110253}
  {\bibfield  {journal} {\bibinfo  {journal} {Phys. Status Solidi A}\ }\textbf
  {\bibinfo {volume} {11}},\ \bibinfo {pages} {K123} (\bibinfo {year}
  {1972})}\BibitemShut {NoStop}%
\bibitem [{\citenamefont {Satterthwaite}\ and\ \citenamefont
  {Toepke}(1970)}]{Satterthwaite1970}%
  \BibitemOpen
  \bibfield  {author} {\bibinfo {author} {\bibfnamefont {C.~B.}\ \bibnamefont
  {Satterthwaite}}\ and\ \bibinfo {author} {\bibfnamefont {I.~L.}\ \bibnamefont
  {Toepke}},\ }\bibfield  {title} {\enquote {\bibinfo {title}
  {Superconductivity of hydrides and deuterides of thorium},}\ }\href {\doibase
  10.1103/PhysRevLett.25.741} {\bibfield  {journal} {\bibinfo  {journal} {Phys.
  Rev. Lett.}\ }\textbf {\bibinfo {volume} {25}},\ \bibinfo {pages} {741}
  (\bibinfo {year} {1970})}\BibitemShut {NoStop}%
\bibitem [{\citenamefont {Goncharenko}\ \emph {et~al.}(2008)\citenamefont
  {Goncharenko}, \citenamefont {Eremets}, \citenamefont {Hanfland},
  \citenamefont {Tse}, \citenamefont {Amboage}, \citenamefont {Yao},\ and\
  \citenamefont {Trojan}}]{goncharenko2008}%
  \BibitemOpen
  \bibfield  {author} {\bibinfo {author} {\bibfnamefont {I.}~\bibnamefont
  {Goncharenko}}, \bibinfo {author} {\bibfnamefont {M.~I.}\ \bibnamefont
  {Eremets}}, \bibinfo {author} {\bibfnamefont {M.}~\bibnamefont {Hanfland}},
  \bibinfo {author} {\bibfnamefont {J.~S.}\ \bibnamefont {Tse}}, \bibinfo
  {author} {\bibfnamefont {M.}~\bibnamefont {Amboage}}, \bibinfo {author}
  {\bibfnamefont {Y.}~\bibnamefont {Yao}}, \ and\ \bibinfo {author}
  {\bibfnamefont {I.~A.}\ \bibnamefont {Trojan}},\ }\bibfield  {title}
  {\enquote {\bibinfo {title} {Pressure-induced hydrogen-dominant metallic
  state in aluminum hydride},}\ }\href {\doibase
  10.1103/PhysRevLett.100.045504} {\bibfield  {journal} {\bibinfo  {journal}
  {Phys. Rev. Lett.}\ }\textbf {\bibinfo {volume} {100}},\ \bibinfo {pages}
  {045504} (\bibinfo {year} {2008})}\BibitemShut {NoStop}%
\bibitem [{\citenamefont {Eremets}\ \emph {et~al.}(2008)\citenamefont
  {Eremets}, \citenamefont {Trojan}, \citenamefont {Medvedev}, \citenamefont
  {Tse},\ and\ \citenamefont {Yao}}]{eremets2008}%
  \BibitemOpen
  \bibfield  {author} {\bibinfo {author} {\bibfnamefont {M.~I.}\ \bibnamefont
  {Eremets}}, \bibinfo {author} {\bibfnamefont {I.~A.}\ \bibnamefont {Trojan}},
  \bibinfo {author} {\bibfnamefont {S.~A.}\ \bibnamefont {Medvedev}}, \bibinfo
  {author} {\bibfnamefont {J.~S.}\ \bibnamefont {Tse}}, \ and\ \bibinfo
  {author} {\bibfnamefont {Y.}~\bibnamefont {Yao}},\ }\bibfield  {title}
  {\enquote {\bibinfo {title} {Superconductivity in hydrogen dominant
  materials: Silane},}\ }\href {\doibase 10.1126/science.1153282} {\bibfield
  {journal} {\bibinfo  {journal} {Science}\ }\textbf {\bibinfo {volume}
  {319}},\ \bibinfo {pages} {1506} (\bibinfo {year} {2008})}\BibitemShut
  {NoStop}%
\bibitem [{\citenamefont {Drozdov}\ \emph {et~al.}(2015)\citenamefont
  {Drozdov}, \citenamefont {Eremets}, \citenamefont {Troyan}, \citenamefont
  {Ksenofontov},\ and\ \citenamefont {Shylin}}]{drozdov2015}%
  \BibitemOpen
  \bibfield  {author} {\bibinfo {author} {\bibfnamefont {A.~P.}\ \bibnamefont
  {Drozdov}}, \bibinfo {author} {\bibfnamefont {M.~I.}\ \bibnamefont
  {Eremets}}, \bibinfo {author} {\bibfnamefont {I.~A.}\ \bibnamefont {Troyan}},
  \bibinfo {author} {\bibfnamefont {V.}~\bibnamefont {Ksenofontov}}, \ and\
  \bibinfo {author} {\bibfnamefont {S.~I.}\ \bibnamefont {Shylin}},\ }\bibfield
   {title} {\enquote {\bibinfo {title} {Conventional superconductivity at 203
  kelvin at high pressures in the sulfur hydride system},}\ }\href
  {http://dx.doi.org/10.1038/nature14964} {\bibfield  {journal} {\bibinfo
  {journal} {Nature}\ }\textbf {\bibinfo {volume} {525}},\ \bibinfo {pages}
  {73} (\bibinfo {year} {2015})}\BibitemShut {NoStop}%
\bibitem [{\citenamefont {Drozdov}\ \emph {et~al.}(2019)\citenamefont
  {Drozdov}, \citenamefont {Kong}, \citenamefont {Minkov}, \citenamefont
  {Besedin}, \citenamefont {Kuzovnikov}, \citenamefont {Mozaffari},
  \citenamefont {Balicas}, \citenamefont {Balakirev}, \citenamefont {Graf},
  \citenamefont {Prakapenka}, \citenamefont {Greenberg}, \citenamefont
  {Knyazev}, \citenamefont {Tkacz},\ and\ \citenamefont
  {Eremets}}]{drozdov2019}%
  \BibitemOpen
  \bibfield  {author} {\bibinfo {author} {\bibfnamefont {A.~P.}\ \bibnamefont
  {Drozdov}}, \bibinfo {author} {\bibfnamefont {P.~P.}\ \bibnamefont {Kong}},
  \bibinfo {author} {\bibfnamefont {V.~S.}\ \bibnamefont {Minkov}}, \bibinfo
  {author} {\bibfnamefont {S.~P.}\ \bibnamefont {Besedin}}, \bibinfo {author}
  {\bibfnamefont {M.~A.}\ \bibnamefont {Kuzovnikov}}, \bibinfo {author}
  {\bibfnamefont {S.}~\bibnamefont {Mozaffari}}, \bibinfo {author}
  {\bibfnamefont {L.}~\bibnamefont {Balicas}}, \bibinfo {author} {\bibfnamefont
  {F.~F.}\ \bibnamefont {Balakirev}}, \bibinfo {author} {\bibfnamefont {D.~E.}\
  \bibnamefont {Graf}}, \bibinfo {author} {\bibfnamefont {V.~B.}\ \bibnamefont
  {Prakapenka}}, \bibinfo {author} {\bibfnamefont {E.}~\bibnamefont
  {Greenberg}}, \bibinfo {author} {\bibfnamefont {D.~A.}\ \bibnamefont
  {Knyazev}}, \bibinfo {author} {\bibfnamefont {M.}~\bibnamefont {Tkacz}}, \
  and\ \bibinfo {author} {\bibfnamefont {M.~I.}\ \bibnamefont {Eremets}},\
  }\bibfield  {title} {\enquote {\bibinfo {title} {Superconductivity at 250 k
  in lanthanum hydride under high pressures},}\ }\href {\doibase
  10.1038/s41586-019-1201-8} {\bibfield  {journal} {\bibinfo  {journal}
  {Nature}\ }\textbf {\bibinfo {volume} {569}},\ \bibinfo {pages} {528}
  (\bibinfo {year} {2019})}\BibitemShut {NoStop}%
\bibitem [{\citenamefont {Kong}\ \emph {et~al.}(2021)\citenamefont {Kong},
  \citenamefont {Minkov}, \citenamefont {Kuzovnikov}, \citenamefont {Drozdov},
  \citenamefont {Besedin}, \citenamefont {Mozaffari}, \citenamefont {Balicas},
  \citenamefont {Balakirev}, \citenamefont {Prakapenka}, \citenamefont
  {Chariton}, \citenamefont {Knyazev}, \citenamefont {Greenberg},\ and\
  \citenamefont {Eremets}}]{kong2021}%
  \BibitemOpen
  \bibfield  {author} {\bibinfo {author} {\bibfnamefont {P.}~\bibnamefont
  {Kong}}, \bibinfo {author} {\bibfnamefont {V.~S.}\ \bibnamefont {Minkov}},
  \bibinfo {author} {\bibfnamefont {M.~A.}\ \bibnamefont {Kuzovnikov}},
  \bibinfo {author} {\bibfnamefont {A.~P.}\ \bibnamefont {Drozdov}}, \bibinfo
  {author} {\bibfnamefont {S.~P.}\ \bibnamefont {Besedin}}, \bibinfo {author}
  {\bibfnamefont {S.}~\bibnamefont {Mozaffari}}, \bibinfo {author}
  {\bibfnamefont {L.}~\bibnamefont {Balicas}}, \bibinfo {author} {\bibfnamefont
  {F.~F.}\ \bibnamefont {Balakirev}}, \bibinfo {author} {\bibfnamefont {V.~B.}\
  \bibnamefont {Prakapenka}}, \bibinfo {author} {\bibfnamefont
  {S.}~\bibnamefont {Chariton}}, \bibinfo {author} {\bibfnamefont {D.~A.}\
  \bibnamefont {Knyazev}}, \bibinfo {author} {\bibfnamefont {E.}~\bibnamefont
  {Greenberg}}, \ and\ \bibinfo {author} {\bibfnamefont {M.~I.}\ \bibnamefont
  {Eremets}},\ }\bibfield  {title} {\enquote {\bibinfo {title}
  {Superconductivity up to 243 k in the yttrium-hydrogen system under high
  pressure},}\ }\href {\doibase 10.1038/s41467-021-25372-2} {\bibfield
  {journal} {\bibinfo  {journal} {Nature Communications}\ }\textbf {\bibinfo
  {volume} {12}},\ \bibinfo {pages} {5075} (\bibinfo {year}
  {2021})}\BibitemShut {NoStop}%
\bibitem [{\citenamefont {Troyan}\ \emph {et~al.}(2021)\citenamefont {Troyan},
  \citenamefont {Semenok}, \citenamefont {Kvashnin}, \citenamefont {Sadakov},
  \citenamefont {Sobolevskiy}, \citenamefont {Pudalov}, \citenamefont
  {Ivanova}, \citenamefont {Prakapenka}, \citenamefont {Greenberg},
  \citenamefont {Gavriliuk}, \citenamefont {Lyubutin}, \citenamefont
  {Struzhkin}, \citenamefont {Bergara}, \citenamefont {Errea}, \citenamefont
  {Bianco}, \citenamefont {Calandra}, \citenamefont {Mauri}, \citenamefont
  {Monacelli}, \citenamefont {Akashi},\ and\ \citenamefont
  {Oganov}}]{troyan2021}%
  \BibitemOpen
  \bibfield  {author} {\bibinfo {author} {\bibfnamefont {I.~A.}\ \bibnamefont
  {Troyan}}, \bibinfo {author} {\bibfnamefont {D.~V.}\ \bibnamefont {Semenok}},
  \bibinfo {author} {\bibfnamefont {A.~G.}\ \bibnamefont {Kvashnin}}, \bibinfo
  {author} {\bibfnamefont {A.~V.}\ \bibnamefont {Sadakov}}, \bibinfo {author}
  {\bibfnamefont {O.~A.}\ \bibnamefont {Sobolevskiy}}, \bibinfo {author}
  {\bibfnamefont {V.~M.}\ \bibnamefont {Pudalov}}, \bibinfo {author}
  {\bibfnamefont {A.~G.}\ \bibnamefont {Ivanova}}, \bibinfo {author}
  {\bibfnamefont {V.~B.}\ \bibnamefont {Prakapenka}}, \bibinfo {author}
  {\bibfnamefont {E.}~\bibnamefont {Greenberg}}, \bibinfo {author}
  {\bibfnamefont {A.~G.}\ \bibnamefont {Gavriliuk}}, \bibinfo {author}
  {\bibfnamefont {I.~S.}\ \bibnamefont {Lyubutin}}, \bibinfo {author}
  {\bibfnamefont {V.~V.}\ \bibnamefont {Struzhkin}}, \bibinfo {author}
  {\bibfnamefont {A.}~\bibnamefont {Bergara}}, \bibinfo {author} {\bibfnamefont
  {I.}~\bibnamefont {Errea}}, \bibinfo {author} {\bibfnamefont
  {R.}~\bibnamefont {Bianco}}, \bibinfo {author} {\bibfnamefont
  {M.}~\bibnamefont {Calandra}}, \bibinfo {author} {\bibfnamefont
  {F.}~\bibnamefont {Mauri}}, \bibinfo {author} {\bibfnamefont
  {L.}~\bibnamefont {Monacelli}}, \bibinfo {author} {\bibfnamefont
  {R.}~\bibnamefont {Akashi}}, \ and\ \bibinfo {author} {\bibfnamefont {A.~R.}\
  \bibnamefont {Oganov}},\ }\bibfield  {title} {\enquote {\bibinfo {title}
  {Anomalous high-temperature superconductivity in yh6},}\ }\href {\doibase
  https://doi.org/10.1002/adma.202006832} {\bibfield  {journal} {\bibinfo
  {journal} {Advanced Materials}\ }\textbf {\bibinfo {volume} {33}},\ \bibinfo
  {pages} {2006832} (\bibinfo {year} {2021})}\BibitemShut {NoStop}%
\bibitem [{\citenamefont {Ma}\ \emph {et~al.}(2022)\citenamefont {Ma},
  \citenamefont {Wang}, \citenamefont {Xie}, \citenamefont {Yang},
  \citenamefont {Wang}, \citenamefont {Zhou}, \citenamefont {Liu},
  \citenamefont {Yu}, \citenamefont {Zhao}, \citenamefont {Wang}, \citenamefont
  {Liu},\ and\ \citenamefont {Ma}}]{ma2022}%
  \BibitemOpen
  \bibfield  {author} {\bibinfo {author} {\bibfnamefont {L.}~\bibnamefont
  {Ma}}, \bibinfo {author} {\bibfnamefont {K.}~\bibnamefont {Wang}}, \bibinfo
  {author} {\bibfnamefont {Y.}~\bibnamefont {Xie}}, \bibinfo {author}
  {\bibfnamefont {X.}~\bibnamefont {Yang}}, \bibinfo {author} {\bibfnamefont
  {Y.}~\bibnamefont {Wang}}, \bibinfo {author} {\bibfnamefont {M.}~\bibnamefont
  {Zhou}}, \bibinfo {author} {\bibfnamefont {H.}~\bibnamefont {Liu}}, \bibinfo
  {author} {\bibfnamefont {X.}~\bibnamefont {Yu}}, \bibinfo {author}
  {\bibfnamefont {Y.}~\bibnamefont {Zhao}}, \bibinfo {author} {\bibfnamefont
  {H.}~\bibnamefont {Wang}}, \bibinfo {author} {\bibfnamefont {G.}~\bibnamefont
  {Liu}}, \ and\ \bibinfo {author} {\bibfnamefont {Y.}~\bibnamefont {Ma}},\
  }\bibfield  {title} {\enquote {\bibinfo {title} {High-temperature
  superconducting phase in clathrate calcium hydride ${\mathrm{cah}}_{6}$ up to
  215 k at a pressure of 172 gpa},}\ }\href {\doibase
  10.1103/PhysRevLett.128.167001} {\bibfield  {journal} {\bibinfo  {journal}
  {Phys. Rev. Lett.}\ }\textbf {\bibinfo {volume} {128}},\ \bibinfo {pages}
  {167001} (\bibinfo {year} {2022})}\BibitemShut {NoStop}%
\bibitem [{\citenamefont {Bi}\ \emph {et~al.}(2019)\citenamefont {Bi},
  \citenamefont {Zarifi}, \citenamefont {Terpstra},\ and\ \citenamefont
  {Zurek}}]{bi2019}%
  \BibitemOpen
  \bibfield  {author} {\bibinfo {author} {\bibfnamefont {T.}~\bibnamefont
  {Bi}}, \bibinfo {author} {\bibfnamefont {N.}~\bibnamefont {Zarifi}}, \bibinfo
  {author} {\bibfnamefont {T.}~\bibnamefont {Terpstra}}, \ and\ \bibinfo
  {author} {\bibfnamefont {E.}~\bibnamefont {Zurek}},\ }\bibfield  {title}
  {\enquote {\bibinfo {title} {The search for superconductivity in high
  pressure hydrides},}\ }in\ \href {\doibase
  https://doi.org/10.1016/B978-0-12-409547-2.11435-0} {\emph {\bibinfo
  {booktitle} {Reference Module in Chemistry, Molecular Sciences and Chemical
  Engineering}}}\ (\bibinfo  {publisher} {Elsevier},\ \bibinfo {year}
  {2019})\BibitemShut {NoStop}%
\bibitem [{\citenamefont {Flores-Livas}\ \emph {et~al.}(2020)\citenamefont
  {Flores-Livas}, \citenamefont {Boeri}, \citenamefont {Sanna}, \citenamefont
  {Profeta}, \citenamefont {Arita},\ and\ \citenamefont
  {Eremets}}]{Flores-Livas2020A}%
  \BibitemOpen
  \bibfield  {author} {\bibinfo {author} {\bibfnamefont {J.~A.}\ \bibnamefont
  {Flores-Livas}}, \bibinfo {author} {\bibfnamefont {L.}~\bibnamefont {Boeri}},
  \bibinfo {author} {\bibfnamefont {A.}~\bibnamefont {Sanna}}, \bibinfo
  {author} {\bibfnamefont {G.}~\bibnamefont {Profeta}}, \bibinfo {author}
  {\bibfnamefont {R.}~\bibnamefont {Arita}}, \ and\ \bibinfo {author}
  {\bibfnamefont {M.}~\bibnamefont {Eremets}},\ }\bibfield  {title} {\enquote
  {\bibinfo {title} {A perspective on conventional high-temperature
  superconductors at high pressure: Methods and materials},}\ }\href {\doibase
  https://doi.org/10.1016/j.physrep.2020.02.003} {\bibfield  {journal}
  {\bibinfo  {journal} {Physics Reports}\ }\textbf {\bibinfo {volume} {856}},\
  \bibinfo {pages} {1} (\bibinfo {year} {2020})}\BibitemShut {NoStop}%
\bibitem [{\citenamefont {Pickard}\ \emph {et~al.}(2020)\citenamefont
  {Pickard}, \citenamefont {Errea},\ and\ \citenamefont
  {Eremets}}]{Pickard2020}%
  \BibitemOpen
  \bibfield  {author} {\bibinfo {author} {\bibfnamefont {C.~J.}\ \bibnamefont
  {Pickard}}, \bibinfo {author} {\bibfnamefont {I.}~\bibnamefont {Errea}}, \
  and\ \bibinfo {author} {\bibfnamefont {M.~I.}\ \bibnamefont {Eremets}},\
  }\bibfield  {title} {\enquote {\bibinfo {title} {Superconducting hydrides
  under pressure},}\ }\href {\doibase 10.1146/annurev-conmatphys-031218-013413}
  {\bibfield  {journal} {\bibinfo  {journal} {Annu. Rev. Condens. Matter
  Phys.}\ }\textbf {\bibinfo {volume} {11}},\ \bibinfo {pages} {57} (\bibinfo
  {year} {2020})}\BibitemShut {NoStop}%
\bibitem [{\citenamefont {Semenok}\ \emph
  {et~al.}(2020{\natexlab{a}})\citenamefont {Semenok}, \citenamefont {Kruglov},
  \citenamefont {Savkin}, \citenamefont {Kvashnin},\ and\ \citenamefont
  {Oganov}}]{semenok2020A}%
  \BibitemOpen
  \bibfield  {author} {\bibinfo {author} {\bibfnamefont {D.~V.}\ \bibnamefont
  {Semenok}}, \bibinfo {author} {\bibfnamefont {I.~A.}\ \bibnamefont
  {Kruglov}}, \bibinfo {author} {\bibfnamefont {I.~A.}\ \bibnamefont {Savkin}},
  \bibinfo {author} {\bibfnamefont {A.~G.}\ \bibnamefont {Kvashnin}}, \ and\
  \bibinfo {author} {\bibfnamefont {A.~R.}\ \bibnamefont {Oganov}},\ }\bibfield
   {title} {\enquote {\bibinfo {title} {On distribution of superconductivity in
  metal hydrides},}\ }\href {\doibase
  https://doi.org/10.1016/j.cossms.2020.100808} {\bibfield  {journal} {\bibinfo
   {journal} {Current Opinion in Solid State and Materials Science}\ }\textbf
  {\bibinfo {volume} {24}},\ \bibinfo {pages} {100808} (\bibinfo {year}
  {2020}{\natexlab{a}})}\BibitemShut {NoStop}%
\bibitem [{\citenamefont {Ishikawa}\ \emph {et~al.}(2019)\citenamefont
  {Ishikawa}, \citenamefont {Miyake},\ and\ \citenamefont
  {Shimizu}}]{Ishikawa2019A}%
  \BibitemOpen
  \bibfield  {author} {\bibinfo {author} {\bibfnamefont {T.}~\bibnamefont
  {Ishikawa}}, \bibinfo {author} {\bibfnamefont {T.}~\bibnamefont {Miyake}}, \
  and\ \bibinfo {author} {\bibfnamefont {K.}~\bibnamefont {Shimizu}},\
  }\bibfield  {title} {\enquote {\bibinfo {title} {Materials informatics based
  on evolutionary algorithms: Application to search for superconducting
  hydrogen compounds},}\ }\href {\doibase 10.1103/PhysRevB.100.174506}
  {\bibfield  {journal} {\bibinfo  {journal} {Phys. Rev. B}\ }\textbf {\bibinfo
  {volume} {100}},\ \bibinfo {pages} {174506} (\bibinfo {year}
  {2019})}\BibitemShut {NoStop}%
\bibitem [{\citenamefont {Belli}\ \emph {et~al.}(2021)\citenamefont {Belli},
  \citenamefont {Novoa}, \citenamefont {Contreras-Garcia},\ and\ \citenamefont
  {Errea}}]{Belli2021}%
  \BibitemOpen
  \bibfield  {author} {\bibinfo {author} {\bibfnamefont {F.}~\bibnamefont
  {Belli}}, \bibinfo {author} {\bibfnamefont {T.}~\bibnamefont {Novoa}},
  \bibinfo {author} {\bibfnamefont {J.}~\bibnamefont {Contreras-Garcia}}, \
  and\ \bibinfo {author} {\bibfnamefont {I.}~\bibnamefont {Errea}},\ }\bibfield
   {title} {\enquote {\bibinfo {title} {Strong correlation between electronic
  bonding network and critical temperature in hydrogen-based
  superconductors},}\ }\href {\doibase 10.1038/s41467-021-25687-0} {\bibfield
  {journal} {\bibinfo  {journal} {Nature Commun.}\ }\textbf {\bibinfo {volume}
  {12}},\ \bibinfo {pages} {5381} (\bibinfo {year} {2021})}\BibitemShut
  {NoStop}%
\bibitem [{\citenamefont {Shen}\ and\ \citenamefont {Mao}(2016)}]{Shen2016A}%
  \BibitemOpen
  \bibfield  {author} {\bibinfo {author} {\bibfnamefont {G.}~\bibnamefont
  {Shen}}\ and\ \bibinfo {author} {\bibfnamefont {H.~K.}\ \bibnamefont {Mao}},\
  }\bibfield  {title} {\enquote {\bibinfo {title} {High-pressure studies with
  x-rays using diamond anvil cells},}\ }\href {\doibase
  10.1088/1361-6633/80/1/016101} {\bibfield  {journal} {\bibinfo  {journal}
  {Reports on Progress in Physics}\ }\textbf {\bibinfo {volume} {80}},\
  \bibinfo {pages} {016101} (\bibinfo {year} {2016})}\BibitemShut {NoStop}%
\bibitem [{\citenamefont {Bassett}(2009)}]{Bassett2009}%
  \BibitemOpen
  \bibfield  {author} {\bibinfo {author} {\bibfnamefont {W.~A.}\ \bibnamefont
  {Bassett}},\ }\bibfield  {title} {\enquote {\bibinfo {title} {Diamond anvil
  cell, 50th birthday},}\ }\href {\doibase 10.1080/08957950802597239}
  {\bibfield  {journal} {\bibinfo  {journal} {High Pressure Research}\ }\textbf
  {\bibinfo {volume} {29}},\ \bibinfo {pages} {163} (\bibinfo {year}
  {2009})}\BibitemShut {NoStop}%
\bibitem [{\citenamefont {Jayaraman}(1983)}]{Jayaraman1983}%
  \BibitemOpen
  \bibfield  {author} {\bibinfo {author} {\bibfnamefont {A.}~\bibnamefont
  {Jayaraman}},\ }\bibfield  {title} {\enquote {\bibinfo {title} {Diamond anvil
  cell and high-pressure physical investigations},}\ }\href {\doibase
  10.1103/RevModPhys.55.65} {\bibfield  {journal} {\bibinfo  {journal} {Rev.
  Mod. Phys.}\ }\textbf {\bibinfo {volume} {55}},\ \bibinfo {pages} {65}
  (\bibinfo {year} {1983})}\BibitemShut {NoStop}%
\bibitem [{\citenamefont {Sakata}\ \emph {et~al.}(2020)\citenamefont {Sakata},
  \citenamefont {Einaga}, \citenamefont {Dezhong}, \citenamefont {Sato},
  \citenamefont {Orimo},\ and\ \citenamefont {Shimizu}}]{sakata2020}%
  \BibitemOpen
  \bibfield  {author} {\bibinfo {author} {\bibfnamefont {M.}~\bibnamefont
  {Sakata}}, \bibinfo {author} {\bibfnamefont {M.}~\bibnamefont {Einaga}},
  \bibinfo {author} {\bibfnamefont {M.}~\bibnamefont {Dezhong}}, \bibinfo
  {author} {\bibfnamefont {T.}~\bibnamefont {Sato}}, \bibinfo {author}
  {\bibfnamefont {S.}~\bibnamefont {Orimo}}, \ and\ \bibinfo {author}
  {\bibfnamefont {K.}~\bibnamefont {Shimizu}},\ }\bibfield  {title} {\enquote
  {\bibinfo {title} {Superconductivity of lanthanum hydride synthesized using
  alh3 as a hydrogen source},}\ }\href {\doibase 10.1088/1361-6668/abb204}
  {\bibfield  {journal} {\bibinfo  {journal} {Supercond. Sci. Technol.}\
  }\textbf {\bibinfo {volume} {33}},\ \bibinfo {pages} {114004} (\bibinfo
  {year} {2020})}\BibitemShut {NoStop}%
\bibitem [{\citenamefont {Bhattacharyya}\ \emph {et~al.}(2024)\citenamefont
  {Bhattacharyya}, \citenamefont {Chen}, \citenamefont {Huang}, \citenamefont
  {Chatterjee}, \citenamefont {Huang}, \citenamefont {Kobrin}, \citenamefont
  {Lyu}, \citenamefont {Smart}, \citenamefont {Block}, \citenamefont {Wang},
  \citenamefont {Wang}, \citenamefont {Wu}, \citenamefont {Hsieh},
  \citenamefont {Ma}, \citenamefont {Mandyam}, \citenamefont {Chen},
  \citenamefont {Davis}, \citenamefont {Geballe}, \citenamefont {Zu},
  \citenamefont {Struzhkin}, \citenamefont {Jeanloz}, \citenamefont {Moore},
  \citenamefont {Cui}, \citenamefont {Galli}, \citenamefont {Halperin},
  \citenamefont {Laumann},\ and\ \citenamefont {Yao}}]{bhattacharyya2024}%
  \BibitemOpen
  \bibfield  {author} {\bibinfo {author} {\bibfnamefont {P.}~\bibnamefont
  {Bhattacharyya}}, \bibinfo {author} {\bibfnamefont {W.}~\bibnamefont {Chen}},
  \bibinfo {author} {\bibfnamefont {X.}~\bibnamefont {Huang}}, \bibinfo
  {author} {\bibfnamefont {S.}~\bibnamefont {Chatterjee}}, \bibinfo {author}
  {\bibfnamefont {B.}~\bibnamefont {Huang}}, \bibinfo {author} {\bibfnamefont
  {B.}~\bibnamefont {Kobrin}}, \bibinfo {author} {\bibfnamefont
  {Y.}~\bibnamefont {Lyu}}, \bibinfo {author} {\bibfnamefont {T.~J.}\
  \bibnamefont {Smart}}, \bibinfo {author} {\bibfnamefont {M.}~\bibnamefont
  {Block}}, \bibinfo {author} {\bibfnamefont {E.}~\bibnamefont {Wang}},
  \bibinfo {author} {\bibfnamefont {Z.}~\bibnamefont {Wang}}, \bibinfo {author}
  {\bibfnamefont {W.}~\bibnamefont {Wu}}, \bibinfo {author} {\bibfnamefont
  {S.}~\bibnamefont {Hsieh}}, \bibinfo {author} {\bibfnamefont
  {H.}~\bibnamefont {Ma}}, \bibinfo {author} {\bibfnamefont {S.}~\bibnamefont
  {Mandyam}}, \bibinfo {author} {\bibfnamefont {B.}~\bibnamefont {Chen}},
  \bibinfo {author} {\bibfnamefont {E.}~\bibnamefont {Davis}}, \bibinfo
  {author} {\bibfnamefont {Z.~M.}\ \bibnamefont {Geballe}}, \bibinfo {author}
  {\bibfnamefont {C.}~\bibnamefont {Zu}}, \bibinfo {author} {\bibfnamefont
  {V.}~\bibnamefont {Struzhkin}}, \bibinfo {author} {\bibfnamefont
  {R.}~\bibnamefont {Jeanloz}}, \bibinfo {author} {\bibfnamefont {J.~E.}\
  \bibnamefont {Moore}}, \bibinfo {author} {\bibfnamefont {T.}~\bibnamefont
  {Cui}}, \bibinfo {author} {\bibfnamefont {G.}~\bibnamefont {Galli}}, \bibinfo
  {author} {\bibfnamefont {B.~I.}\ \bibnamefont {Halperin}}, \bibinfo {author}
  {\bibfnamefont {C.~R.}\ \bibnamefont {Laumann}}, \ and\ \bibinfo {author}
  {\bibfnamefont {N.~Y.}\ \bibnamefont {Yao}},\ }\bibfield  {title} {\enquote
  {\bibinfo {title} {Imaging the meissner effect in hydride superconductors
  using quantum sensors},}\ }\href {\doibase 10.1038/s41586-024-07026-7}
  {\bibfield  {journal} {\bibinfo  {journal} {Nature}\ }\textbf {\bibinfo
  {volume} {627}},\ \bibinfo {pages} {73} (\bibinfo {year} {2024})}\BibitemShut
  {NoStop}%
\bibitem [{\citenamefont {Minkov}\ \emph {et~al.}(0243)\citenamefont {Minkov},
  \citenamefont {Ksenofontov}, \citenamefont {Bud'ko}, \citenamefont
  {Talantsev},\ and\ \citenamefont {Eremets}}]{minkov2023}%
  \BibitemOpen
  \bibfield  {author} {\bibinfo {author} {\bibfnamefont {V.~S.}\ \bibnamefont
  {Minkov}}, \bibinfo {author} {\bibfnamefont {V.}~\bibnamefont {Ksenofontov}},
  \bibinfo {author} {\bibfnamefont {S.~L.}\ \bibnamefont {Bud'ko}}, \bibinfo
  {author} {\bibfnamefont {E.~F.}\ \bibnamefont {Talantsev}}, \ and\ \bibinfo
  {author} {\bibfnamefont {M.~I.}\ \bibnamefont {Eremets}},\ }\bibfield
  {title} {\enquote {\bibinfo {title} {Magnetic flux trapping in hydrogen-rich
  high-temperature superconductors},}\ }\href {\doibase
  10.1038/s41567-023-02089-1} {\bibfield  {journal} {\bibinfo  {journal} {Nat.
  Phys.}\ }\textbf {\bibinfo {volume} {19}},\ \bibinfo {pages} {1293} (\bibinfo
  {year} {20243})}\BibitemShut {NoStop}%
\bibitem [{\citenamefont {Gebreyohannes}\ \emph
  {et~al.}(2022{\natexlab{a}})\citenamefont {Gebreyohannes}, \citenamefont
  {Geffe},\ and\ \citenamefont {Singh}}]{gebreyohannes2022}%
  \BibitemOpen
  \bibfield  {author} {\bibinfo {author} {\bibfnamefont {M.~G.}\ \bibnamefont
  {Gebreyohannes}}, \bibinfo {author} {\bibfnamefont {C.~A.}\ \bibnamefont
  {Geffe}}, \ and\ \bibinfo {author} {\bibfnamefont {P.}~\bibnamefont
  {Singh}},\ }\bibfield  {title} {\enquote {\bibinfo {title} {Computational
  prediction of new stable superconducting magnesium hydrides at
  high-pressures},}\ }\href {\doibase
  https://doi.org/10.1016/j.physc.2022.1354052} {\bibfield  {journal} {\bibinfo
   {journal} {Physica C: Superconductivity and its Applications}\ }\textbf
  {\bibinfo {volume} {599}},\ \bibinfo {pages} {1354052} (\bibinfo {year}
  {2022}{\natexlab{a}})}\BibitemShut {NoStop}%
\bibitem [{\citenamefont {Wu}\ \emph {et~al.}(2023)\citenamefont {Wu},
  \citenamefont {Sun}, \citenamefont {Durajski}, \citenamefont {Zheng},
  \citenamefont {Antropov}, \citenamefont {Ho},\ and\ \citenamefont
  {Wu}}]{PhysRevMaterials.7.L101801}%
  \BibitemOpen
  \bibfield  {author} {\bibinfo {author} {\bibfnamefont {Z.}~\bibnamefont
  {Wu}}, \bibinfo {author} {\bibfnamefont {Y.}~\bibnamefont {Sun}}, \bibinfo
  {author} {\bibfnamefont {A.~P.}\ \bibnamefont {Durajski}}, \bibinfo {author}
  {\bibfnamefont {F.}~\bibnamefont {Zheng}}, \bibinfo {author} {\bibfnamefont
  {V.}~\bibnamefont {Antropov}}, \bibinfo {author} {\bibfnamefont {K.-M.}\
  \bibnamefont {Ho}}, \ and\ \bibinfo {author} {\bibfnamefont {S.}~\bibnamefont
  {Wu}},\ }\bibfield  {title} {\enquote {\bibinfo {title} {Effect of doping on
  the phase stability and superconductivity in
  $\mathrm{La}{\mathrm{h}}_{10}$},}\ }\href {\doibase
  10.1103/PhysRevMaterials.7.L101801} {\bibfield  {journal} {\bibinfo
  {journal} {Phys. Rev. Mater.}\ }\textbf {\bibinfo {volume} {7}},\ \bibinfo
  {pages} {L101801} (\bibinfo {year} {2023})}\BibitemShut {NoStop}%
\bibitem [{zen()}]{zenodo}%
  \BibitemOpen
  \href {https://zenodo.org/records/XXXX} {}\bibinfo {note} {The database is
  available at: \url{https://zenodo.org/records/XXXX}}\BibitemShut {NoStop}%
\bibitem [{\citenamefont {Abe}\ and\ \citenamefont {Ashcroft}(2011)}]{abe2011}%
  \BibitemOpen
  \bibfield  {author} {\bibinfo {author} {\bibfnamefont {K.}~\bibnamefont
  {Abe}}\ and\ \bibinfo {author} {\bibfnamefont {N.~W.}\ \bibnamefont
  {Ashcroft}},\ }\bibfield  {title} {\enquote {\bibinfo {title} {Crystalline
  diborane at high pressures},}\ }\href {\doibase 10.1103/PhysRevB.84.104118}
  {\bibfield  {journal} {\bibinfo  {journal} {Phys. Rev. B}\ }\textbf {\bibinfo
  {volume} {84}},\ \bibinfo {pages} {104118} (\bibinfo {year}
  {2011})}\BibitemShut {NoStop}%
\bibitem [{\citenamefont {Abe}\ and\ \citenamefont {Ashcroft}(2013)}]{abe2013}%
  \BibitemOpen
  \bibfield  {author} {\bibinfo {author} {\bibfnamefont {K.}~\bibnamefont
  {Abe}}\ and\ \bibinfo {author} {\bibfnamefont {N.~W.}\ \bibnamefont
  {Ashcroft}},\ }\bibfield  {title} {\enquote {\bibinfo {title} {Quantum
  disproportionation: The high hydrides at elevated pressures},}\ }\href
  {\doibase 10.1103/PhysRevB.88.174110} {\bibfield  {journal} {\bibinfo
  {journal} {Phys. Rev. B}\ }\textbf {\bibinfo {volume} {88}},\ \bibinfo
  {pages} {174110} (\bibinfo {year} {2013})}\BibitemShut {NoStop}%
\bibitem [{\citenamefont {Abe}\ and\ \citenamefont {Ashcroft}(2015)}]{abe2015}%
  \BibitemOpen
  \bibfield  {author} {\bibinfo {author} {\bibfnamefont {K.}~\bibnamefont
  {Abe}}\ and\ \bibinfo {author} {\bibfnamefont {N.~W.}\ \bibnamefont
  {Ashcroft}},\ }\bibfield  {title} {\enquote {\bibinfo {title} {Stabilization
  and highly metallic properties of heavy group-v hydrides at high
  pressures},}\ }\href {\doibase 10.1103/PhysRevB.92.224109} {\bibfield
  {journal} {\bibinfo  {journal} {Phys. Rev. B}\ }\textbf {\bibinfo {volume}
  {92}},\ \bibinfo {pages} {224109} (\bibinfo {year} {2015})}\BibitemShut
  {NoStop}%
\bibitem [{\citenamefont {Abe}(2017)}]{abe2017}%
  \BibitemOpen
  \bibfield  {author} {\bibinfo {author} {\bibfnamefont {K.}~\bibnamefont
  {Abe}},\ }\bibfield  {title} {\enquote {\bibinfo {title} {Hydrogen-rich
  scandium compounds at high pressures},}\ }\href {\doibase
  10.1103/PhysRevB.96.144108} {\bibfield  {journal} {\bibinfo  {journal} {Phys.
  Rev. B}\ }\textbf {\bibinfo {volume} {96}},\ \bibinfo {pages} {144108}
  (\bibinfo {year} {2017})}\BibitemShut {NoStop}%
\bibitem [{\citenamefont {Abe}(2018{\natexlab{a}})}]{abe2018}%
  \BibitemOpen
  \bibfield  {author} {\bibinfo {author} {\bibfnamefont {K.}~\bibnamefont
  {Abe}},\ }\bibfield  {title} {\enquote {\bibinfo {title} {High-pressure
  properties of dense metallic zirconium hydrides studied by ab initio
  calculations},}\ }\href {\doibase 10.1103/PhysRevB.98.134103} {\bibfield
  {journal} {\bibinfo  {journal} {Phys. Rev. B}\ }\textbf {\bibinfo {volume}
  {98}},\ \bibinfo {pages} {134103} (\bibinfo {year}
  {2018}{\natexlab{a}})}\BibitemShut {NoStop}%
\bibitem [{\citenamefont {Abe}(2018{\natexlab{b}})}]{abe2018A}%
  \BibitemOpen
  \bibfield  {author} {\bibinfo {author} {\bibfnamefont {K.}~\bibnamefont
  {Abe}},\ }\bibfield  {title} {\enquote {\bibinfo {title} {High-pressure
  properties of dense metallic zirconium hydrides studied by ab initio
  calculations},}\ }\href {\doibase 10.1103/PhysRevB.98.134103} {\bibfield
  {journal} {\bibinfo  {journal} {Phys. Rev. B}\ }\textbf {\bibinfo {volume}
  {98}},\ \bibinfo {pages} {134103} (\bibinfo {year}
  {2018}{\natexlab{b}})}\BibitemShut {NoStop}%
\bibitem [{\citenamefont {Abe}(2019)}]{abe2019}%
  \BibitemOpen
  \bibfield  {author} {\bibinfo {author} {\bibfnamefont {K.}~\bibnamefont
  {Abe}},\ }\bibfield  {title} {\enquote {\bibinfo {title} {Ab initio study of
  metallic aluminum hydrides at high pressures},}\ }\href {\doibase
  10.1103/PhysRevB.100.174105} {\bibfield  {journal} {\bibinfo  {journal}
  {Phys. Rev. B}\ }\textbf {\bibinfo {volume} {100}},\ \bibinfo {pages}
  {174105} (\bibinfo {year} {2019})}\BibitemShut {NoStop}%
\bibitem [{\citenamefont {Abe}(2021)}]{abe2021}%
  \BibitemOpen
  \bibfield  {author} {\bibinfo {author} {\bibfnamefont {K.}~\bibnamefont
  {Abe}},\ }\bibfield  {title} {\enquote {\bibinfo {title} {Metallic silicon
  subhydrides at high pressures studied by ab initio calculations},}\ }\href
  {\doibase 10.1103/PhysRevB.103.134118} {\bibfield  {journal} {\bibinfo
  {journal} {Phys. Rev. B}\ }\textbf {\bibinfo {volume} {103}},\ \bibinfo
  {pages} {134118} (\bibinfo {year} {2021})}\BibitemShut {NoStop}%
\bibitem [{\citenamefont {Akashi}\ \emph {et~al.}(2015)\citenamefont {Akashi},
  \citenamefont {Kawamura}, \citenamefont {Tsuneyuki}, \citenamefont {Nomura},\
  and\ \citenamefont {Arita}}]{akashi2015}%
  \BibitemOpen
  \bibfield  {author} {\bibinfo {author} {\bibfnamefont {R.}~\bibnamefont
  {Akashi}}, \bibinfo {author} {\bibfnamefont {M.}~\bibnamefont {Kawamura}},
  \bibinfo {author} {\bibfnamefont {S.}~\bibnamefont {Tsuneyuki}}, \bibinfo
  {author} {\bibfnamefont {Y.}~\bibnamefont {Nomura}}, \ and\ \bibinfo {author}
  {\bibfnamefont {R.}~\bibnamefont {Arita}},\ }\bibfield  {title} {\enquote
  {\bibinfo {title} {First-principles study of the pressure and
  crystal-structure dependences of the superconducting transition temperature
  in compressed sulfur hydrides},}\ }\href {\doibase
  10.1103/PhysRevB.91.224513} {\bibfield  {journal} {\bibinfo  {journal} {Phys.
  Rev. B}\ }\textbf {\bibinfo {volume} {91}},\ \bibinfo {pages} {224513}
  (\bibinfo {year} {2015})}\BibitemShut {NoStop}%
\bibitem [{\citenamefont {Bi}\ \emph {et~al.}(2017)\citenamefont {Bi},
  \citenamefont {Miller}, \citenamefont {Shamp},\ and\ \citenamefont
  {Zurek}}]{bi2017}%
  \BibitemOpen
  \bibfield  {author} {\bibinfo {author} {\bibfnamefont {T.}~\bibnamefont
  {Bi}}, \bibinfo {author} {\bibfnamefont {D.~P.}\ \bibnamefont {Miller}},
  \bibinfo {author} {\bibfnamefont {A.}~\bibnamefont {Shamp}}, \ and\ \bibinfo
  {author} {\bibfnamefont {E.}~\bibnamefont {Zurek}},\ }\bibfield  {title}
  {\enquote {\bibinfo {title} {Superconducting phases of phosphorus hydride
  under pressure: Stabilization by mobile molecular hydrogen},}\ }\href
  {\doibase https://doi.org/10.1002/ange.201701660} {\bibfield  {journal}
  {\bibinfo  {journal} {Angewandte Chemie}\ }\textbf {\bibinfo {volume}
  {129}},\ \bibinfo {pages} {10326} (\bibinfo {year} {2017})}\BibitemShut
  {NoStop}%
\bibitem [{\citenamefont {Bi}\ and\ \citenamefont {Zurek}(2021)}]{bi2021}%
  \BibitemOpen
  \bibfield  {author} {\bibinfo {author} {\bibfnamefont {T.}~\bibnamefont
  {Bi}}\ and\ \bibinfo {author} {\bibfnamefont {E.}~\bibnamefont {Zurek}},\
  }\bibfield  {title} {\enquote {\bibinfo {title} {Electronic structure and
  superconductivity of compressed metal tetrahydrides},}\ }\href {\doibase
  https://doi.org/10.1002/chem.202102679} {\bibfield  {journal} {\bibinfo
  {journal} {Chemistry – A European Journal}\ }\textbf {\bibinfo {volume}
  {27}},\ \bibinfo {pages} {14858} (\bibinfo {year} {2021})}\BibitemShut
  {NoStop}%
\bibitem [{\citenamefont {Chang}\ \emph {et~al.}(2020)\citenamefont {Chang},
  \citenamefont {Silayi}, \citenamefont {Papaconstantopoulos},\ and\
  \citenamefont {Mehl}}]{chang2020}%
  \BibitemOpen
  \bibfield  {author} {\bibinfo {author} {\bibfnamefont {P.-H.}\ \bibnamefont
  {Chang}}, \bibinfo {author} {\bibfnamefont {S.}~\bibnamefont {Silayi}},
  \bibinfo {author} {\bibfnamefont {D.}~\bibnamefont {Papaconstantopoulos}}, \
  and\ \bibinfo {author} {\bibfnamefont {M.}~\bibnamefont {Mehl}},\ }\bibfield
  {title} {\enquote {\bibinfo {title} {Pressure-induced high-temperature
  superconductivity in hypothetical h3x (x=as, se, br, sb, te and i) in the h3s
  structure with im3¯m symmetry},}\ }\href {\doibase
  https://doi.org/10.1016/j.jpcs.2019.109315} {\bibfield  {journal} {\bibinfo
  {journal} {Journal of Physics and Chemistry of Solids}\ }\textbf {\bibinfo
  {volume} {139}},\ \bibinfo {pages} {109315} (\bibinfo {year}
  {2020})}\BibitemShut {NoStop}%
\bibitem [{\citenamefont {Chen}\ \emph {et~al.}(2008)\citenamefont {Chen},
  \citenamefont {Wang}, \citenamefont {Struzhkin}, \citenamefont {Mao},
  \citenamefont {Hemley},\ and\ \citenamefont {Lin}}]{chen2008}%
  \BibitemOpen
  \bibfield  {author} {\bibinfo {author} {\bibfnamefont {X.-J.}\ \bibnamefont
  {Chen}}, \bibinfo {author} {\bibfnamefont {J.-L.}\ \bibnamefont {Wang}},
  \bibinfo {author} {\bibfnamefont {V.~V.}\ \bibnamefont {Struzhkin}}, \bibinfo
  {author} {\bibfnamefont {H.-k.}\ \bibnamefont {Mao}}, \bibinfo {author}
  {\bibfnamefont {R.~J.}\ \bibnamefont {Hemley}}, \ and\ \bibinfo {author}
  {\bibfnamefont {H.-Q.}\ \bibnamefont {Lin}},\ }\bibfield  {title} {\enquote
  {\bibinfo {title} {Superconducting behavior in compressed solid
  ${\mathrm{sih}}_{4}$ with a layered structure},}\ }\href {\doibase
  10.1103/PhysRevLett.101.077002} {\bibfield  {journal} {\bibinfo  {journal}
  {Phys. Rev. Lett.}\ }\textbf {\bibinfo {volume} {101}},\ \bibinfo {pages}
  {077002} (\bibinfo {year} {2008})}\BibitemShut {NoStop}%
\bibitem [{\citenamefont {Chen}\ \emph {et~al.}(2014)\citenamefont {Chen},
  \citenamefont {Tian}, \citenamefont {Duan}, \citenamefont {Bao},
  \citenamefont {Jin}, \citenamefont {Liu},\ and\ \citenamefont
  {Cui}}]{chen2014}%
  \BibitemOpen
  \bibfield  {author} {\bibinfo {author} {\bibfnamefont {C.}~\bibnamefont
  {Chen}}, \bibinfo {author} {\bibfnamefont {F.}~\bibnamefont {Tian}}, \bibinfo
  {author} {\bibfnamefont {D.}~\bibnamefont {Duan}}, \bibinfo {author}
  {\bibfnamefont {K.}~\bibnamefont {Bao}}, \bibinfo {author} {\bibfnamefont
  {X.}~\bibnamefont {Jin}}, \bibinfo {author} {\bibfnamefont {B.}~\bibnamefont
  {Liu}}, \ and\ \bibinfo {author} {\bibfnamefont {T.}~\bibnamefont {Cui}},\
  }\bibfield  {title} {\enquote {\bibinfo {title} {{Pressure induced phase
  transition in MH2 (M = V, Nb)}},}\ }\href {\doibase 10.1063/1.4866179}
  {\bibfield  {journal} {\bibinfo  {journal} {The Journal of Chemical Physics}\
  }\textbf {\bibinfo {volume} {140}},\ \bibinfo {pages} {114703} (\bibinfo
  {year} {2014})}\BibitemShut {NoStop}%
\bibitem [{\citenamefont {Chen}\ \emph {et~al.}(2015)\citenamefont {Chen},
  \citenamefont {Xu}, \citenamefont {Sun},\ and\ \citenamefont
  {Wang}}]{chen2015}%
  \BibitemOpen
  \bibfield  {author} {\bibinfo {author} {\bibfnamefont {C.}~\bibnamefont
  {Chen}}, \bibinfo {author} {\bibfnamefont {Y.}~\bibnamefont {Xu}}, \bibinfo
  {author} {\bibfnamefont {X.}~\bibnamefont {Sun}}, \ and\ \bibinfo {author}
  {\bibfnamefont {S.}~\bibnamefont {Wang}},\ }\bibfield  {title} {\enquote
  {\bibinfo {title} {Novel superconducting phases of hcl and hbr under high
  pressure: An ab initio study},}\ }\href {\doibase 10.1021/acs.jpcc.5b01653}
  {\bibfield  {journal} {\bibinfo  {journal} {The Journal of Physical Chemistry
  C}\ }\textbf {\bibinfo {volume} {119}},\ \bibinfo {pages} {17039} (\bibinfo
  {year} {2015})}\BibitemShut {NoStop}%
\bibitem [{\citenamefont {Chen}\ \emph
  {et~al.}(2021{\natexlab{a}})\citenamefont {Chen}, \citenamefont {Semenok},
  \citenamefont {Kvashnin}, \citenamefont {Huang}, \citenamefont {Kruglov},
  \citenamefont {Galasso}, \citenamefont {Song}, \citenamefont {Duan},
  \citenamefont {Goncharov}, \citenamefont {Prakapenka}, \citenamefont
  {Oganov},\ and\ \citenamefont {Cui}}]{chen2021}%
  \BibitemOpen
  \bibfield  {author} {\bibinfo {author} {\bibfnamefont {W.}~\bibnamefont
  {Chen}}, \bibinfo {author} {\bibfnamefont {D.~V.}\ \bibnamefont {Semenok}},
  \bibinfo {author} {\bibfnamefont {A.~G.}\ \bibnamefont {Kvashnin}}, \bibinfo
  {author} {\bibfnamefont {X.}~\bibnamefont {Huang}}, \bibinfo {author}
  {\bibfnamefont {I.~A.}\ \bibnamefont {Kruglov}}, \bibinfo {author}
  {\bibfnamefont {M.}~\bibnamefont {Galasso}}, \bibinfo {author} {\bibfnamefont
  {H.}~\bibnamefont {Song}}, \bibinfo {author} {\bibfnamefont {D.}~\bibnamefont
  {Duan}}, \bibinfo {author} {\bibfnamefont {A.~F.}\ \bibnamefont {Goncharov}},
  \bibinfo {author} {\bibfnamefont {V.~B.}\ \bibnamefont {Prakapenka}},
  \bibinfo {author} {\bibfnamefont {A.~R.}\ \bibnamefont {Oganov}}, \ and\
  \bibinfo {author} {\bibfnamefont {T.}~\bibnamefont {Cui}},\ }\bibfield
  {title} {\enquote {\bibinfo {title} {Synthesis of molecular metallic barium
  superhydride: pseudocubic bah12},}\ }\href {\doibase
  10.1038/s41467-020-20103-5} {\bibfield  {journal} {\bibinfo  {journal}
  {Nature Communications}\ }\textbf {\bibinfo {volume} {12}},\ \bibinfo {pages}
  {273} (\bibinfo {year} {2021}{\natexlab{a}})}\BibitemShut {NoStop}%
\bibitem [{\citenamefont {Chen}\ \emph
  {et~al.}(2021{\natexlab{b}})\citenamefont {Chen}, \citenamefont {Conway},
  \citenamefont {Sun}, \citenamefont {Kuang}, \citenamefont {Lu},\ and\
  \citenamefont {Hermann}}]{chen2021A}%
  \BibitemOpen
  \bibfield  {author} {\bibinfo {author} {\bibfnamefont {B.}~\bibnamefont
  {Chen}}, \bibinfo {author} {\bibfnamefont {L.~J.}\ \bibnamefont {Conway}},
  \bibinfo {author} {\bibfnamefont {W.}~\bibnamefont {Sun}}, \bibinfo {author}
  {\bibfnamefont {X.}~\bibnamefont {Kuang}}, \bibinfo {author} {\bibfnamefont
  {C.}~\bibnamefont {Lu}}, \ and\ \bibinfo {author} {\bibfnamefont
  {A.}~\bibnamefont {Hermann}},\ }\bibfield  {title} {\enquote {\bibinfo
  {title} {Phase stability and superconductivity of lead hydrides at high
  pressure},}\ }\href {\doibase 10.1103/PhysRevB.103.035131} {\bibfield
  {journal} {\bibinfo  {journal} {Phys. Rev. B}\ }\textbf {\bibinfo {volume}
  {103}},\ \bibinfo {pages} {035131} (\bibinfo {year}
  {2021}{\natexlab{b}})}\BibitemShut {NoStop}%
\bibitem [{\citenamefont {Chen}\ \emph
  {et~al.}(2021{\natexlab{c}})\citenamefont {Chen}, \citenamefont {Huang},
  \citenamefont {Tsuei},\ and\ \citenamefont {Jeng}}]{chen2021B}%
  \BibitemOpen
  \bibfield  {author} {\bibinfo {author} {\bibfnamefont {C.-H.}\ \bibnamefont
  {Chen}}, \bibinfo {author} {\bibfnamefont {A.}~\bibnamefont {Huang}},
  \bibinfo {author} {\bibfnamefont {C.~C.}\ \bibnamefont {Tsuei}}, \ and\
  \bibinfo {author} {\bibfnamefont {H.-T.}\ \bibnamefont {Jeng}},\ }\bibfield
  {title} {\enquote {\bibinfo {title} {Possible high-tc superconductivity at
  50 gpa in sodium hydride with clathrate structure},}\ }\href {\doibase
  10.1088/1367-2630/ac1df3} {\bibfield  {journal} {\bibinfo  {journal} {New
  Journal of Physics}\ }\textbf {\bibinfo {volume} {23}},\ \bibinfo {pages}
  {093007} (\bibinfo {year} {2021}{\natexlab{c}})}\BibitemShut {NoStop}%
\bibitem [{\citenamefont {Chen}\ \emph
  {et~al.}(2021{\natexlab{d}})\citenamefont {Chen}, \citenamefont {Semenok},
  \citenamefont {Huang}, \citenamefont {Shu}, \citenamefont {Li}, \citenamefont
  {Duan}, \citenamefont {Cui},\ and\ \citenamefont {Oganov}}]{chen2021C}%
  \BibitemOpen
  \bibfield  {author} {\bibinfo {author} {\bibfnamefont {W.}~\bibnamefont
  {Chen}}, \bibinfo {author} {\bibfnamefont {D.~V.}\ \bibnamefont {Semenok}},
  \bibinfo {author} {\bibfnamefont {X.}~\bibnamefont {Huang}}, \bibinfo
  {author} {\bibfnamefont {H.}~\bibnamefont {Shu}}, \bibinfo {author}
  {\bibfnamefont {X.}~\bibnamefont {Li}}, \bibinfo {author} {\bibfnamefont
  {D.}~\bibnamefont {Duan}}, \bibinfo {author} {\bibfnamefont {T.}~\bibnamefont
  {Cui}}, \ and\ \bibinfo {author} {\bibfnamefont {A.~R.}\ \bibnamefont
  {Oganov}},\ }\bibfield  {title} {\enquote {\bibinfo {title} {High-temperature
  superconducting phases in cerium superhydride with a ${T}_{c}$ up to 115 k
  below a pressure of 1 megabar},}\ }\href {\doibase
  10.1103/PhysRevLett.127.117001} {\bibfield  {journal} {\bibinfo  {journal}
  {Phys. Rev. Lett.}\ }\textbf {\bibinfo {volume} {127}},\ \bibinfo {pages}
  {117001} (\bibinfo {year} {2021}{\natexlab{d}})}\BibitemShut {NoStop}%
\bibitem [{\citenamefont {Chen}\ \emph {et~al.}(2022)\citenamefont {Chen},
  \citenamefont {Liu}, \citenamefont {Lin}, \citenamefont {Jiang},
  \citenamefont {Du}, \citenamefont {Zhang}, \citenamefont {Song},
  \citenamefont {Xie}, \citenamefont {Cui},\ and\ \citenamefont
  {Duan}}]{chen2022}%
  \BibitemOpen
  \bibfield  {author} {\bibinfo {author} {\bibfnamefont {Y.}~\bibnamefont
  {Chen}}, \bibinfo {author} {\bibfnamefont {Z.}~\bibnamefont {Liu}}, \bibinfo
  {author} {\bibfnamefont {Z.}~\bibnamefont {Lin}}, \bibinfo {author}
  {\bibfnamefont {Q.}~\bibnamefont {Jiang}}, \bibinfo {author} {\bibfnamefont
  {M.}~\bibnamefont {Du}}, \bibinfo {author} {\bibfnamefont {Z.}~\bibnamefont
  {Zhang}}, \bibinfo {author} {\bibfnamefont {H.}~\bibnamefont {Song}},
  \bibinfo {author} {\bibfnamefont {H.}~\bibnamefont {Xie}}, \bibinfo {author}
  {\bibfnamefont {T.}~\bibnamefont {Cui}}, \ and\ \bibinfo {author}
  {\bibfnamefont {D.}~\bibnamefont {Duan}},\ }\bibfield  {title} {\enquote
  {\bibinfo {title} {High t c superconductivity in layered hydrides xh15 (x =
  ca, sr, y, la) under high pressures},}\ }\href {\doibase
  10.1007/s11467-022-1182-1} {\bibfield  {journal} {\bibinfo  {journal}
  {Frontiers of Physics}\ }\textbf {\bibinfo {volume} {17}},\ \bibinfo {pages}
  {63502} (\bibinfo {year} {2022})}\BibitemShut {NoStop}%
\bibitem [{\citenamefont {Cheng}\ \emph {et~al.}(2015)\citenamefont {Cheng},
  \citenamefont {Zhang}, \citenamefont {Wang}, \citenamefont {Zhong},
  \citenamefont {Yang}, \citenamefont {Chen},\ and\ \citenamefont
  {Lin}}]{cheng2015}%
  \BibitemOpen
  \bibfield  {author} {\bibinfo {author} {\bibfnamefont {Y.}~\bibnamefont
  {Cheng}}, \bibinfo {author} {\bibfnamefont {C.}~\bibnamefont {Zhang}},
  \bibinfo {author} {\bibfnamefont {T.}~\bibnamefont {Wang}}, \bibinfo {author}
  {\bibfnamefont {G.}~\bibnamefont {Zhong}}, \bibinfo {author} {\bibfnamefont
  {C.}~\bibnamefont {Yang}}, \bibinfo {author} {\bibfnamefont {X.~J.}\
  \bibnamefont {Chen}}, \ and\ \bibinfo {author} {\bibfnamefont {H.~Q.}\
  \bibnamefont {Lin}},\ }\bibfield  {title} {\enquote {\bibinfo {title}
  {Pressure-induced superconductivity in h2-containing hydride pbh4(h2)2},}\
  }\href {\doibase 10.1038/srep16475} {\bibfield  {journal} {\bibinfo
  {journal} {Scientific Reports}\ }\textbf {\bibinfo {volume} {5}},\ \bibinfo
  {pages} {16475} (\bibinfo {year} {2015})}\BibitemShut {NoStop}%
\bibitem [{\citenamefont {Cui}\ \emph {et~al.}(2015)\citenamefont {Cui},
  \citenamefont {Shi}, \citenamefont {Liu}, \citenamefont {Yao}, \citenamefont
  {Wang}, \citenamefont {Iitaka},\ and\ \citenamefont {Ma}}]{cui2015}%
  \BibitemOpen
  \bibfield  {author} {\bibinfo {author} {\bibfnamefont {W.}~\bibnamefont
  {Cui}}, \bibinfo {author} {\bibfnamefont {J.}~\bibnamefont {Shi}}, \bibinfo
  {author} {\bibfnamefont {H.}~\bibnamefont {Liu}}, \bibinfo {author}
  {\bibfnamefont {Y.}~\bibnamefont {Yao}}, \bibinfo {author} {\bibfnamefont
  {H.}~\bibnamefont {Wang}}, \bibinfo {author} {\bibfnamefont {T.}~\bibnamefont
  {Iitaka}}, \ and\ \bibinfo {author} {\bibfnamefont {Y.}~\bibnamefont {Ma}},\
  }\bibfield  {title} {\enquote {\bibinfo {title} {Hydrogen segregation and its
  roles in structural stability and metallization: Silane under pressure},}\
  }\href {\doibase 10.1038/srep13039} {\bibfield  {journal} {\bibinfo
  {journal} {Scientific Reports}\ }\textbf {\bibinfo {volume} {5}},\ \bibinfo
  {pages} {13039} (\bibinfo {year} {2015})}\BibitemShut {NoStop}%
\bibitem [{\citenamefont {Esfahani}\ \emph {et~al.}(2016)\citenamefont
  {Esfahani}, \citenamefont {Wang}, \citenamefont {Oganov}, \citenamefont
  {Dong}, \citenamefont {Zhu}, \citenamefont {Wang}, \citenamefont {Rakitin},\
  and\ \citenamefont {Zhou}}]{davariesfahani2016}%
  \BibitemOpen
  \bibfield  {author} {\bibinfo {author} {\bibfnamefont {M.~M.~D.}\
  \bibnamefont {Esfahani}}, \bibinfo {author} {\bibfnamefont {Z.}~\bibnamefont
  {Wang}}, \bibinfo {author} {\bibfnamefont {A.~R.}\ \bibnamefont {Oganov}},
  \bibinfo {author} {\bibfnamefont {H.}~\bibnamefont {Dong}}, \bibinfo {author}
  {\bibfnamefont {Q.}~\bibnamefont {Zhu}}, \bibinfo {author} {\bibfnamefont
  {S.}~\bibnamefont {Wang}}, \bibinfo {author} {\bibfnamefont {M.~S.}\
  \bibnamefont {Rakitin}}, \ and\ \bibinfo {author} {\bibfnamefont {X.~F.}\
  \bibnamefont {Zhou}},\ }\bibfield  {title} {\enquote {\bibinfo {title}
  {Superconductivity of novel tin hydrides (snnhm) under pressure},}\ }\href
  {\doibase 10.1038/srep22873} {\bibfield  {journal} {\bibinfo  {journal}
  {Scientific Reports}\ }\textbf {\bibinfo {volume} {6}},\ \bibinfo {pages}
  {22873} (\bibinfo {year} {2016})}\BibitemShut {NoStop}%
\bibitem [{\citenamefont {Davari~Esfahani}\ \emph {et~al.}(2017)\citenamefont
  {Davari~Esfahani}, \citenamefont {Oganov}, \citenamefont {Niu},\ and\
  \citenamefont {Zhang}}]{davariesfahani2017}%
  \BibitemOpen
  \bibfield  {author} {\bibinfo {author} {\bibfnamefont {M.~M.}\ \bibnamefont
  {Davari~Esfahani}}, \bibinfo {author} {\bibfnamefont {A.~R.}\ \bibnamefont
  {Oganov}}, \bibinfo {author} {\bibfnamefont {H.}~\bibnamefont {Niu}}, \ and\
  \bibinfo {author} {\bibfnamefont {J.}~\bibnamefont {Zhang}},\ }\bibfield
  {title} {\enquote {\bibinfo {title} {Superconductivity and unexpected
  chemistry of germanium hydrides under pressure},}\ }\href {\doibase
  10.1103/PhysRevB.95.134506} {\bibfield  {journal} {\bibinfo  {journal} {Phys.
  Rev. B}\ }\textbf {\bibinfo {volume} {95}},\ \bibinfo {pages} {134506}
  (\bibinfo {year} {2017})}\BibitemShut {NoStop}%
\bibitem [{\citenamefont {Du}\ \emph {et~al.}(2021)\citenamefont {Du},
  \citenamefont {Zhang}, \citenamefont {Song}, \citenamefont {Yu},
  \citenamefont {Cui}, \citenamefont {Kresin},\ and\ \citenamefont
  {Duan}}]{du2021}%
  \BibitemOpen
  \bibfield  {author} {\bibinfo {author} {\bibfnamefont {M.}~\bibnamefont
  {Du}}, \bibinfo {author} {\bibfnamefont {Z.}~\bibnamefont {Zhang}}, \bibinfo
  {author} {\bibfnamefont {H.}~\bibnamefont {Song}}, \bibinfo {author}
  {\bibfnamefont {H.}~\bibnamefont {Yu}}, \bibinfo {author} {\bibfnamefont
  {T.}~\bibnamefont {Cui}}, \bibinfo {author} {\bibfnamefont {V.~Z.}\
  \bibnamefont {Kresin}}, \ and\ \bibinfo {author} {\bibfnamefont
  {D.}~\bibnamefont {Duan}},\ }\bibfield  {title} {\enquote {\bibinfo {title}
  {High-temperature superconductivity in transition metallic hydrides mh11 (m =
  mo{,} w{,} nb{,} and ta) under high pressure},}\ }\href {\doibase
  10.1039/D0CP06435A} {\bibfield  {journal} {\bibinfo  {journal} {Phys. Chem.
  Chem. Phys.}\ }\textbf {\bibinfo {volume} {23}},\ \bibinfo {pages} {6717}
  (\bibinfo {year} {2021})}\BibitemShut {NoStop}%
\bibitem [{\citenamefont {Duan}\ \emph {et~al.}(2010)\citenamefont {Duan},
  \citenamefont {Tian}, \citenamefont {He}, \citenamefont {Meng}, \citenamefont
  {Wang}, \citenamefont {Chen}, \citenamefont {Zhao}, \citenamefont {Liu},\
  and\ \citenamefont {Cui}}]{duan2010}%
  \BibitemOpen
  \bibfield  {author} {\bibinfo {author} {\bibfnamefont {D.}~\bibnamefont
  {Duan}}, \bibinfo {author} {\bibfnamefont {F.}~\bibnamefont {Tian}}, \bibinfo
  {author} {\bibfnamefont {Z.}~\bibnamefont {He}}, \bibinfo {author}
  {\bibfnamefont {X.}~\bibnamefont {Meng}}, \bibinfo {author} {\bibfnamefont
  {L.}~\bibnamefont {Wang}}, \bibinfo {author} {\bibfnamefont {C.}~\bibnamefont
  {Chen}}, \bibinfo {author} {\bibfnamefont {X.}~\bibnamefont {Zhao}}, \bibinfo
  {author} {\bibfnamefont {B.}~\bibnamefont {Liu}}, \ and\ \bibinfo {author}
  {\bibfnamefont {T.}~\bibnamefont {Cui}},\ }\bibfield  {title} {\enquote
  {\bibinfo {title} {{Hydrogen bond symmetrization and superconducting phase of
  HBr and HCl under high pressure: An ab initio study}},}\ }\href {\doibase
  10.1063/1.3471446} {\bibfield  {journal} {\bibinfo  {journal} {The Journal of
  Chemical Physics}\ }\textbf {\bibinfo {volume} {133}},\ \bibinfo {pages}
  {074509} (\bibinfo {year} {2010})}\BibitemShut {NoStop}%
\bibitem [{\citenamefont {Duan}\ \emph {et~al.}(2014)\citenamefont {Duan},
  \citenamefont {Liu}, \citenamefont {Tian}, \citenamefont {Li}, \citenamefont
  {Huang}, \citenamefont {Zhao}, \citenamefont {Yu}, \citenamefont {Liu},
  \citenamefont {Tian},\ and\ \citenamefont {Cui}}]{duan2014}%
  \BibitemOpen
  \bibfield  {author} {\bibinfo {author} {\bibfnamefont {D.}~\bibnamefont
  {Duan}}, \bibinfo {author} {\bibfnamefont {Y.}~\bibnamefont {Liu}}, \bibinfo
  {author} {\bibfnamefont {F.}~\bibnamefont {Tian}}, \bibinfo {author}
  {\bibfnamefont {D.}~\bibnamefont {Li}}, \bibinfo {author} {\bibfnamefont
  {X.}~\bibnamefont {Huang}}, \bibinfo {author} {\bibfnamefont
  {Z.}~\bibnamefont {Zhao}}, \bibinfo {author} {\bibfnamefont {H.}~\bibnamefont
  {Yu}}, \bibinfo {author} {\bibfnamefont {B.}~\bibnamefont {Liu}}, \bibinfo
  {author} {\bibfnamefont {W.}~\bibnamefont {Tian}}, \ and\ \bibinfo {author}
  {\bibfnamefont {T.}~\bibnamefont {Cui}},\ }\bibfield  {title} {\enquote
  {\bibinfo {title} {Pressure-induced metallization of dense (h 2 s) 2 h 2 with
  high-t c superconductivity},}\ }\href {\doibase 10.1038/srep06968} {\bibfield
   {journal} {\bibinfo  {journal} {Scientific Reports}\ }\textbf {\bibinfo
  {volume} {4}},\ \bibinfo {pages} {6968} (\bibinfo {year} {2014})}\BibitemShut
  {NoStop}%
\bibitem [{\citenamefont {Duan}\ \emph {et~al.}(2015)\citenamefont {Duan},
  \citenamefont {Tian}, \citenamefont {Liu}, \citenamefont {Huang},
  \citenamefont {Li}, \citenamefont {Yu}, \citenamefont {Ma}, \citenamefont
  {Liu},\ and\ \citenamefont {Cui}}]{duan2015}%
  \BibitemOpen
  \bibfield  {author} {\bibinfo {author} {\bibfnamefont {D.}~\bibnamefont
  {Duan}}, \bibinfo {author} {\bibfnamefont {F.}~\bibnamefont {Tian}}, \bibinfo
  {author} {\bibfnamefont {Y.}~\bibnamefont {Liu}}, \bibinfo {author}
  {\bibfnamefont {X.}~\bibnamefont {Huang}}, \bibinfo {author} {\bibfnamefont
  {D.}~\bibnamefont {Li}}, \bibinfo {author} {\bibfnamefont {H.}~\bibnamefont
  {Yu}}, \bibinfo {author} {\bibfnamefont {Y.}~\bibnamefont {Ma}}, \bibinfo
  {author} {\bibfnamefont {B.}~\bibnamefont {Liu}}, \ and\ \bibinfo {author}
  {\bibfnamefont {T.}~\bibnamefont {Cui}},\ }\bibfield  {title} {\enquote
  {\bibinfo {title} {Enhancement of tc in the atomic phase of iodine-doped
  hydrogen at high pressures},}\ }\href {\doibase 10.1039/C5CP05218A}
  {\bibfield  {journal} {\bibinfo  {journal} {Phys. Chem. Chem. Phys.}\
  }\textbf {\bibinfo {volume} {17}},\ \bibinfo {pages} {32335} (\bibinfo {year}
  {2015})}\BibitemShut {NoStop}%
\bibitem [{\citenamefont {Duda}\ \emph {et~al.}(2018)\citenamefont {Duda},
  \citenamefont {Szewczyk}, \citenamefont {Jarosik}, \citenamefont
  {Szczesniak}, \citenamefont {Sowinska},\ and\ \citenamefont
  {Szczesniak}}]{duda2018}%
  \BibitemOpen
  \bibfield  {author} {\bibinfo {author} {\bibfnamefont {A.}~\bibnamefont
  {Duda}}, \bibinfo {author} {\bibfnamefont {K.}~\bibnamefont {Szewczyk}},
  \bibinfo {author} {\bibfnamefont {M.}~\bibnamefont {Jarosik}}, \bibinfo
  {author} {\bibfnamefont {K.}~\bibnamefont {Szczesniak}}, \bibinfo {author}
  {\bibfnamefont {M.}~\bibnamefont {Sowinska}}, \ and\ \bibinfo {author}
  {\bibfnamefont {D.}~\bibnamefont {Szczesniak}},\ }\bibfield  {title}
  {\enquote {\bibinfo {title} {Characterization of the superconducting state in
  hafnium hydride under high pressure},}\ }\href {\doibase
  https://doi.org/10.1016/j.physb.2017.10.107} {\bibfield  {journal} {\bibinfo
  {journal} {Physica B: Condensed Matter}\ }\textbf {\bibinfo {volume} {536}},\
  \bibinfo {pages} {275} (\bibinfo {year} {2018})}\BibitemShut {NoStop}%
\bibitem [{\citenamefont {Durajski}\ and\ \citenamefont
  {Szczesniak}(2018)}]{durajski2018}%
  \BibitemOpen
  \bibfield  {author} {\bibinfo {author} {\bibfnamefont {A.~P.}\ \bibnamefont
  {Durajski}}\ and\ \bibinfo {author} {\bibfnamefont {R.}~\bibnamefont
  {Szczesniak}},\ }\bibfield  {title} {\enquote {\bibinfo {title} {{Structural,
  electronic, vibrational, and superconducting properties of hydrogenated
  chlorine}},}\ }\href {\doibase 10.1063/1.5031202} {\bibfield  {journal}
  {\bibinfo  {journal} {The Journal of Chemical Physics}\ }\textbf {\bibinfo
  {volume} {149}},\ \bibinfo {pages} {074101} (\bibinfo {year}
  {2018})}\BibitemShut {NoStop}%
\bibitem [{\citenamefont {Einaga}\ \emph {et~al.}(2016)\citenamefont {Einaga},
  \citenamefont {Sakata}, \citenamefont {Ishikawa}, \citenamefont {Shimizu},
  \citenamefont {Eremets}, \citenamefont {Drozdov}, \citenamefont {Troyan},
  \citenamefont {Hirao},\ and\ \citenamefont {Ohishi}}]{einaga2016}%
  \BibitemOpen
  \bibfield  {author} {\bibinfo {author} {\bibfnamefont {M.}~\bibnamefont
  {Einaga}}, \bibinfo {author} {\bibfnamefont {M.}~\bibnamefont {Sakata}},
  \bibinfo {author} {\bibfnamefont {T.}~\bibnamefont {Ishikawa}}, \bibinfo
  {author} {\bibfnamefont {K.}~\bibnamefont {Shimizu}}, \bibinfo {author}
  {\bibfnamefont {M.~I.}\ \bibnamefont {Eremets}}, \bibinfo {author}
  {\bibfnamefont {A.~P.}\ \bibnamefont {Drozdov}}, \bibinfo {author}
  {\bibfnamefont {I.~A.}\ \bibnamefont {Troyan}}, \bibinfo {author}
  {\bibfnamefont {N.}~\bibnamefont {Hirao}}, \ and\ \bibinfo {author}
  {\bibfnamefont {Y.}~\bibnamefont {Ohishi}},\ }\bibfield  {title} {\enquote
  {\bibinfo {title} {Crystal structure of the superconducting phase of sulfur
  hydride},}\ }\href {\doibase 10.1038/nphys3760} {\bibfield  {journal}
  {\bibinfo  {journal} {Nature Physics}\ }\textbf {\bibinfo {volume} {12}},\
  \bibinfo {pages} {835} (\bibinfo {year} {2016})}\BibitemShut {NoStop}%
\bibitem [{\citenamefont {Errea}\ \emph {et~al.}(2013)\citenamefont {Errea},
  \citenamefont {Calandra},\ and\ \citenamefont {Mauri}}]{errea2013}%
  \BibitemOpen
  \bibfield  {author} {\bibinfo {author} {\bibfnamefont {I.}~\bibnamefont
  {Errea}}, \bibinfo {author} {\bibfnamefont {M.}~\bibnamefont {Calandra}}, \
  and\ \bibinfo {author} {\bibfnamefont {F.}~\bibnamefont {Mauri}},\ }\bibfield
   {title} {\enquote {\bibinfo {title} {First-principles theory of
  anharmonicity and the inverse isotope effect in superconducting
  palladium-hydride compounds},}\ }\href {\doibase
  10.1103/PhysRevLett.111.177002} {\bibfield  {journal} {\bibinfo  {journal}
  {Phys. Rev. Lett.}\ }\textbf {\bibinfo {volume} {111}},\ \bibinfo {pages}
  {177002} (\bibinfo {year} {2013})}\BibitemShut {NoStop}%
\bibitem [{\citenamefont {Errea}\ \emph {et~al.}(2016)\citenamefont {Errea},
  \citenamefont {Calandra}, \citenamefont {Pickard}, \citenamefont {Nelson},
  \citenamefont {Needs}, \citenamefont {Li}, \citenamefont {Liu}, \citenamefont
  {Zhang}, \citenamefont {Ma},\ and\ \citenamefont {Mauri}}]{errea2016}%
  \BibitemOpen
  \bibfield  {author} {\bibinfo {author} {\bibfnamefont {I.}~\bibnamefont
  {Errea}}, \bibinfo {author} {\bibfnamefont {M.}~\bibnamefont {Calandra}},
  \bibinfo {author} {\bibfnamefont {C.~J.}\ \bibnamefont {Pickard}}, \bibinfo
  {author} {\bibfnamefont {J.~R.}\ \bibnamefont {Nelson}}, \bibinfo {author}
  {\bibfnamefont {R.~J.}\ \bibnamefont {Needs}}, \bibinfo {author}
  {\bibfnamefont {Y.}~\bibnamefont {Li}}, \bibinfo {author} {\bibfnamefont
  {H.}~\bibnamefont {Liu}}, \bibinfo {author} {\bibfnamefont {Y.}~\bibnamefont
  {Zhang}}, \bibinfo {author} {\bibfnamefont {Y.}~\bibnamefont {Ma}}, \ and\
  \bibinfo {author} {\bibfnamefont {F.}~\bibnamefont {Mauri}},\ }\bibfield
  {title} {\enquote {\bibinfo {title} {Quantum hydrogen-bond symmetrization in
  the superconducting hydrogen sulfide system},}\ }\href {\doibase
  10.1038/nature17175} {\bibfield  {journal} {\bibinfo  {journal} {Nature}\
  }\textbf {\bibinfo {volume} {532}},\ \bibinfo {pages} {81} (\bibinfo {year}
  {2016})}\BibitemShut {NoStop}%
\bibitem [{\citenamefont {Feng}\ \emph {et~al.}(2015)\citenamefont {Feng},
  \citenamefont {Zhang}, \citenamefont {Gao}, \citenamefont {Liu},\ and\
  \citenamefont {Wang}}]{feng2015}%
  \BibitemOpen
  \bibfield  {author} {\bibinfo {author} {\bibfnamefont {X.}~\bibnamefont
  {Feng}}, \bibinfo {author} {\bibfnamefont {J.}~\bibnamefont {Zhang}},
  \bibinfo {author} {\bibfnamefont {G.}~\bibnamefont {Gao}}, \bibinfo {author}
  {\bibfnamefont {H.}~\bibnamefont {Liu}}, \ and\ \bibinfo {author}
  {\bibfnamefont {H.}~\bibnamefont {Wang}},\ }\bibfield  {title} {\enquote
  {\bibinfo {title} {Compressed sodalite-like mgh6 as a potential
  high-temperature superconductor},}\ }\href {\doibase 10.1039/C5RA11459D}
  {\bibfield  {journal} {\bibinfo  {journal} {RSC Adv.}\ }\textbf {\bibinfo
  {volume} {5}},\ \bibinfo {pages} {59292} (\bibinfo {year}
  {2015})}\BibitemShut {NoStop}%
\bibitem [{\citenamefont {Flores-Livas}\ \emph {et~al.}(2012)\citenamefont
  {Flores-Livas}, \citenamefont {Amsler}, \citenamefont {Lenosky},
  \citenamefont {Lehtovaara}, \citenamefont {Botti}, \citenamefont {Marques},\
  and\ \citenamefont {Goedecker}}]{flores-livas2012}%
  \BibitemOpen
  \bibfield  {author} {\bibinfo {author} {\bibfnamefont {J.~A.}\ \bibnamefont
  {Flores-Livas}}, \bibinfo {author} {\bibfnamefont {M.}~\bibnamefont
  {Amsler}}, \bibinfo {author} {\bibfnamefont {T.~J.}\ \bibnamefont {Lenosky}},
  \bibinfo {author} {\bibfnamefont {L.}~\bibnamefont {Lehtovaara}}, \bibinfo
  {author} {\bibfnamefont {S.}~\bibnamefont {Botti}}, \bibinfo {author}
  {\bibfnamefont {M.~A.~L.}\ \bibnamefont {Marques}}, \ and\ \bibinfo {author}
  {\bibfnamefont {S.}~\bibnamefont {Goedecker}},\ }\bibfield  {title} {\enquote
  {\bibinfo {title} {High-pressure structures of disilane and their
  superconducting properties},}\ }\href {\doibase
  10.1103/PhysRevLett.108.117004} {\bibfield  {journal} {\bibinfo  {journal}
  {Phys. Rev. Lett.}\ }\textbf {\bibinfo {volume} {108}},\ \bibinfo {pages}
  {117004} (\bibinfo {year} {2012})}\BibitemShut {NoStop}%
\bibitem [{\citenamefont {Flores-Livas}\ \emph
  {et~al.}(2016{\natexlab{a}})\citenamefont {Flores-Livas}, \citenamefont
  {Amsler}, \citenamefont {Heil}, \citenamefont {Sanna}, \citenamefont {Boeri},
  \citenamefont {Profeta}, \citenamefont {Wolverton}, \citenamefont
  {Goedecker},\ and\ \citenamefont {Gross}}]{flores-livas2016}%
  \BibitemOpen
  \bibfield  {author} {\bibinfo {author} {\bibfnamefont {J.~A.}\ \bibnamefont
  {Flores-Livas}}, \bibinfo {author} {\bibfnamefont {M.}~\bibnamefont
  {Amsler}}, \bibinfo {author} {\bibfnamefont {C.}~\bibnamefont {Heil}},
  \bibinfo {author} {\bibfnamefont {A.}~\bibnamefont {Sanna}}, \bibinfo
  {author} {\bibfnamefont {L.}~\bibnamefont {Boeri}}, \bibinfo {author}
  {\bibfnamefont {G.}~\bibnamefont {Profeta}}, \bibinfo {author} {\bibfnamefont
  {C.}~\bibnamefont {Wolverton}}, \bibinfo {author} {\bibfnamefont
  {S.}~\bibnamefont {Goedecker}}, \ and\ \bibinfo {author} {\bibfnamefont
  {E.~K.~U.}\ \bibnamefont {Gross}},\ }\bibfield  {title} {\enquote {\bibinfo
  {title} {Superconductivity in metastable phases of phosphorus-hydride
  compounds under high pressure},}\ }\href {\doibase
  10.1103/PhysRevB.93.020508} {\bibfield  {journal} {\bibinfo  {journal} {Phys.
  Rev. B}\ }\textbf {\bibinfo {volume} {93}},\ \bibinfo {pages} {020508}
  (\bibinfo {year} {2016}{\natexlab{a}})}\BibitemShut {NoStop}%
\bibitem [{\citenamefont {Flores-Livas}\ \emph
  {et~al.}(2016{\natexlab{b}})\citenamefont {Flores-Livas}, \citenamefont
  {Sanna},\ and\ \citenamefont {Gross}}]{flores-livas2016A}%
  \BibitemOpen
  \bibfield  {author} {\bibinfo {author} {\bibfnamefont {J.~A.}\ \bibnamefont
  {Flores-Livas}}, \bibinfo {author} {\bibfnamefont {A.}~\bibnamefont {Sanna}},
  \ and\ \bibinfo {author} {\bibfnamefont {E.~K.}\ \bibnamefont {Gross}},\
  }\bibfield  {title} {\enquote {\bibinfo {title} {High temperature
  superconductivity in sulfur and selenium hydrides at high pressure},}\ }\href
  {\doibase 10.1140/epjb/e2016-70020-0} {\bibfield  {journal} {\bibinfo
  {journal} {European Physical Journal B}\ }\textbf {\bibinfo {volume} {89}},\
  \bibinfo {pages} {63} (\bibinfo {year} {2016}{\natexlab{b}})}\BibitemShut
  {NoStop}%
\bibitem [{\citenamefont {Fu}\ \emph {et~al.}(2016)\citenamefont {Fu},
  \citenamefont {Du}, \citenamefont {Zhang}, \citenamefont {Peng},
  \citenamefont {Zhang}, \citenamefont {Pickard}, \citenamefont {Needs},
  \citenamefont {Singh}, \citenamefont {Zheng},\ and\ \citenamefont
  {Ma}}]{fu2016}%
  \BibitemOpen
  \bibfield  {author} {\bibinfo {author} {\bibfnamefont {Y.}~\bibnamefont
  {Fu}}, \bibinfo {author} {\bibfnamefont {X.}~\bibnamefont {Du}}, \bibinfo
  {author} {\bibfnamefont {L.}~\bibnamefont {Zhang}}, \bibinfo {author}
  {\bibfnamefont {F.}~\bibnamefont {Peng}}, \bibinfo {author} {\bibfnamefont
  {M.}~\bibnamefont {Zhang}}, \bibinfo {author} {\bibfnamefont {C.~J.}\
  \bibnamefont {Pickard}}, \bibinfo {author} {\bibfnamefont {R.~J.}\
  \bibnamefont {Needs}}, \bibinfo {author} {\bibfnamefont {D.~J.}\ \bibnamefont
  {Singh}}, \bibinfo {author} {\bibfnamefont {W.}~\bibnamefont {Zheng}}, \ and\
  \bibinfo {author} {\bibfnamefont {Y.}~\bibnamefont {Ma}},\ }\bibfield
  {title} {\enquote {\bibinfo {title} {High-pressure phase stability and
  superconductivity of pnictogen hydrides and chemical trends for compressed
  hydrides},}\ }\href {\doibase 10.1021/acs.chemmater.5b04638} {\bibfield
  {journal} {\bibinfo  {journal} {Chemistry of Materials}\ }\textbf {\bibinfo
  {volume} {28}},\ \bibinfo {pages} {1746} (\bibinfo {year}
  {2016})}\BibitemShut {NoStop}%
\bibitem [{\citenamefont {Gao}\ \emph {et~al.}(2008)\citenamefont {Gao},
  \citenamefont {Oganov}, \citenamefont {Bergara}, \citenamefont
  {Martinez-Canales}, \citenamefont {Cui}, \citenamefont {Iitaka},
  \citenamefont {Ma},\ and\ \citenamefont {Zou}}]{gao2008}%
  \BibitemOpen
  \bibfield  {author} {\bibinfo {author} {\bibfnamefont {G.}~\bibnamefont
  {Gao}}, \bibinfo {author} {\bibfnamefont {A.~R.}\ \bibnamefont {Oganov}},
  \bibinfo {author} {\bibfnamefont {A.}~\bibnamefont {Bergara}}, \bibinfo
  {author} {\bibfnamefont {M.}~\bibnamefont {Martinez-Canales}}, \bibinfo
  {author} {\bibfnamefont {T.}~\bibnamefont {Cui}}, \bibinfo {author}
  {\bibfnamefont {T.}~\bibnamefont {Iitaka}}, \bibinfo {author} {\bibfnamefont
  {Y.}~\bibnamefont {Ma}}, \ and\ \bibinfo {author} {\bibfnamefont
  {G.}~\bibnamefont {Zou}},\ }\bibfield  {title} {\enquote {\bibinfo {title}
  {Superconducting high pressure phase of germane},}\ }\href {\doibase
  10.1103/PhysRevLett.101.107002} {\bibfield  {journal} {\bibinfo  {journal}
  {Phys. Rev. Lett.}\ }\textbf {\bibinfo {volume} {101}},\ \bibinfo {pages}
  {107002} (\bibinfo {year} {2008})}\BibitemShut {NoStop}%
\bibitem [{\citenamefont {Gao}\ \emph {et~al.}(2010)\citenamefont {Gao},
  \citenamefont {Oganov}, \citenamefont {Li}, \citenamefont {Li}, \citenamefont
  {Wang}, \citenamefont {Cui}, \citenamefont {Ma}, \citenamefont {Bergara},
  \citenamefont {Lyakhov}, \citenamefont {Iitaka},\ and\ \citenamefont
  {Zou}}]{gao2010}%
  \BibitemOpen
  \bibfield  {author} {\bibinfo {author} {\bibfnamefont {G.}~\bibnamefont
  {Gao}}, \bibinfo {author} {\bibfnamefont {A.~R.}\ \bibnamefont {Oganov}},
  \bibinfo {author} {\bibfnamefont {P.}~\bibnamefont {Li}}, \bibinfo {author}
  {\bibfnamefont {Z.}~\bibnamefont {Li}}, \bibinfo {author} {\bibfnamefont
  {H.}~\bibnamefont {Wang}}, \bibinfo {author} {\bibfnamefont {T.}~\bibnamefont
  {Cui}}, \bibinfo {author} {\bibfnamefont {Y.}~\bibnamefont {Ma}}, \bibinfo
  {author} {\bibfnamefont {A.}~\bibnamefont {Bergara}}, \bibinfo {author}
  {\bibfnamefont {A.~O.}\ \bibnamefont {Lyakhov}}, \bibinfo {author}
  {\bibfnamefont {T.}~\bibnamefont {Iitaka}}, \ and\ \bibinfo {author}
  {\bibfnamefont {G.}~\bibnamefont {Zou}},\ }\bibfield  {title} {\enquote
  {\bibinfo {title} {High-pressure crystal structures and superconductivity of
  stannane (snh<sub>4</sub>)},}\ }\href {\doibase 10.1073/pnas.0908342107}
  {\bibfield  {journal} {\bibinfo  {journal} {Proceedings of the National
  Academy of Sciences}\ }\textbf {\bibinfo {volume} {107}},\ \bibinfo {pages}
  {1317} (\bibinfo {year} {2010})}\BibitemShut {NoStop}%
\bibitem [{\citenamefont {Gao}\ \emph {et~al.}(2011)\citenamefont {Gao},
  \citenamefont {Wang}, \citenamefont {Bergara}, \citenamefont {Li},
  \citenamefont {Liu},\ and\ \citenamefont {Ma}}]{gao2011}%
  \BibitemOpen
  \bibfield  {author} {\bibinfo {author} {\bibfnamefont {G.}~\bibnamefont
  {Gao}}, \bibinfo {author} {\bibfnamefont {H.}~\bibnamefont {Wang}}, \bibinfo
  {author} {\bibfnamefont {A.}~\bibnamefont {Bergara}}, \bibinfo {author}
  {\bibfnamefont {Y.}~\bibnamefont {Li}}, \bibinfo {author} {\bibfnamefont
  {G.}~\bibnamefont {Liu}}, \ and\ \bibinfo {author} {\bibfnamefont
  {Y.}~\bibnamefont {Ma}},\ }\bibfield  {title} {\enquote {\bibinfo {title}
  {Metallic and superconducting gallane under high pressure},}\ }\href
  {\doibase 10.1103/PhysRevB.84.064118} {\bibfield  {journal} {\bibinfo
  {journal} {Phys. Rev. B}\ }\textbf {\bibinfo {volume} {84}},\ \bibinfo
  {pages} {064118} (\bibinfo {year} {2011})}\BibitemShut {NoStop}%
\bibitem [{\citenamefont {Gao}\ \emph {et~al.}(2013)\citenamefont {Gao},
  \citenamefont {Hoffmann}, \citenamefont {Ashcroft}, \citenamefont {Liu},
  \citenamefont {Bergara},\ and\ \citenamefont {Ma}}]{gao2013}%
  \BibitemOpen
  \bibfield  {author} {\bibinfo {author} {\bibfnamefont {G.}~\bibnamefont
  {Gao}}, \bibinfo {author} {\bibfnamefont {R.}~\bibnamefont {Hoffmann}},
  \bibinfo {author} {\bibfnamefont {N.~W.}\ \bibnamefont {Ashcroft}}, \bibinfo
  {author} {\bibfnamefont {H.}~\bibnamefont {Liu}}, \bibinfo {author}
  {\bibfnamefont {A.}~\bibnamefont {Bergara}}, \ and\ \bibinfo {author}
  {\bibfnamefont {Y.}~\bibnamefont {Ma}},\ }\bibfield  {title} {\enquote
  {\bibinfo {title} {Theoretical study of the ground-state structures and
  properties of niobium hydrides under pressure},}\ }\href {\doibase
  10.1103/PhysRevB.88.184104} {\bibfield  {journal} {\bibinfo  {journal} {Phys.
  Rev. B}\ }\textbf {\bibinfo {volume} {88}},\ \bibinfo {pages} {184104}
  (\bibinfo {year} {2013})}\BibitemShut {NoStop}%
\bibitem [{\citenamefont {Gao}\ \emph {et~al.}(2021)\citenamefont {Gao},
  \citenamefont {Cui}, \citenamefont {Chen}, \citenamefont {Wang},
  \citenamefont {Hao}, \citenamefont {Shi}, \citenamefont {Liu}, \citenamefont
  {Botti}, \citenamefont {Marques},\ and\ \citenamefont {Li}}]{gao2021A}%
  \BibitemOpen
  \bibfield  {author} {\bibinfo {author} {\bibfnamefont {K.}~\bibnamefont
  {Gao}}, \bibinfo {author} {\bibfnamefont {W.}~\bibnamefont {Cui}}, \bibinfo
  {author} {\bibfnamefont {J.}~\bibnamefont {Chen}}, \bibinfo {author}
  {\bibfnamefont {Q.}~\bibnamefont {Wang}}, \bibinfo {author} {\bibfnamefont
  {J.}~\bibnamefont {Hao}}, \bibinfo {author} {\bibfnamefont {J.}~\bibnamefont
  {Shi}}, \bibinfo {author} {\bibfnamefont {C.}~\bibnamefont {Liu}}, \bibinfo
  {author} {\bibfnamefont {S.}~\bibnamefont {Botti}}, \bibinfo {author}
  {\bibfnamefont {M.~A.~L.}\ \bibnamefont {Marques}}, \ and\ \bibinfo {author}
  {\bibfnamefont {Y.}~\bibnamefont {Li}},\ }\bibfield  {title} {\enquote
  {\bibinfo {title} {Superconducting hydrogen tubes in hafnium hydrides at high
  pressure},}\ }\href {\doibase 10.1103/PhysRevB.104.214511} {\bibfield
  {journal} {\bibinfo  {journal} {Phys. Rev. B}\ }\textbf {\bibinfo {volume}
  {104}},\ \bibinfo {pages} {214511} (\bibinfo {year} {2021})}\BibitemShut
  {NoStop}%
\bibitem [{\citenamefont {Gebreyohannes}\ \emph
  {et~al.}(2022{\natexlab{b}})\citenamefont {Gebreyohannes}, \citenamefont
  {Geffe},\ and\ \citenamefont {Singh}}]{gebreyohannes2022A}%
  \BibitemOpen
  \bibfield  {author} {\bibinfo {author} {\bibfnamefont {M.~G.}\ \bibnamefont
  {Gebreyohannes}}, \bibinfo {author} {\bibfnamefont {C.~A.}\ \bibnamefont
  {Geffe}}, \ and\ \bibinfo {author} {\bibfnamefont {P.}~\bibnamefont
  {Singh}},\ }\bibfield  {title} {\enquote {\bibinfo {title} {Computational
  study of pressurized tetragonal magnesium hydride (mgh4) as a potential
  candidate for high-temperature superconducting material},}\ }\href {\doibase
  10.1088/2053-1591/ac5e22} {\bibfield  {journal} {\bibinfo  {journal}
  {Materials Research Express}\ }\textbf {\bibinfo {volume} {9}},\ \bibinfo
  {pages} {036001} (\bibinfo {year} {2022}{\natexlab{b}})}\BibitemShut
  {NoStop}%
\bibitem [{\citenamefont {Gu}\ \emph {et~al.}(2017)\citenamefont {Gu},
  \citenamefont {Lu}, \citenamefont {Xia}, \citenamefont {Sun},\ and\
  \citenamefont {Xing}}]{gu2017}%
  \BibitemOpen
  \bibfield  {author} {\bibinfo {author} {\bibfnamefont {Q.}~\bibnamefont
  {Gu}}, \bibinfo {author} {\bibfnamefont {P.}~\bibnamefont {Lu}}, \bibinfo
  {author} {\bibfnamefont {K.}~\bibnamefont {Xia}}, \bibinfo {author}
  {\bibfnamefont {J.}~\bibnamefont {Sun}}, \ and\ \bibinfo {author}
  {\bibfnamefont {D.}~\bibnamefont {Xing}},\ }\bibfield  {title} {\enquote
  {\bibinfo {title} {High-temperature superconducting phase of hbr under
  pressure predicted by first-principles calculations},}\ }\href {\doibase
  10.1103/PhysRevB.96.064517} {\bibfield  {journal} {\bibinfo  {journal} {Phys.
  Rev. B}\ }\textbf {\bibinfo {volume} {96}},\ \bibinfo {pages} {064517}
  (\bibinfo {year} {2017})}\BibitemShut {NoStop}%
\bibitem [{\citenamefont {Hai}\ \emph {et~al.}(2021)\citenamefont {Hai},
  \citenamefont {Lu}, \citenamefont {Tian}, \citenamefont {Jiang},
  \citenamefont {Yang}, \citenamefont {Li}, \citenamefont {Yan}, \citenamefont
  {Zhang}, \citenamefont {Chen},\ and\ \citenamefont {Zhong}}]{hai2021}%
  \BibitemOpen
  \bibfield  {author} {\bibinfo {author} {\bibfnamefont {Y.-L.}\ \bibnamefont
  {Hai}}, \bibinfo {author} {\bibfnamefont {N.}~\bibnamefont {Lu}}, \bibinfo
  {author} {\bibfnamefont {H.-L.}\ \bibnamefont {Tian}}, \bibinfo {author}
  {\bibfnamefont {M.-J.}\ \bibnamefont {Jiang}}, \bibinfo {author}
  {\bibfnamefont {W.}~\bibnamefont {Yang}}, \bibinfo {author} {\bibfnamefont
  {W.-J.}\ \bibnamefont {Li}}, \bibinfo {author} {\bibfnamefont {X.-W.}\
  \bibnamefont {Yan}}, \bibinfo {author} {\bibfnamefont {C.}~\bibnamefont
  {Zhang}}, \bibinfo {author} {\bibfnamefont {X.-J.}\ \bibnamefont {Chen}}, \
  and\ \bibinfo {author} {\bibfnamefont {G.-H.}\ \bibnamefont {Zhong}},\
  }\bibfield  {title} {\enquote {\bibinfo {title} {Cage structure and near
  room-temperature superconductivity in tbhn (n = 1–12)},}\ }\href {\doibase
  10.1021/acs.jpcc.1c00645} {\bibfield  {journal} {\bibinfo  {journal} {The
  Journal of Physical Chemistry C}\ }\textbf {\bibinfo {volume} {125}},\
  \bibinfo {pages} {3640} (\bibinfo {year} {2021})}\BibitemShut {NoStop}%
\bibitem [{\citenamefont {zhong Han}\ \emph {et~al.}(2017)\citenamefont {zhong
  Han}, \citenamefont {Lu}, \citenamefont {Wang}, \citenamefont {ling Hou},\
  and\ \citenamefont {hong Shao}}]{han2017}%
  \BibitemOpen
  \bibfield  {author} {\bibinfo {author} {\bibfnamefont {Z.}~\bibnamefont
  {zhong Han}}, \bibinfo {author} {\bibfnamefont {Y.}~\bibnamefont {Lu}},
  \bibinfo {author} {\bibfnamefont {W.}~\bibnamefont {Wang}}, \bibinfo {author}
  {\bibfnamefont {Z.}~\bibnamefont {ling Hou}}, \ and\ \bibinfo {author}
  {\bibfnamefont {X.}~\bibnamefont {hong Shao}},\ }\bibfield  {title} {\enquote
  {\bibinfo {title} {The novel structure and superconductivity of zirconium
  hydride},}\ }\href {\doibase https://doi.org/10.1016/j.commatsci.2017.03.021}
  {\bibfield  {journal} {\bibinfo  {journal} {Computational Materials Science}\
  }\textbf {\bibinfo {volume} {134}},\ \bibinfo {pages} {38} (\bibinfo {year}
  {2017})}\BibitemShut {NoStop}%
\bibitem [{\citenamefont {He}\ \emph {et~al.}(2023)\citenamefont {He},
  \citenamefont {Zhang}, \citenamefont {Li}, \citenamefont {Zhang},
  \citenamefont {Min}, \citenamefont {Zhang}, \citenamefont {Lu}, \citenamefont
  {Zhao}, \citenamefont {Shi}, \citenamefont {Peng}, \citenamefont {Wang},
  \citenamefont {Feng}, \citenamefont {Song}, \citenamefont {Wang},
  \citenamefont {Prakapenka}, \citenamefont {Chariton}, \citenamefont {Liu},\
  and\ \citenamefont {Jin}}]{he2023}%
  \BibitemOpen
  \bibfield  {author} {\bibinfo {author} {\bibfnamefont {X.}~\bibnamefont
  {He}}, \bibinfo {author} {\bibfnamefont {C.~L.}\ \bibnamefont {Zhang}},
  \bibinfo {author} {\bibfnamefont {Z.~W.}\ \bibnamefont {Li}}, \bibinfo
  {author} {\bibfnamefont {S.~J.}\ \bibnamefont {Zhang}}, \bibinfo {author}
  {\bibfnamefont {B.~S.}\ \bibnamefont {Min}}, \bibinfo {author} {\bibfnamefont
  {J.}~\bibnamefont {Zhang}}, \bibinfo {author} {\bibfnamefont
  {K.}~\bibnamefont {Lu}}, \bibinfo {author} {\bibfnamefont {J.~F.}\
  \bibnamefont {Zhao}}, \bibinfo {author} {\bibfnamefont {L.~C.}\ \bibnamefont
  {Shi}}, \bibinfo {author} {\bibfnamefont {Y.}~\bibnamefont {Peng}}, \bibinfo
  {author} {\bibfnamefont {X.~C.}\ \bibnamefont {Wang}}, \bibinfo {author}
  {\bibfnamefont {S.~M.}\ \bibnamefont {Feng}}, \bibinfo {author}
  {\bibfnamefont {J.}~\bibnamefont {Song}}, \bibinfo {author} {\bibfnamefont
  {L.~H.}\ \bibnamefont {Wang}}, \bibinfo {author} {\bibfnamefont {V.~B.}\
  \bibnamefont {Prakapenka}}, \bibinfo {author} {\bibfnamefont
  {S.}~\bibnamefont {Chariton}}, \bibinfo {author} {\bibfnamefont {H.~Z.}\
  \bibnamefont {Liu}}, \ and\ \bibinfo {author} {\bibfnamefont {C.~Q.}\
  \bibnamefont {Jin}},\ }\bibfield  {title} {\enquote {\bibinfo {title}
  {Superconductivity observed in tantalum polyhydride at high pressure},}\
  }\href {\doibase 10.1088/0256-307X/40/5/057404} {\bibfield  {journal}
  {\bibinfo  {journal} {Chinese Physics Letters}\ }\textbf {\bibinfo {volume}
  {40}},\ \bibinfo {pages} {057404} (\bibinfo {year} {2023})}\BibitemShut
  {NoStop}%
\bibitem [{\citenamefont {Heil}\ \emph {et~al.}(2019)\citenamefont {Heil},
  \citenamefont {di~Cataldo}, \citenamefont {Bachelet},\ and\ \citenamefont
  {Boeri}}]{heil2019}%
  \BibitemOpen
  \bibfield  {author} {\bibinfo {author} {\bibfnamefont {C.}~\bibnamefont
  {Heil}}, \bibinfo {author} {\bibfnamefont {S.}~\bibnamefont {di~Cataldo}},
  \bibinfo {author} {\bibfnamefont {G.~B.}\ \bibnamefont {Bachelet}}, \ and\
  \bibinfo {author} {\bibfnamefont {L.}~\bibnamefont {Boeri}},\ }\bibfield
  {title} {\enquote {\bibinfo {title} {Superconductivity in sodalite-like
  yttrium hydride clathrates},}\ }\href {\doibase 10.1103/PhysRevB.99.220502}
  {\bibfield  {journal} {\bibinfo  {journal} {Phys. Rev. B}\ }\textbf {\bibinfo
  {volume} {99}},\ \bibinfo {pages} {220502} (\bibinfo {year}
  {2019})}\BibitemShut {NoStop}%
\bibitem [{\citenamefont {Hong}\ \emph {et~al.}(2022)\citenamefont {Hong},
  \citenamefont {Shan}, \citenamefont {Yang}, \citenamefont {Yue},
  \citenamefont {Yang}, \citenamefont {Liu}, \citenamefont {Sun}, \citenamefont
  {Dai}, \citenamefont {Yu}, \citenamefont {Yin}, \citenamefont {Yu},
  \citenamefont {Cheng},\ and\ \citenamefont {Zhao}}]{hong2022}%
  \BibitemOpen
  \bibfield  {author} {\bibinfo {author} {\bibfnamefont {F.}~\bibnamefont
  {Hong}}, \bibinfo {author} {\bibfnamefont {P.}~\bibnamefont {Shan}}, \bibinfo
  {author} {\bibfnamefont {L.}~\bibnamefont {Yang}}, \bibinfo {author}
  {\bibfnamefont {B.}~\bibnamefont {Yue}}, \bibinfo {author} {\bibfnamefont
  {P.}~\bibnamefont {Yang}}, \bibinfo {author} {\bibfnamefont {Z.}~\bibnamefont
  {Liu}}, \bibinfo {author} {\bibfnamefont {J.}~\bibnamefont {Sun}}, \bibinfo
  {author} {\bibfnamefont {J.}~\bibnamefont {Dai}}, \bibinfo {author}
  {\bibfnamefont {H.}~\bibnamefont {Yu}}, \bibinfo {author} {\bibfnamefont
  {Y.}~\bibnamefont {Yin}}, \bibinfo {author} {\bibfnamefont {X.}~\bibnamefont
  {Yu}}, \bibinfo {author} {\bibfnamefont {J.}~\bibnamefont {Cheng}}, \ and\
  \bibinfo {author} {\bibfnamefont {Z.}~\bibnamefont {Zhao}},\ }\bibfield
  {title} {\enquote {\bibinfo {title} {Possible superconductivity at 70 k in
  tin hydride snhx under high pressure},}\ }\href {\doibase
  https://doi.org/10.1016/j.mtphys.2021.100596} {\bibfield  {journal} {\bibinfo
   {journal} {Materials Today Physics}\ }\textbf {\bibinfo {volume} {22}},\
  \bibinfo {pages} {100596} (\bibinfo {year} {2022})}\BibitemShut {NoStop}%
\bibitem [{\citenamefont {Hooper}\ \emph {et~al.}(2013)\citenamefont {Hooper},
  \citenamefont {Altintas}, \citenamefont {Shamp},\ and\ \citenamefont
  {Zurek}}]{hooper2013}%
  \BibitemOpen
  \bibfield  {author} {\bibinfo {author} {\bibfnamefont {J.}~\bibnamefont
  {Hooper}}, \bibinfo {author} {\bibfnamefont {B.}~\bibnamefont {Altintas}},
  \bibinfo {author} {\bibfnamefont {A.}~\bibnamefont {Shamp}}, \ and\ \bibinfo
  {author} {\bibfnamefont {E.}~\bibnamefont {Zurek}},\ }\bibfield  {title}
  {\enquote {\bibinfo {title} {Polyhydrides of the alkaline earth metals: A
  look at the extremes under pressure},}\ }\href {\doibase 10.1021/jp311571n}
  {\bibfield  {journal} {\bibinfo  {journal} {The Journal of Physical Chemistry
  C}\ }\textbf {\bibinfo {volume} {117}},\ \bibinfo {pages} {2982} (\bibinfo
  {year} {2013})}\BibitemShut {NoStop}%
\bibitem [{\citenamefont {Hou}\ \emph {et~al.}(2015{\natexlab{a}})\citenamefont
  {Hou}, \citenamefont {Zhao}, \citenamefont {Tian}, \citenamefont {Li},
  \citenamefont {Duan}, \citenamefont {Zhao}, \citenamefont {Chu},
  \citenamefont {Liu},\ and\ \citenamefont {Cui}}]{hou2015}%
  \BibitemOpen
  \bibfield  {author} {\bibinfo {author} {\bibfnamefont {P.}~\bibnamefont
  {Hou}}, \bibinfo {author} {\bibfnamefont {X.}~\bibnamefont {Zhao}}, \bibinfo
  {author} {\bibfnamefont {F.}~\bibnamefont {Tian}}, \bibinfo {author}
  {\bibfnamefont {D.}~\bibnamefont {Li}}, \bibinfo {author} {\bibfnamefont
  {D.}~\bibnamefont {Duan}}, \bibinfo {author} {\bibfnamefont {Z.}~\bibnamefont
  {Zhao}}, \bibinfo {author} {\bibfnamefont {B.}~\bibnamefont {Chu}}, \bibinfo
  {author} {\bibfnamefont {B.}~\bibnamefont {Liu}}, \ and\ \bibinfo {author}
  {\bibfnamefont {T.}~\bibnamefont {Cui}},\ }\bibfield  {title} {\enquote
  {\bibinfo {title} {High pressure structures and superconductivity of alh3(h2)
  predicted by first principles},}\ }\href {\doibase 10.1039/C4RA14990D}
  {\bibfield  {journal} {\bibinfo  {journal} {RSC Adv.}\ }\textbf {\bibinfo
  {volume} {5}},\ \bibinfo {pages} {5096} (\bibinfo {year}
  {2015}{\natexlab{a}})}\BibitemShut {NoStop}%
\bibitem [{\citenamefont {Hou}\ \emph {et~al.}(2015{\natexlab{b}})\citenamefont
  {Hou}, \citenamefont {Tian}, \citenamefont {Li}, \citenamefont {Zhao},
  \citenamefont {Duan}, \citenamefont {Zhang}, \citenamefont {Sha},
  \citenamefont {Liu},\ and\ \citenamefont {Cui}}]{hou2015A}%
  \BibitemOpen
  \bibfield  {author} {\bibinfo {author} {\bibfnamefont {P.}~\bibnamefont
  {Hou}}, \bibinfo {author} {\bibfnamefont {F.}~\bibnamefont {Tian}}, \bibinfo
  {author} {\bibfnamefont {D.}~\bibnamefont {Li}}, \bibinfo {author}
  {\bibfnamefont {Z.}~\bibnamefont {Zhao}}, \bibinfo {author} {\bibfnamefont
  {D.}~\bibnamefont {Duan}}, \bibinfo {author} {\bibfnamefont {H.}~\bibnamefont
  {Zhang}}, \bibinfo {author} {\bibfnamefont {X.}~\bibnamefont {Sha}}, \bibinfo
  {author} {\bibfnamefont {B.}~\bibnamefont {Liu}}, \ and\ \bibinfo {author}
  {\bibfnamefont {T.}~\bibnamefont {Cui}},\ }\bibfield  {title} {\enquote
  {\bibinfo {title} {Ab initio study of germanium-hydride compounds under high
  pressure},}\ }\href {\doibase 10.1039/C4RA13183E} {\bibfield  {journal}
  {\bibinfo  {journal} {RSC Adv.}\ }\textbf {\bibinfo {volume} {5}},\ \bibinfo
  {pages} {19432} (\bibinfo {year} {2015}{\natexlab{b}})}\BibitemShut {NoStop}%
\bibitem [{\citenamefont {Hu}\ \emph {et~al.}(2013)\citenamefont {Hu},
  \citenamefont {Oganov}, \citenamefont {Zhu}, \citenamefont {Qian},
  \citenamefont {Frapper}, \citenamefont {Lyakhov},\ and\ \citenamefont
  {Zhou}}]{hu2013}%
  \BibitemOpen
  \bibfield  {author} {\bibinfo {author} {\bibfnamefont {C.-H.}\ \bibnamefont
  {Hu}}, \bibinfo {author} {\bibfnamefont {A.~R.}\ \bibnamefont {Oganov}},
  \bibinfo {author} {\bibfnamefont {Q.}~\bibnamefont {Zhu}}, \bibinfo {author}
  {\bibfnamefont {G.-R.}\ \bibnamefont {Qian}}, \bibinfo {author}
  {\bibfnamefont {G.}~\bibnamefont {Frapper}}, \bibinfo {author} {\bibfnamefont
  {A.~O.}\ \bibnamefont {Lyakhov}}, \ and\ \bibinfo {author} {\bibfnamefont
  {H.-Y.}\ \bibnamefont {Zhou}},\ }\bibfield  {title} {\enquote {\bibinfo
  {title} {Pressure-induced stabilization and insulator-superconductor
  transition of bh},}\ }\href {\doibase 10.1103/PhysRevLett.110.165504}
  {\bibfield  {journal} {\bibinfo  {journal} {Phys. Rev. Lett.}\ }\textbf
  {\bibinfo {volume} {110}},\ \bibinfo {pages} {165504} (\bibinfo {year}
  {2013})}\BibitemShut {NoStop}%
\bibitem [{\citenamefont {Huo}\ \emph {et~al.}(2022)\citenamefont {Huo},
  \citenamefont {Zhuang}, \citenamefont {Jin}, \citenamefont {An},
  \citenamefont {Liu}, \citenamefont {Song},\ and\ \citenamefont
  {Cui}}]{huo2022}%
  \BibitemOpen
  \bibfield  {author} {\bibinfo {author} {\bibfnamefont {Z.}~\bibnamefont
  {Huo}}, \bibinfo {author} {\bibfnamefont {Q.}~\bibnamefont {Zhuang}},
  \bibinfo {author} {\bibfnamefont {X.}~\bibnamefont {Jin}}, \bibinfo {author}
  {\bibfnamefont {L.}~\bibnamefont {An}}, \bibinfo {author} {\bibfnamefont
  {Y.}~\bibnamefont {Liu}}, \bibinfo {author} {\bibfnamefont {L.}~\bibnamefont
  {Song}}, \ and\ \bibinfo {author} {\bibfnamefont {T.}~\bibnamefont {Cui}},\
  }\bibfield  {title} {\enquote {\bibinfo {title} {Effect of hydrogen content
  on superconductivity in la–h compounds},}\ }\href {\doibase
  https://doi.org/10.1016/j.rinp.2022.106060} {\bibfield  {journal} {\bibinfo
  {journal} {Results in Physics}\ }\textbf {\bibinfo {volume} {43}},\ \bibinfo
  {pages} {106060} (\bibinfo {year} {2022})}\BibitemShut {NoStop}%
\bibitem [{\citenamefont {Hutcheon}\ \emph {et~al.}(2020)\citenamefont
  {Hutcheon}, \citenamefont {Shipley},\ and\ \citenamefont
  {Needs}}]{hutcheon2020}%
  \BibitemOpen
  \bibfield  {author} {\bibinfo {author} {\bibfnamefont {M.~J.}\ \bibnamefont
  {Hutcheon}}, \bibinfo {author} {\bibfnamefont {A.~M.}\ \bibnamefont
  {Shipley}}, \ and\ \bibinfo {author} {\bibfnamefont {R.~J.}\ \bibnamefont
  {Needs}},\ }\bibfield  {title} {\enquote {\bibinfo {title} {Predicting novel
  superconducting hydrides using machine learning approaches},}\ }\href
  {\doibase 10.1103/PhysRevB.101.144505} {\bibfield  {journal} {\bibinfo
  {journal} {Phys. Rev. B}\ }\textbf {\bibinfo {volume} {101}},\ \bibinfo
  {pages} {144505} (\bibinfo {year} {2020})}\BibitemShut {NoStop}%
\bibitem [{\citenamefont {Ishikawa}\ \emph {et~al.}(2016)\citenamefont
  {Ishikawa}, \citenamefont {Nakanishi}, \citenamefont {Shimizu}, \citenamefont
  {Katayama-Yoshida}, \citenamefont {Oda},\ and\ \citenamefont
  {Suzuki}}]{ishikawa2016}%
  \BibitemOpen
  \bibfield  {author} {\bibinfo {author} {\bibfnamefont {T.}~\bibnamefont
  {Ishikawa}}, \bibinfo {author} {\bibfnamefont {A.}~\bibnamefont {Nakanishi}},
  \bibinfo {author} {\bibfnamefont {K.}~\bibnamefont {Shimizu}}, \bibinfo
  {author} {\bibfnamefont {H.}~\bibnamefont {Katayama-Yoshida}}, \bibinfo
  {author} {\bibfnamefont {T.}~\bibnamefont {Oda}}, \ and\ \bibinfo {author}
  {\bibfnamefont {N.}~\bibnamefont {Suzuki}},\ }\bibfield  {title} {\enquote
  {\bibinfo {title} {Superconducting h5s2 phase in sulfur-hydrogen system under
  high-pressure},}\ }\href {\doibase 10.1038/srep23160} {\bibfield  {journal}
  {\bibinfo  {journal} {Scientific Reports}\ }\textbf {\bibinfo {volume} {6}},\
  \bibinfo {pages} {23160} (\bibinfo {year} {2016})}\BibitemShut {NoStop}%
\bibitem [{\citenamefont {Jeon}\ \emph {et~al.}(2022)\citenamefont {Jeon},
  \citenamefont {Wang}, \citenamefont {Liu}, \citenamefont {Bok}, \citenamefont
  {Bang},\ and\ \citenamefont {Cho}}]{jeon2022}%
  \BibitemOpen
  \bibfield  {author} {\bibinfo {author} {\bibfnamefont {H.}~\bibnamefont
  {Jeon}}, \bibinfo {author} {\bibfnamefont {C.}~\bibnamefont {Wang}}, \bibinfo
  {author} {\bibfnamefont {S.}~\bibnamefont {Liu}}, \bibinfo {author}
  {\bibfnamefont {J.~M.}\ \bibnamefont {Bok}}, \bibinfo {author} {\bibfnamefont
  {Y.}~\bibnamefont {Bang}}, \ and\ \bibinfo {author} {\bibfnamefont {J.-H.}\
  \bibnamefont {Cho}},\ }\bibfield  {title} {\enquote {\bibinfo {title}
  {Electron–phonon coupling and superconductivity in an alkaline earth
  hydride cah6 at high pressures},}\ }\href {\doibase 10.1088/1367-2630/ac8a0c}
  {\bibfield  {journal} {\bibinfo  {journal} {New Journal of Physics}\ }\textbf
  {\bibinfo {volume} {24}},\ \bibinfo {pages} {083048} (\bibinfo {year}
  {2022})}\BibitemShut {NoStop}%
\bibitem [{\citenamefont {Jin}\ \emph {et~al.}(2010)\citenamefont {Jin},
  \citenamefont {Meng}, \citenamefont {He}, \citenamefont {Ma}, \citenamefont
  {Liu}, \citenamefont {Cui}, \citenamefont {Zou},\ and\ \citenamefont {kwang
  Mao}}]{jin2010}%
  \BibitemOpen
  \bibfield  {author} {\bibinfo {author} {\bibfnamefont {X.}~\bibnamefont
  {Jin}}, \bibinfo {author} {\bibfnamefont {X.}~\bibnamefont {Meng}}, \bibinfo
  {author} {\bibfnamefont {Z.}~\bibnamefont {He}}, \bibinfo {author}
  {\bibfnamefont {Y.}~\bibnamefont {Ma}}, \bibinfo {author} {\bibfnamefont
  {B.}~\bibnamefont {Liu}}, \bibinfo {author} {\bibfnamefont {T.}~\bibnamefont
  {Cui}}, \bibinfo {author} {\bibfnamefont {G.}~\bibnamefont {Zou}}, \ and\
  \bibinfo {author} {\bibfnamefont {H.}~\bibnamefont {kwang Mao}},\ }\bibfield
  {title} {\enquote {\bibinfo {title} {Superconducting high-pressure phases of
  disilane},}\ }\href {\doibase 10.1073/pnas.1005242107} {\bibfield  {journal}
  {\bibinfo  {journal} {Proceedings of the National Academy of Sciences}\
  }\textbf {\bibinfo {volume} {107}},\ \bibinfo {pages} {9969} (\bibinfo {year}
  {2010})}\BibitemShut {NoStop}%
\bibitem [{\citenamefont {Kim}\ \emph {et~al.}(2010)\citenamefont {Kim},
  \citenamefont {Scheicher}, \citenamefont {kwang Mao}, \citenamefont {Kang},\
  and\ \citenamefont {Ahuja}}]{kim2010}%
  \BibitemOpen
  \bibfield  {author} {\bibinfo {author} {\bibfnamefont {D.~Y.}\ \bibnamefont
  {Kim}}, \bibinfo {author} {\bibfnamefont {R.~H.}\ \bibnamefont {Scheicher}},
  \bibinfo {author} {\bibfnamefont {H.}~\bibnamefont {kwang Mao}}, \bibinfo
  {author} {\bibfnamefont {T.~W.}\ \bibnamefont {Kang}}, \ and\ \bibinfo
  {author} {\bibfnamefont {R.}~\bibnamefont {Ahuja}},\ }\bibfield  {title}
  {\enquote {\bibinfo {title} {General trend for pressurized superconducting
  hydrogen-dense materials},}\ }\href {\doibase 10.1073/pnas.0914462107}
  {\bibfield  {journal} {\bibinfo  {journal} {Proceedings of the National
  Academy of Sciences}\ }\textbf {\bibinfo {volume} {107}},\ \bibinfo {pages}
  {2793} (\bibinfo {year} {2010})}\BibitemShut {NoStop}%
\bibitem [{\citenamefont {Kim}\ \emph {et~al.}(2011)\citenamefont {Kim},
  \citenamefont {Scheicher}, \citenamefont {Pickard}, \citenamefont {Needs},\
  and\ \citenamefont {Ahuja}}]{kim2011}%
  \BibitemOpen
  \bibfield  {author} {\bibinfo {author} {\bibfnamefont {D.~Y.}\ \bibnamefont
  {Kim}}, \bibinfo {author} {\bibfnamefont {R.~H.}\ \bibnamefont {Scheicher}},
  \bibinfo {author} {\bibfnamefont {C.~J.}\ \bibnamefont {Pickard}}, \bibinfo
  {author} {\bibfnamefont {R.~J.}\ \bibnamefont {Needs}}, \ and\ \bibinfo
  {author} {\bibfnamefont {R.}~\bibnamefont {Ahuja}},\ }\bibfield  {title}
  {\enquote {\bibinfo {title} {Predicted formation of superconducting
  platinum-hydride crystals under pressure in the presence of molecular
  hydrogen},}\ }\href {\doibase 10.1103/PhysRevLett.107.117002} {\bibfield
  {journal} {\bibinfo  {journal} {Phys. Rev. Lett.}\ }\textbf {\bibinfo
  {volume} {107}},\ \bibinfo {pages} {117002} (\bibinfo {year}
  {2011})}\BibitemShut {NoStop}%
\bibitem [{\citenamefont {Kruglov}\ \emph {et~al.}(2018)\citenamefont
  {Kruglov}, \citenamefont {Kvashnin}, \citenamefont {Goncharov}, \citenamefont
  {Oganov}, \citenamefont {Lobanov}, \citenamefont {Holtgrewe}, \citenamefont
  {Jiang}, \citenamefont {Prakapenka}, \citenamefont {Greenberg},\ and\
  \citenamefont {Yanilkin}}]{kruglov2018}%
  \BibitemOpen
  \bibfield  {author} {\bibinfo {author} {\bibfnamefont {I.~A.}\ \bibnamefont
  {Kruglov}}, \bibinfo {author} {\bibfnamefont {A.~G.}\ \bibnamefont
  {Kvashnin}}, \bibinfo {author} {\bibfnamefont {A.~F.}\ \bibnamefont
  {Goncharov}}, \bibinfo {author} {\bibfnamefont {A.~R.}\ \bibnamefont
  {Oganov}}, \bibinfo {author} {\bibfnamefont {S.~S.}\ \bibnamefont {Lobanov}},
  \bibinfo {author} {\bibfnamefont {N.}~\bibnamefont {Holtgrewe}}, \bibinfo
  {author} {\bibfnamefont {S.}~\bibnamefont {Jiang}}, \bibinfo {author}
  {\bibfnamefont {V.~B.}\ \bibnamefont {Prakapenka}}, \bibinfo {author}
  {\bibfnamefont {E.}~\bibnamefont {Greenberg}}, \ and\ \bibinfo {author}
  {\bibfnamefont {A.~V.}\ \bibnamefont {Yanilkin}},\ }\bibfield  {title}
  {\enquote {\bibinfo {title} {Uranium polyhydrides at moderate pressures:
  Prediction, synthesis, and expected superconductivity},}\ }\href {\doibase
  10.1126/sciadv.aat9776} {\bibfield  {journal} {\bibinfo  {journal} {Science
  Advances}\ }\textbf {\bibinfo {volume} {4}},\ \bibinfo {pages} {eaat9776}
  (\bibinfo {year} {2018})}\BibitemShut {NoStop}%
\bibitem [{\citenamefont {Kruglov}\ \emph {et~al.}(2020)\citenamefont
  {Kruglov}, \citenamefont {Semenok}, \citenamefont {Song}, \citenamefont
  {Szczesniak}, \citenamefont {Wrona}, \citenamefont {Akashi}, \citenamefont
  {Davari~Esfahani}, \citenamefont {Duan}, \citenamefont {Cui}, \citenamefont
  {Kvashnin},\ and\ \citenamefont {Oganov}}]{kruglov2020}%
  \BibitemOpen
  \bibfield  {author} {\bibinfo {author} {\bibfnamefont {I.~A.}\ \bibnamefont
  {Kruglov}}, \bibinfo {author} {\bibfnamefont {D.~V.}\ \bibnamefont
  {Semenok}}, \bibinfo {author} {\bibfnamefont {H.}~\bibnamefont {Song}},
  \bibinfo {author} {\bibfnamefont {R.}~\bibnamefont {Szczesniak}}, \bibinfo
  {author} {\bibfnamefont {I.~A.}\ \bibnamefont {Wrona}}, \bibinfo {author}
  {\bibfnamefont {R.}~\bibnamefont {Akashi}}, \bibinfo {author} {\bibfnamefont
  {M.~M.}\ \bibnamefont {Davari~Esfahani}}, \bibinfo {author} {\bibfnamefont
  {D.}~\bibnamefont {Duan}}, \bibinfo {author} {\bibfnamefont {T.}~\bibnamefont
  {Cui}}, \bibinfo {author} {\bibfnamefont {A.~G.}\ \bibnamefont {Kvashnin}}, \
  and\ \bibinfo {author} {\bibfnamefont {A.~R.}\ \bibnamefont {Oganov}},\
  }\bibfield  {title} {\enquote {\bibinfo {title} {Superconductivity of
  ${\mathrm{lah}}_{10}$ and ${\mathrm{lah}}_{16}$ polyhydrides},}\ }\href
  {\doibase 10.1103/PhysRevB.101.024508} {\bibfield  {journal} {\bibinfo
  {journal} {Phys. Rev. B}\ }\textbf {\bibinfo {volume} {101}},\ \bibinfo
  {pages} {024508} (\bibinfo {year} {2020})}\BibitemShut {NoStop}%
\bibitem [{\citenamefont {Kuzovnikov}\ and\ \citenamefont
  {Tkacz}(2019)}]{kuzovnikov2019}%
  \BibitemOpen
  \bibfield  {author} {\bibinfo {author} {\bibfnamefont {M.~A.}\ \bibnamefont
  {Kuzovnikov}}\ and\ \bibinfo {author} {\bibfnamefont {M.}~\bibnamefont
  {Tkacz}},\ }\bibfield  {title} {\enquote {\bibinfo {title} {High-pressure
  synthesis of novel polyhydrides of zr and hf with a th4h15-type structure},}\
  }\href {\doibase 10.1021/acs.jpcc.9b07918} {\bibfield  {journal} {\bibinfo
  {journal} {The Journal of Physical Chemistry C}\ }\textbf {\bibinfo {volume}
  {123}},\ \bibinfo {pages} {30059} (\bibinfo {year} {2019})}\BibitemShut
  {NoStop}%
\bibitem [{\citenamefont {Kvashnin}\ \emph
  {et~al.}(2018{\natexlab{a}})\citenamefont {Kvashnin}, \citenamefont
  {Semenok}, \citenamefont {Kruglov}, \citenamefont {Wrona},\ and\
  \citenamefont {Oganov}}]{kvashnin2018}%
  \BibitemOpen
  \bibfield  {author} {\bibinfo {author} {\bibfnamefont {A.~G.}\ \bibnamefont
  {Kvashnin}}, \bibinfo {author} {\bibfnamefont {D.~V.}\ \bibnamefont
  {Semenok}}, \bibinfo {author} {\bibfnamefont {I.~A.}\ \bibnamefont
  {Kruglov}}, \bibinfo {author} {\bibfnamefont {I.~A.}\ \bibnamefont {Wrona}},
  \ and\ \bibinfo {author} {\bibfnamefont {A.~R.}\ \bibnamefont {Oganov}},\
  }\bibfield  {title} {\enquote {\bibinfo {title} {High-temperature
  superconductivity in a th–h system under pressure conditions},}\ }\href
  {\doibase 10.1021/acsami.8b17100} {\bibfield  {journal} {\bibinfo  {journal}
  {ACS Applied Materials \& Interfaces}\ }\textbf {\bibinfo {volume} {10}},\
  \bibinfo {pages} {43809} (\bibinfo {year} {2018}{\natexlab{a}})}\BibitemShut
  {NoStop}%
\bibitem [{\citenamefont {Kvashnin}\ \emph
  {et~al.}(2018{\natexlab{b}})\citenamefont {Kvashnin}, \citenamefont
  {Kruglov}, \citenamefont {Semenok},\ and\ \citenamefont
  {Oganov}}]{kvashnin2018A}%
  \BibitemOpen
  \bibfield  {author} {\bibinfo {author} {\bibfnamefont {A.~G.}\ \bibnamefont
  {Kvashnin}}, \bibinfo {author} {\bibfnamefont {I.~A.}\ \bibnamefont
  {Kruglov}}, \bibinfo {author} {\bibfnamefont {D.~V.}\ \bibnamefont
  {Semenok}}, \ and\ \bibinfo {author} {\bibfnamefont {A.~R.}\ \bibnamefont
  {Oganov}},\ }\bibfield  {title} {\enquote {\bibinfo {title} {Iron
  superhydrides feh5 and feh6: Stability, electronic properties, and
  superconductivity},}\ }\href {\doibase 10.1021/acs.jpcc.8b01270} {\bibfield
  {journal} {\bibinfo  {journal} {The Journal of Physical Chemistry C}\
  }\textbf {\bibinfo {volume} {122}},\ \bibinfo {pages} {4731} (\bibinfo {year}
  {2018}{\natexlab{b}})}\BibitemShut {NoStop}%
\bibitem [{\citenamefont {Li}\ \emph {et~al.}(2010)\citenamefont {Li},
  \citenamefont {Gao}, \citenamefont {Xie}, \citenamefont {Ma}, \citenamefont
  {Cui},\ and\ \citenamefont {Zou}}]{li2010}%
  \BibitemOpen
  \bibfield  {author} {\bibinfo {author} {\bibfnamefont {Y.}~\bibnamefont
  {Li}}, \bibinfo {author} {\bibfnamefont {G.}~\bibnamefont {Gao}}, \bibinfo
  {author} {\bibfnamefont {Y.}~\bibnamefont {Xie}}, \bibinfo {author}
  {\bibfnamefont {Y.}~\bibnamefont {Ma}}, \bibinfo {author} {\bibfnamefont
  {T.}~\bibnamefont {Cui}}, \ and\ \bibinfo {author} {\bibfnamefont
  {G.}~\bibnamefont {Zou}},\ }\bibfield  {title} {\enquote {\bibinfo {title}
  {Superconductivity at \&\#x223c;100\&\#xa0;k in dense
  sih<sub>4</sub>(h<sub>2</sub>)<sub>2</sub> predicted by first principles},}\
  }\href {\doibase 10.1073/pnas.1007354107} {\bibfield  {journal} {\bibinfo
  {journal} {Proceedings of the National Academy of Sciences}\ }\textbf
  {\bibinfo {volume} {107}},\ \bibinfo {pages} {15708} (\bibinfo {year}
  {2010})}\BibitemShut {NoStop}%
\bibitem [{\citenamefont {Li}\ \emph {et~al.}(2014)\citenamefont {Li},
  \citenamefont {Hao}, \citenamefont {Liu}, \citenamefont {Li},\ and\
  \citenamefont {Ma}}]{li2014}%
  \BibitemOpen
  \bibfield  {author} {\bibinfo {author} {\bibfnamefont {Y.}~\bibnamefont
  {Li}}, \bibinfo {author} {\bibfnamefont {J.}~\bibnamefont {Hao}}, \bibinfo
  {author} {\bibfnamefont {H.}~\bibnamefont {Liu}}, \bibinfo {author}
  {\bibfnamefont {Y.}~\bibnamefont {Li}}, \ and\ \bibinfo {author}
  {\bibfnamefont {Y.}~\bibnamefont {Ma}},\ }\bibfield  {title} {\enquote
  {\bibinfo {title} {{The metallization and superconductivity of dense hydrogen
  sulfide}},}\ }\href {\doibase 10.1063/1.4874158} {\bibfield  {journal}
  {\bibinfo  {journal} {The Journal of Chemical Physics}\ }\textbf {\bibinfo
  {volume} {140}},\ \bibinfo {pages} {174712} (\bibinfo {year}
  {2014})}\BibitemShut {NoStop}%
\bibitem [{\citenamefont {Li}\ \emph {et~al.}(2015)\citenamefont {Li},
  \citenamefont {Hao}, \citenamefont {Liu}, \citenamefont {Tse}, \citenamefont
  {Wang},\ and\ \citenamefont {Ma}}]{li2015}%
  \BibitemOpen
  \bibfield  {author} {\bibinfo {author} {\bibfnamefont {Y.}~\bibnamefont
  {Li}}, \bibinfo {author} {\bibfnamefont {J.}~\bibnamefont {Hao}}, \bibinfo
  {author} {\bibfnamefont {H.}~\bibnamefont {Liu}}, \bibinfo {author}
  {\bibfnamefont {J.~S.}\ \bibnamefont {Tse}}, \bibinfo {author} {\bibfnamefont
  {Y.}~\bibnamefont {Wang}}, \ and\ \bibinfo {author} {\bibfnamefont
  {Y.}~\bibnamefont {Ma}},\ }\bibfield  {title} {\enquote {\bibinfo {title}
  {Pressure-stabilized superconductive yttrium hydrides},}\ }\href {\doibase
  10.1038/srep09948} {\bibfield  {journal} {\bibinfo  {journal} {Scientific
  Reports}\ }\textbf {\bibinfo {volume} {5}},\ \bibinfo {pages} {9948}
  (\bibinfo {year} {2015})}\BibitemShut {NoStop}%
\bibitem [{\citenamefont {Li}\ \emph {et~al.}(2016)\citenamefont {Li},
  \citenamefont {Liu},\ and\ \citenamefont {Peng}}]{li2016}%
  \BibitemOpen
  \bibfield  {author} {\bibinfo {author} {\bibfnamefont {X.}~\bibnamefont
  {Li}}, \bibinfo {author} {\bibfnamefont {H.}~\bibnamefont {Liu}}, \ and\
  \bibinfo {author} {\bibfnamefont {F.}~\bibnamefont {Peng}},\ }\bibfield
  {title} {\enquote {\bibinfo {title} {Crystal structures and superconductivity
  of technetium hydrides under pressure},}\ }\href {\doibase
  10.1039/C6CP05702K} {\bibfield  {journal} {\bibinfo  {journal} {Phys. Chem.
  Chem. Phys.}\ }\textbf {\bibinfo {volume} {18}},\ \bibinfo {pages} {28791}
  (\bibinfo {year} {2016})}\BibitemShut {NoStop}%
\bibitem [{\citenamefont {Li}\ and\ \citenamefont {Peng}(2017)}]{li2017}%
  \BibitemOpen
  \bibfield  {author} {\bibinfo {author} {\bibfnamefont {X.}~\bibnamefont
  {Li}}\ and\ \bibinfo {author} {\bibfnamefont {F.}~\bibnamefont {Peng}},\
  }\bibfield  {title} {\enquote {\bibinfo {title} {Superconductivity of
  pressure-stabilized vanadium hydrides},}\ }\href {\doibase
  10.1021/acs.inorgchem.7b01686} {\bibfield  {journal} {\bibinfo  {journal}
  {Inorganic Chemistry}\ }\textbf {\bibinfo {volume} {56}},\ \bibinfo {pages}
  {13759} (\bibinfo {year} {2017})}\BibitemShut {NoStop}%
\bibitem [{\citenamefont {Li}\ \emph {et~al.}(2017)\citenamefont {Li},
  \citenamefont {Hu},\ and\ \citenamefont {Huang}}]{li2017A}%
  \BibitemOpen
  \bibfield  {author} {\bibinfo {author} {\bibfnamefont {X.-F.}\ \bibnamefont
  {Li}}, \bibinfo {author} {\bibfnamefont {Z.-Y.}\ \bibnamefont {Hu}}, \ and\
  \bibinfo {author} {\bibfnamefont {B.}~\bibnamefont {Huang}},\ }\bibfield
  {title} {\enquote {\bibinfo {title} {Phase diagram and superconductivity of
  compressed zirconium hydrides},}\ }\href {\doibase 10.1039/C6CP08036G}
  {\bibfield  {journal} {\bibinfo  {journal} {Phys. Chem. Chem. Phys.}\
  }\textbf {\bibinfo {volume} {19}},\ \bibinfo {pages} {3538} (\bibinfo {year}
  {2017})}\BibitemShut {NoStop}%
\bibitem [{\citenamefont {Li}\ \emph {et~al.}(2019{\natexlab{a}})\citenamefont
  {Li}, \citenamefont {Miao}, \citenamefont {Ti}, \citenamefont {Liu},
  \citenamefont {Chen}, \citenamefont {Shi},\ and\ \citenamefont
  {Gregoryanz}}]{li2019}%
  \BibitemOpen
  \bibfield  {author} {\bibinfo {author} {\bibfnamefont {B.}~\bibnamefont
  {Li}}, \bibinfo {author} {\bibfnamefont {Z.}~\bibnamefont {Miao}}, \bibinfo
  {author} {\bibfnamefont {L.}~\bibnamefont {Ti}}, \bibinfo {author}
  {\bibfnamefont {S.}~\bibnamefont {Liu}}, \bibinfo {author} {\bibfnamefont
  {J.}~\bibnamefont {Chen}}, \bibinfo {author} {\bibfnamefont {Z.}~\bibnamefont
  {Shi}}, \ and\ \bibinfo {author} {\bibfnamefont {E.}~\bibnamefont
  {Gregoryanz}},\ }\bibfield  {title} {\enquote {\bibinfo {title} {{Predicted
  high-temperature superconductivity in cerium hydrides at high pressures}},}\
  }\href {\doibase 10.1063/1.5130583} {\bibfield  {journal} {\bibinfo
  {journal} {Journal of Applied Physics}\ }\textbf {\bibinfo {volume} {126}},\
  \bibinfo {pages} {235901} (\bibinfo {year} {2019}{\natexlab{a}})}\BibitemShut
  {NoStop}%
\bibitem [{\citenamefont {Li}\ \emph {et~al.}(2019{\natexlab{b}})\citenamefont
  {Li}, \citenamefont {Li}, \citenamefont {Wang}, \citenamefont {Liu},
  \citenamefont {Li},\ and\ \citenamefont {Liu}}]{li2019A}%
  \BibitemOpen
  \bibfield  {author} {\bibinfo {author} {\bibfnamefont {H.}~\bibnamefont
  {Li}}, \bibinfo {author} {\bibfnamefont {X.}~\bibnamefont {Li}}, \bibinfo
  {author} {\bibfnamefont {H.}~\bibnamefont {Wang}}, \bibinfo {author}
  {\bibfnamefont {G.}~\bibnamefont {Liu}}, \bibinfo {author} {\bibfnamefont
  {Y.}~\bibnamefont {Li}}, \ and\ \bibinfo {author} {\bibfnamefont
  {H.}~\bibnamefont {Liu}},\ }\bibfield  {title} {\enquote {\bibinfo {title}
  {Superconducting tah5 at high pressure},}\ }\href {\doibase
  10.1088/1367-2630/ab5a9a} {\bibfield  {journal} {\bibinfo  {journal} {New
  Journal of Physics}\ }\textbf {\bibinfo {volume} {21}},\ \bibinfo {pages}
  {123009} (\bibinfo {year} {2019}{\natexlab{b}})}\BibitemShut {NoStop}%
\bibitem [{\citenamefont {Li}\ \emph {et~al.}(2020)\citenamefont {Li},
  \citenamefont {Sun}, \citenamefont {Liu}, \citenamefont {Wang},\ and\
  \citenamefont {Liu}}]{li2020}%
  \BibitemOpen
  \bibfield  {author} {\bibinfo {author} {\bibfnamefont {H.}~\bibnamefont
  {Li}}, \bibinfo {author} {\bibfnamefont {Y.}~\bibnamefont {Sun}}, \bibinfo
  {author} {\bibfnamefont {G.}~\bibnamefont {Liu}}, \bibinfo {author}
  {\bibfnamefont {H.}~\bibnamefont {Wang}}, \ and\ \bibinfo {author}
  {\bibfnamefont {H.}~\bibnamefont {Liu}},\ }\bibfield  {title} {\enquote
  {\bibinfo {title} {Superconducting thorium hydrides under high pressure},}\
  }\href {\doibase https://doi.org/10.1016/j.ssc.2020.113820} {\bibfield
  {journal} {\bibinfo  {journal} {Solid State Communications}\ }\textbf
  {\bibinfo {volume} {309}},\ \bibinfo {pages} {113820} (\bibinfo {year}
  {2020})}\BibitemShut {NoStop}%
\bibitem [{\citenamefont {Li}\ \emph {et~al.}(2022)\citenamefont {Li},
  \citenamefont {Zhang}, \citenamefont {Wang}, \citenamefont {Zhang},
  \citenamefont {Jia}, \citenamefont {Feng}, \citenamefont {Lu}, \citenamefont
  {Zhao}, \citenamefont {Zhang}, \citenamefont {Min}, \citenamefont {Long},
  \citenamefont {Yu}, \citenamefont {Wang}, \citenamefont {Ye}, \citenamefont
  {Zhang}, \citenamefont {Prakapenka}, \citenamefont {Chariton}, \citenamefont
  {Ginsberg}, \citenamefont {Bass}, \citenamefont {Yuan}, \citenamefont {Liu},\
  and\ \citenamefont {Jin}}]{li2022}%
  \BibitemOpen
  \bibfield  {author} {\bibinfo {author} {\bibfnamefont {Z.}~\bibnamefont
  {Li}}, \bibinfo {author} {\bibfnamefont {C.}~\bibnamefont {Zhang}}, \bibinfo
  {author} {\bibfnamefont {X.}~\bibnamefont {Wang}}, \bibinfo {author}
  {\bibfnamefont {S.}~\bibnamefont {Zhang}}, \bibinfo {author} {\bibfnamefont
  {Y.}~\bibnamefont {Jia}}, \bibinfo {author} {\bibfnamefont {S.}~\bibnamefont
  {Feng}}, \bibinfo {author} {\bibfnamefont {K.}~\bibnamefont {Lu}}, \bibinfo
  {author} {\bibfnamefont {J.}~\bibnamefont {Zhao}}, \bibinfo {author}
  {\bibfnamefont {J.}~\bibnamefont {Zhang}}, \bibinfo {author} {\bibfnamefont
  {B.}~\bibnamefont {Min}}, \bibinfo {author} {\bibfnamefont {Y.}~\bibnamefont
  {Long}}, \bibinfo {author} {\bibfnamefont {R.}~\bibnamefont {Yu}}, \bibinfo
  {author} {\bibfnamefont {L.}~\bibnamefont {Wang}}, \bibinfo {author}
  {\bibfnamefont {M.}~\bibnamefont {Ye}}, \bibinfo {author} {\bibfnamefont
  {Z.}~\bibnamefont {Zhang}}, \bibinfo {author} {\bibfnamefont
  {V.}~\bibnamefont {Prakapenka}}, \bibinfo {author} {\bibfnamefont
  {S.}~\bibnamefont {Chariton}}, \bibinfo {author} {\bibfnamefont {P.~A.}\
  \bibnamefont {Ginsberg}}, \bibinfo {author} {\bibfnamefont {J.}~\bibnamefont
  {Bass}}, \bibinfo {author} {\bibfnamefont {S.}~\bibnamefont {Yuan}}, \bibinfo
  {author} {\bibfnamefont {H.}~\bibnamefont {Liu}}, \ and\ \bibinfo {author}
  {\bibfnamefont {C.}~\bibnamefont {Jin}},\ }\bibfield  {title} {\enquote
  {\bibinfo {title} {Superconductivity above 200 k discovered in superhydrides
  of calcium},}\ }\href {\doibase 10.1038/s41467-022-30454-w} {\bibfield
  {journal} {\bibinfo  {journal} {Nature Communications}\ }\textbf {\bibinfo
  {volume} {13}},\ \bibinfo {pages} {2863} (\bibinfo {year}
  {2022})}\BibitemShut {NoStop}%
\bibitem [{\citenamefont {Li}\ \emph {et~al.}(2023)\citenamefont {Li},
  \citenamefont {He}, \citenamefont {Zhang}, \citenamefont {Lu}, \citenamefont
  {Min}, \citenamefont {Zhang}, \citenamefont {Zhang}, \citenamefont {Zhao},
  \citenamefont {Shi}, \citenamefont {Peng}, \citenamefont {Feng},
  \citenamefont {Deng}, \citenamefont {Song}, \citenamefont {Liu},
  \citenamefont {Wang}, \citenamefont {Yu}, \citenamefont {Wang}, \citenamefont
  {Li}, \citenamefont {Bass}, \citenamefont {Prakapenka}, \citenamefont
  {Chariton}, \citenamefont {Liu},\ and\ \citenamefont {Jin}}]{li2023}%
  \BibitemOpen
  \bibfield  {author} {\bibinfo {author} {\bibfnamefont {Z.}~\bibnamefont
  {Li}}, \bibinfo {author} {\bibfnamefont {X.}~\bibnamefont {He}}, \bibinfo
  {author} {\bibfnamefont {C.}~\bibnamefont {Zhang}}, \bibinfo {author}
  {\bibfnamefont {K.}~\bibnamefont {Lu}}, \bibinfo {author} {\bibfnamefont
  {B.}~\bibnamefont {Min}}, \bibinfo {author} {\bibfnamefont {J.}~\bibnamefont
  {Zhang}}, \bibinfo {author} {\bibfnamefont {S.}~\bibnamefont {Zhang}},
  \bibinfo {author} {\bibfnamefont {J.}~\bibnamefont {Zhao}}, \bibinfo {author}
  {\bibfnamefont {L.}~\bibnamefont {Shi}}, \bibinfo {author} {\bibfnamefont
  {Y.}~\bibnamefont {Peng}}, \bibinfo {author} {\bibfnamefont {S.}~\bibnamefont
  {Feng}}, \bibinfo {author} {\bibfnamefont {Z.}~\bibnamefont {Deng}}, \bibinfo
  {author} {\bibfnamefont {J.}~\bibnamefont {Song}}, \bibinfo {author}
  {\bibfnamefont {Q.}~\bibnamefont {Liu}}, \bibinfo {author} {\bibfnamefont
  {X.}~\bibnamefont {Wang}}, \bibinfo {author} {\bibfnamefont {R.}~\bibnamefont
  {Yu}}, \bibinfo {author} {\bibfnamefont {L.}~\bibnamefont {Wang}}, \bibinfo
  {author} {\bibfnamefont {Y.}~\bibnamefont {Li}}, \bibinfo {author}
  {\bibfnamefont {J.~D.}\ \bibnamefont {Bass}}, \bibinfo {author}
  {\bibfnamefont {V.}~\bibnamefont {Prakapenka}}, \bibinfo {author}
  {\bibfnamefont {S.}~\bibnamefont {Chariton}}, \bibinfo {author}
  {\bibfnamefont {H.}~\bibnamefont {Liu}}, \ and\ \bibinfo {author}
  {\bibfnamefont {C.}~\bibnamefont {Jin}},\ }\bibfield  {title} {\enquote
  {\bibinfo {title} {Superconductivity above 70 k observed in lutetium
  polyhydrides},}\ }\href {\doibase 10.1007/s11433-023-2101-9} {\bibfield
  {journal} {\bibinfo  {journal} {Science China: Physics, Mechanics and
  Astronomy}\ }\textbf {\bibinfo {volume} {66}},\ \bibinfo {pages} {267411}
  (\bibinfo {year} {2023})}\BibitemShut {NoStop}%
\bibitem [{\citenamefont {Liao}\ \emph {et~al.}(2020)\citenamefont {Liao},
  \citenamefont {Liu}, \citenamefont {Zhang}, \citenamefont {Guo},\ and\
  \citenamefont {Ke}}]{liao2020}%
  \BibitemOpen
  \bibfield  {author} {\bibinfo {author} {\bibfnamefont {Z.}~\bibnamefont
  {Liao}}, \bibinfo {author} {\bibfnamefont {C.}~\bibnamefont {Liu}}, \bibinfo
  {author} {\bibfnamefont {Y.}~\bibnamefont {Zhang}}, \bibinfo {author}
  {\bibfnamefont {Y.}~\bibnamefont {Guo}}, \ and\ \bibinfo {author}
  {\bibfnamefont {X.}~\bibnamefont {Ke}},\ }\bibfield  {title} {\enquote
  {\bibinfo {title} {{First-principles study on crystal structures and
  superconductivity of molybdenum hydrides under high pressure}},}\ }\href
  {\doibase 10.1063/5.0005873} {\bibfield  {journal} {\bibinfo  {journal}
  {Journal of Applied Physics}\ }\textbf {\bibinfo {volume} {128}},\ \bibinfo
  {pages} {105901} (\bibinfo {year} {2020})}\BibitemShut {NoStop}%
\bibitem [{\citenamefont {Liu}\ \emph {et~al.}(2015{\natexlab{a}})\citenamefont
  {Liu}, \citenamefont {Duan}, \citenamefont {Tian}, \citenamefont {Liu},
  \citenamefont {Wang}, \citenamefont {Huang}, \citenamefont {Li},
  \citenamefont {Ma}, \citenamefont {Liu},\ and\ \citenamefont
  {Cui}}]{liu2015}%
  \BibitemOpen
  \bibfield  {author} {\bibinfo {author} {\bibfnamefont {Y.}~\bibnamefont
  {Liu}}, \bibinfo {author} {\bibfnamefont {D.}~\bibnamefont {Duan}}, \bibinfo
  {author} {\bibfnamefont {F.}~\bibnamefont {Tian}}, \bibinfo {author}
  {\bibfnamefont {H.}~\bibnamefont {Liu}}, \bibinfo {author} {\bibfnamefont
  {C.}~\bibnamefont {Wang}}, \bibinfo {author} {\bibfnamefont {X.}~\bibnamefont
  {Huang}}, \bibinfo {author} {\bibfnamefont {D.}~\bibnamefont {Li}}, \bibinfo
  {author} {\bibfnamefont {Y.}~\bibnamefont {Ma}}, \bibinfo {author}
  {\bibfnamefont {B.}~\bibnamefont {Liu}}, \ and\ \bibinfo {author}
  {\bibfnamefont {T.}~\bibnamefont {Cui}},\ }\bibfield  {title} {\enquote
  {\bibinfo {title} {Pressure-induced structures and properties in indium
  hydrides},}\ }\href {\doibase 10.1021/acs.inorgchem.5b01684} {\bibfield
  {journal} {\bibinfo  {journal} {Inorganic Chemistry}\ }\textbf {\bibinfo
  {volume} {54}},\ \bibinfo {pages} {9924} (\bibinfo {year}
  {2015}{\natexlab{a}})}\BibitemShut {NoStop}%
\bibitem [{\citenamefont {Liu}\ \emph {et~al.}(2015{\natexlab{b}})\citenamefont
  {Liu}, \citenamefont {Duan}, \citenamefont {Huang}, \citenamefont {Tian},
  \citenamefont {Li}, \citenamefont {Sha}, \citenamefont {Wang}, \citenamefont
  {Zhang}, \citenamefont {Yang}, \citenamefont {Liu},\ and\ \citenamefont
  {Cui}}]{liu2015A}%
  \BibitemOpen
  \bibfield  {author} {\bibinfo {author} {\bibfnamefont {Y.}~\bibnamefont
  {Liu}}, \bibinfo {author} {\bibfnamefont {D.}~\bibnamefont {Duan}}, \bibinfo
  {author} {\bibfnamefont {X.}~\bibnamefont {Huang}}, \bibinfo {author}
  {\bibfnamefont {F.}~\bibnamefont {Tian}}, \bibinfo {author} {\bibfnamefont
  {D.}~\bibnamefont {Li}}, \bibinfo {author} {\bibfnamefont {X.}~\bibnamefont
  {Sha}}, \bibinfo {author} {\bibfnamefont {C.}~\bibnamefont {Wang}}, \bibinfo
  {author} {\bibfnamefont {H.}~\bibnamefont {Zhang}}, \bibinfo {author}
  {\bibfnamefont {T.}~\bibnamefont {Yang}}, \bibinfo {author} {\bibfnamefont
  {B.}~\bibnamefont {Liu}}, \ and\ \bibinfo {author} {\bibfnamefont
  {T.}~\bibnamefont {Cui}},\ }\bibfield  {title} {\enquote {\bibinfo {title}
  {Structures and properties of osmium hydrides under pressure from first
  principle calculation},}\ }\href {\doibase 10.1021/acs.jpcc.5b03791}
  {\bibfield  {journal} {\bibinfo  {journal} {The Journal of Physical Chemistry
  C}\ }\textbf {\bibinfo {volume} {119}},\ \bibinfo {pages} {15905} (\bibinfo
  {year} {2015}{\natexlab{b}})}\BibitemShut {NoStop}%
\bibitem [{\citenamefont {Liu}\ \emph {et~al.}(2015{\natexlab{c}})\citenamefont
  {Liu}, \citenamefont {Duan}, \citenamefont {Tian}, \citenamefont {Wang},
  \citenamefont {Wu}, \citenamefont {Ma}, \citenamefont {Yu}, \citenamefont
  {Li}, \citenamefont {Liu},\ and\ \citenamefont {Cui}}]{liu2015B}%
  \BibitemOpen
  \bibfield  {author} {\bibinfo {author} {\bibfnamefont {Y.}~\bibnamefont
  {Liu}}, \bibinfo {author} {\bibfnamefont {D.}~\bibnamefont {Duan}}, \bibinfo
  {author} {\bibfnamefont {F.}~\bibnamefont {Tian}}, \bibinfo {author}
  {\bibfnamefont {C.}~\bibnamefont {Wang}}, \bibinfo {author} {\bibfnamefont
  {G.}~\bibnamefont {Wu}}, \bibinfo {author} {\bibfnamefont {Y.}~\bibnamefont
  {Ma}}, \bibinfo {author} {\bibfnamefont {H.}~\bibnamefont {Yu}}, \bibinfo
  {author} {\bibfnamefont {D.}~\bibnamefont {Li}}, \bibinfo {author}
  {\bibfnamefont {B.}~\bibnamefont {Liu}}, \ and\ \bibinfo {author}
  {\bibfnamefont {T.}~\bibnamefont {Cui}},\ }\bibfield  {title} {\enquote
  {\bibinfo {title} {Prediction of stoichiometric pohn compounds: crystal
  structures and properties},}\ }\href {\doibase 10.1039/C5RA19223D} {\bibfield
   {journal} {\bibinfo  {journal} {RSC Adv.}\ }\textbf {\bibinfo {volume}
  {5}},\ \bibinfo {pages} {103445} (\bibinfo {year}
  {2015}{\natexlab{c}})}\BibitemShut {NoStop}%
\bibitem [{\citenamefont {Liu}\ \emph {et~al.}(2015{\natexlab{d}})\citenamefont
  {Liu}, \citenamefont {Huang}, \citenamefont {Duan}, \citenamefont {Tian},
  \citenamefont {Liu}, \citenamefont {Li}, \citenamefont {Zhao}, \citenamefont
  {Sha}, \citenamefont {Yu}, \citenamefont {Zhang}, \citenamefont {Liu},\ and\
  \citenamefont {Cui}}]{liu2015C}%
  \BibitemOpen
  \bibfield  {author} {\bibinfo {author} {\bibfnamefont {Y.}~\bibnamefont
  {Liu}}, \bibinfo {author} {\bibfnamefont {X.}~\bibnamefont {Huang}}, \bibinfo
  {author} {\bibfnamefont {D.}~\bibnamefont {Duan}}, \bibinfo {author}
  {\bibfnamefont {F.}~\bibnamefont {Tian}}, \bibinfo {author} {\bibfnamefont
  {H.}~\bibnamefont {Liu}}, \bibinfo {author} {\bibfnamefont {D.}~\bibnamefont
  {Li}}, \bibinfo {author} {\bibfnamefont {Z.}~\bibnamefont {Zhao}}, \bibinfo
  {author} {\bibfnamefont {X.}~\bibnamefont {Sha}}, \bibinfo {author}
  {\bibfnamefont {H.}~\bibnamefont {Yu}}, \bibinfo {author} {\bibfnamefont
  {H.}~\bibnamefont {Zhang}}, \bibinfo {author} {\bibfnamefont
  {B.}~\bibnamefont {Liu}}, \ and\ \bibinfo {author} {\bibfnamefont
  {T.}~\bibnamefont {Cui}},\ }\bibfield  {title} {\enquote {\bibinfo {title}
  {First-principles study on the structural and electronic properties of
  metallic hfh2 under pressure},}\ }\href {\doibase 10.1038/srep11381}
  {\bibfield  {journal} {\bibinfo  {journal} {Scientific Reports}\ }\textbf
  {\bibinfo {volume} {5}},\ \bibinfo {pages} {11381} (\bibinfo {year}
  {2015}{\natexlab{d}})}\BibitemShut {NoStop}%
\bibitem [{\citenamefont {Liu}\ \emph {et~al.}(2016{\natexlab{a}})\citenamefont
  {Liu}, \citenamefont {Duan}, \citenamefont {Tian}, \citenamefont {Wang},
  \citenamefont {Ma}, \citenamefont {Li}, \citenamefont {Huang}, \citenamefont
  {Liu},\ and\ \citenamefont {Cui}}]{liu2016}%
  \BibitemOpen
  \bibfield  {author} {\bibinfo {author} {\bibfnamefont {Y.}~\bibnamefont
  {Liu}}, \bibinfo {author} {\bibfnamefont {D.}~\bibnamefont {Duan}}, \bibinfo
  {author} {\bibfnamefont {F.}~\bibnamefont {Tian}}, \bibinfo {author}
  {\bibfnamefont {C.}~\bibnamefont {Wang}}, \bibinfo {author} {\bibfnamefont
  {Y.}~\bibnamefont {Ma}}, \bibinfo {author} {\bibfnamefont {D.}~\bibnamefont
  {Li}}, \bibinfo {author} {\bibfnamefont {X.}~\bibnamefont {Huang}}, \bibinfo
  {author} {\bibfnamefont {B.}~\bibnamefont {Liu}}, \ and\ \bibinfo {author}
  {\bibfnamefont {T.}~\bibnamefont {Cui}},\ }\bibfield  {title} {\enquote
  {\bibinfo {title} {Stability and properties of the ru–h system at high
  pressure},}\ }\href {\doibase 10.1039/C5CP06617D} {\bibfield  {journal}
  {\bibinfo  {journal} {Phys. Chem. Chem. Phys.}\ }\textbf {\bibinfo {volume}
  {18}},\ \bibinfo {pages} {1516} (\bibinfo {year}
  {2016}{\natexlab{a}})}\BibitemShut {NoStop}%
\bibitem [{\citenamefont {Liu}\ \emph {et~al.}(2016{\natexlab{b}})\citenamefont
  {Liu}, \citenamefont {Li}, \citenamefont {Gao}, \citenamefont {Tse},\ and\
  \citenamefont {Naumov}}]{liu2016A}%
  \BibitemOpen
  \bibfield  {author} {\bibinfo {author} {\bibfnamefont {H.}~\bibnamefont
  {Liu}}, \bibinfo {author} {\bibfnamefont {Y.}~\bibnamefont {Li}}, \bibinfo
  {author} {\bibfnamefont {G.}~\bibnamefont {Gao}}, \bibinfo {author}
  {\bibfnamefont {J.~S.}\ \bibnamefont {Tse}}, \ and\ \bibinfo {author}
  {\bibfnamefont {I.~I.}\ \bibnamefont {Naumov}},\ }\bibfield  {title}
  {\enquote {\bibinfo {title} {Crystal structure and superconductivity of ph3
  at high pressures},}\ }\href {\doibase 10.1021/acs.jpcc.5b12009} {\bibfield
  {journal} {\bibinfo  {journal} {The Journal of Physical Chemistry C}\
  }\textbf {\bibinfo {volume} {120}},\ \bibinfo {pages} {3458} (\bibinfo {year}
  {2016}{\natexlab{b}})}\BibitemShut {NoStop}%
\bibitem [{\citenamefont {Liu}\ \emph {et~al.}(2017{\natexlab{a}})\citenamefont
  {Liu}, \citenamefont {Naumov}, \citenamefont {Hoffmann}, \citenamefont
  {Ashcroft},\ and\ \citenamefont {Hemley}}]{liu2017}%
  \BibitemOpen
  \bibfield  {author} {\bibinfo {author} {\bibfnamefont {H.}~\bibnamefont
  {Liu}}, \bibinfo {author} {\bibfnamefont {I.~I.}\ \bibnamefont {Naumov}},
  \bibinfo {author} {\bibfnamefont {R.}~\bibnamefont {Hoffmann}}, \bibinfo
  {author} {\bibfnamefont {N.~W.}\ \bibnamefont {Ashcroft}}, \ and\ \bibinfo
  {author} {\bibfnamefont {R.~J.}\ \bibnamefont {Hemley}},\ }\bibfield  {title}
  {\enquote {\bibinfo {title} {Potential high-<i>t<sub>c</sub></i>
  superconducting lanthanum and yttrium hydrides at high pressure},}\ }\href
  {\doibase 10.1073/pnas.1704505114} {\bibfield  {journal} {\bibinfo  {journal}
  {Proceedings of the National Academy of Sciences}\ }\textbf {\bibinfo
  {volume} {114}},\ \bibinfo {pages} {6990} (\bibinfo {year}
  {2017}{\natexlab{a}})}\BibitemShut {NoStop}%
\bibitem [{\citenamefont {Liu}\ \emph {et~al.}(2017{\natexlab{b}})\citenamefont
  {Liu}, \citenamefont {Sun}, \citenamefont {Wang},\ and\ \citenamefont
  {Lu}}]{liu2017A}%
  \BibitemOpen
  \bibfield  {author} {\bibinfo {author} {\bibfnamefont {L.-L.}\ \bibnamefont
  {Liu}}, \bibinfo {author} {\bibfnamefont {H.-J.}\ \bibnamefont {Sun}},
  \bibinfo {author} {\bibfnamefont {C.~Z.}\ \bibnamefont {Wang}}, \ and\
  \bibinfo {author} {\bibfnamefont {W.-C.}\ \bibnamefont {Lu}},\ }\bibfield
  {title} {\enquote {\bibinfo {title} {High-pressure structures of yttrium
  hydrides},}\ }\href {\doibase 10.1088/1361-648X/aa787d} {\bibfield  {journal}
  {\bibinfo  {journal} {Journal of Physics: Condensed Matter}\ }\textbf
  {\bibinfo {volume} {29}},\ \bibinfo {pages} {325401} (\bibinfo {year}
  {2017}{\natexlab{b}})}\BibitemShut {NoStop}%
\bibitem [{\citenamefont {Liu}\ \emph {et~al.}(2018)\citenamefont {Liu},
  \citenamefont {Naumov}, \citenamefont {Geballe}, \citenamefont {Somayazulu},
  \citenamefont {Tse},\ and\ \citenamefont {Hemley}}]{liu2018}%
  \BibitemOpen
  \bibfield  {author} {\bibinfo {author} {\bibfnamefont {H.}~\bibnamefont
  {Liu}}, \bibinfo {author} {\bibfnamefont {I.~I.}\ \bibnamefont {Naumov}},
  \bibinfo {author} {\bibfnamefont {Z.~M.}\ \bibnamefont {Geballe}}, \bibinfo
  {author} {\bibfnamefont {M.}~\bibnamefont {Somayazulu}}, \bibinfo {author}
  {\bibfnamefont {J.~S.}\ \bibnamefont {Tse}}, \ and\ \bibinfo {author}
  {\bibfnamefont {R.~J.}\ \bibnamefont {Hemley}},\ }\bibfield  {title}
  {\enquote {\bibinfo {title} {Dynamics and superconductivity in compressed
  lanthanum superhydride},}\ }\href {\doibase 10.1103/PhysRevB.98.100102}
  {\bibfield  {journal} {\bibinfo  {journal} {Phys. Rev. B}\ }\textbf {\bibinfo
  {volume} {98}},\ \bibinfo {pages} {100102} (\bibinfo {year}
  {2018})}\BibitemShut {NoStop}%
\bibitem [{\citenamefont {Liu}\ \emph {et~al.}(2023)\citenamefont {Liu},
  \citenamefont {Cui}, \citenamefont {Shi}, \citenamefont {Hao},\ and\
  \citenamefont {Li}}]{liu2023}%
  \BibitemOpen
  \bibfield  {author} {\bibinfo {author} {\bibfnamefont {M.}~\bibnamefont
  {Liu}}, \bibinfo {author} {\bibfnamefont {W.}~\bibnamefont {Cui}}, \bibinfo
  {author} {\bibfnamefont {J.}~\bibnamefont {Shi}}, \bibinfo {author}
  {\bibfnamefont {J.}~\bibnamefont {Hao}}, \ and\ \bibinfo {author}
  {\bibfnamefont {Y.}~\bibnamefont {Li}},\ }\bibfield  {title} {\enquote
  {\bibinfo {title} {Superconducting h7 chain in gallium hydrides at high
  pressure},}\ }\href {\doibase 10.1039/D2CP05690A} {\bibfield  {journal}
  {\bibinfo  {journal} {Phys. Chem. Chem. Phys.}\ }\textbf {\bibinfo {volume}
  {25}},\ \bibinfo {pages} {7223} (\bibinfo {year} {2023})}\BibitemShut
  {NoStop}%
\bibitem [{\citenamefont {Lonie}\ \emph {et~al.}(2013)\citenamefont {Lonie},
  \citenamefont {Hooper}, \citenamefont {Altintas},\ and\ \citenamefont
  {Zurek}}]{lonie2013}%
  \BibitemOpen
  \bibfield  {author} {\bibinfo {author} {\bibfnamefont {D.~C.}\ \bibnamefont
  {Lonie}}, \bibinfo {author} {\bibfnamefont {J.}~\bibnamefont {Hooper}},
  \bibinfo {author} {\bibfnamefont {B.}~\bibnamefont {Altintas}}, \ and\
  \bibinfo {author} {\bibfnamefont {E.}~\bibnamefont {Zurek}},\ }\bibfield
  {title} {\enquote {\bibinfo {title} {Metallization of magnesium polyhydrides
  under pressure},}\ }\href {\doibase 10.1103/PhysRevB.87.054107} {\bibfield
  {journal} {\bibinfo  {journal} {Phys. Rev. B}\ }\textbf {\bibinfo {volume}
  {87}},\ \bibinfo {pages} {054107} (\bibinfo {year} {2013})}\BibitemShut
  {NoStop}%
\bibitem [{\citenamefont {Lu}\ \emph {et~al.}(2015)\citenamefont {Lu},
  \citenamefont {Wu}, \citenamefont {Liu}, \citenamefont {Tse},\ and\
  \citenamefont {Yang}}]{lu2015}%
  \BibitemOpen
  \bibfield  {author} {\bibinfo {author} {\bibfnamefont {S.}~\bibnamefont
  {Lu}}, \bibinfo {author} {\bibfnamefont {M.}~\bibnamefont {Wu}}, \bibinfo
  {author} {\bibfnamefont {H.}~\bibnamefont {Liu}}, \bibinfo {author}
  {\bibfnamefont {J.~S.}\ \bibnamefont {Tse}}, \ and\ \bibinfo {author}
  {\bibfnamefont {B.}~\bibnamefont {Yang}},\ }\bibfield  {title} {\enquote
  {\bibinfo {title} {Prediction of novel crystal structures and
  superconductivity of compressed hbr},}\ }\href {\doibase 10.1039/C5RA06998J}
  {\bibfield  {journal} {\bibinfo  {journal} {RSC Adv.}\ }\textbf {\bibinfo
  {volume} {5}},\ \bibinfo {pages} {45812} (\bibinfo {year}
  {2015})}\BibitemShut {NoStop}%
\bibitem [{\citenamefont {Majumdar}\ \emph {et~al.}(2017)\citenamefont
  {Majumdar}, \citenamefont {Tse}, \citenamefont {Wu},\ and\ \citenamefont
  {Yao}}]{majumdar2017}%
  \BibitemOpen
  \bibfield  {author} {\bibinfo {author} {\bibfnamefont {A.}~\bibnamefont
  {Majumdar}}, \bibinfo {author} {\bibfnamefont {J.~S.}\ \bibnamefont {Tse}},
  \bibinfo {author} {\bibfnamefont {M.}~\bibnamefont {Wu}}, \ and\ \bibinfo
  {author} {\bibfnamefont {Y.}~\bibnamefont {Yao}},\ }\bibfield  {title}
  {\enquote {\bibinfo {title} {Superconductivity in ${\mathrm{feh}}_{5}$},}\
  }\href {\doibase 10.1103/PhysRevB.96.201107} {\bibfield  {journal} {\bibinfo
  {journal} {Phys. Rev. B}\ }\textbf {\bibinfo {volume} {96}},\ \bibinfo
  {pages} {201107} (\bibinfo {year} {2017})}\BibitemShut {NoStop}%
\bibitem [{\citenamefont {Martinez-Canales}\ \emph {et~al.}(2009)\citenamefont
  {Martinez-Canales}, \citenamefont {Oganov}, \citenamefont {Ma}, \citenamefont
  {Yan}, \citenamefont {Lyakhov},\ and\ \citenamefont
  {Bergara}}]{martinez-canales2009}%
  \BibitemOpen
  \bibfield  {author} {\bibinfo {author} {\bibfnamefont {M.}~\bibnamefont
  {Martinez-Canales}}, \bibinfo {author} {\bibfnamefont {A.~R.}\ \bibnamefont
  {Oganov}}, \bibinfo {author} {\bibfnamefont {Y.}~\bibnamefont {Ma}}, \bibinfo
  {author} {\bibfnamefont {Y.}~\bibnamefont {Yan}}, \bibinfo {author}
  {\bibfnamefont {A.~O.}\ \bibnamefont {Lyakhov}}, \ and\ \bibinfo {author}
  {\bibfnamefont {A.}~\bibnamefont {Bergara}},\ }\bibfield  {title} {\enquote
  {\bibinfo {title} {Novel structures and superconductivity of silane under
  pressure},}\ }\href {\doibase 10.1103/PhysRevLett.102.087005} {\bibfield
  {journal} {\bibinfo  {journal} {Phys. Rev. Lett.}\ }\textbf {\bibinfo
  {volume} {102}},\ \bibinfo {pages} {087005} (\bibinfo {year}
  {2009})}\BibitemShut {NoStop}%
\bibitem [{\citenamefont {Matsuoka}\ \emph {et~al.}(2019)\citenamefont
  {Matsuoka}, \citenamefont {Hishida}, \citenamefont {Kuno}, \citenamefont
  {Hirao}, \citenamefont {Ohishi}, \citenamefont {Sasaki}, \citenamefont
  {Takahama},\ and\ \citenamefont {Shimizu}}]{matsuoka2019}%
  \BibitemOpen
  \bibfield  {author} {\bibinfo {author} {\bibfnamefont {T.}~\bibnamefont
  {Matsuoka}}, \bibinfo {author} {\bibfnamefont {M.}~\bibnamefont {Hishida}},
  \bibinfo {author} {\bibfnamefont {K.}~\bibnamefont {Kuno}}, \bibinfo {author}
  {\bibfnamefont {N.}~\bibnamefont {Hirao}}, \bibinfo {author} {\bibfnamefont
  {Y.}~\bibnamefont {Ohishi}}, \bibinfo {author} {\bibfnamefont
  {S.}~\bibnamefont {Sasaki}}, \bibinfo {author} {\bibfnamefont
  {K.}~\bibnamefont {Takahama}}, \ and\ \bibinfo {author} {\bibfnamefont
  {K.}~\bibnamefont {Shimizu}},\ }\bibfield  {title} {\enquote {\bibinfo
  {title} {Superconductivity of platinum hydride},}\ }\href {\doibase
  10.1103/PhysRevB.99.144511} {\bibfield  {journal} {\bibinfo  {journal} {Phys.
  Rev. B}\ }\textbf {\bibinfo {volume} {99}},\ \bibinfo {pages} {144511}
  (\bibinfo {year} {2019})}\BibitemShut {NoStop}%
\bibitem [{\citenamefont {Mishra}\ \emph {et~al.}(2021)\citenamefont {Mishra},
  \citenamefont {Modak},\ and\ \citenamefont {Verma}}]{mishra2021}%
  \BibitemOpen
  \bibfield  {author} {\bibinfo {author} {\bibfnamefont {A.~K.}\ \bibnamefont
  {Mishra}}, \bibinfo {author} {\bibfnamefont {P.}~\bibnamefont {Modak}}, \
  and\ \bibinfo {author} {\bibfnamefont {A.~K.}\ \bibnamefont {Verma}},\
  }\bibfield  {title} {\enquote {\bibinfo {title} {Effect of metal atoms on the
  superconducting properties of decahydrides, mh10 (m = ga, nb and mo)},}\
  }\href {\doibase https://doi.org/10.1016/j.physb.2021.413217} {\bibfield
  {journal} {\bibinfo  {journal} {Physica B: Condensed Matter}\ }\textbf
  {\bibinfo {volume} {619}},\ \bibinfo {pages} {413217} (\bibinfo {year}
  {2021})}\BibitemShut {NoStop}%
\bibitem [{\citenamefont {Ning}\ \emph {et~al.}(2017)\citenamefont {Ning},
  \citenamefont {Yang}, \citenamefont {Zang},\ and\ \citenamefont
  {Lu}}]{ning2017}%
  \BibitemOpen
  \bibfield  {author} {\bibinfo {author} {\bibfnamefont {Y.-L.}\ \bibnamefont
  {Ning}}, \bibinfo {author} {\bibfnamefont {W.-H.}\ \bibnamefont {Yang}},
  \bibinfo {author} {\bibfnamefont {Q.-J.}\ \bibnamefont {Zang}}, \ and\
  \bibinfo {author} {\bibfnamefont {W.-C.}\ \bibnamefont {Lu}},\ }\bibfield
  {title} {\enquote {\bibinfo {title} {Stability and superconducting properties
  of gah5 at high pressure},}\ }\href {\doibase
  https://doi.org/10.1016/j.physb.2017.08.051} {\bibfield  {journal} {\bibinfo
  {journal} {Physica B: Condensed Matter}\ }\textbf {\bibinfo {volume} {525}},\
  \bibinfo {pages} {36} (\bibinfo {year} {2017})}\BibitemShut {NoStop}%
\bibitem [{\citenamefont {Papaconstantopoulos}\ \emph
  {et~al.}(2018)\citenamefont {Papaconstantopoulos}, \citenamefont {Mehl},\
  and\ \citenamefont {Liu}}]{papaconstantopoulos2018}%
  \BibitemOpen
  \bibfield  {author} {\bibinfo {author} {\bibfnamefont {D.~A.}\ \bibnamefont
  {Papaconstantopoulos}}, \bibinfo {author} {\bibfnamefont {M.~J.}\
  \bibnamefont {Mehl}}, \ and\ \bibinfo {author} {\bibfnamefont
  {H.}~\bibnamefont {Liu}},\ }\bibfield  {title} {\enquote {\bibinfo {title}
  {Stability and high-temperature superconductivity in hydrogenated
  chlorine},}\ }\href {\doibase 10.1007/s40509-017-0136-8} {\bibfield
  {journal} {\bibinfo  {journal} {Quantum Studies: Mathematics and
  Foundations}\ }\textbf {\bibinfo {volume} {5}},\ \bibinfo {pages} {23}
  (\bibinfo {year} {2018})}\BibitemShut {NoStop}%
\bibitem [{\citenamefont {Papaconstantopoulos}\ \emph
  {et~al.}(2020)\citenamefont {Papaconstantopoulos}, \citenamefont {Mehl},\
  and\ \citenamefont {Chang}}]{papaconstantopoulos2020}%
  \BibitemOpen
  \bibfield  {author} {\bibinfo {author} {\bibfnamefont {D.~A.}\ \bibnamefont
  {Papaconstantopoulos}}, \bibinfo {author} {\bibfnamefont {M.~J.}\
  \bibnamefont {Mehl}}, \ and\ \bibinfo {author} {\bibfnamefont {P.-H.}\
  \bibnamefont {Chang}},\ }\bibfield  {title} {\enquote {\bibinfo {title}
  {High-temperature superconductivity in ${\mathrm{lah}}_{10}$},}\ }\href
  {\doibase 10.1103/PhysRevB.101.060506} {\bibfield  {journal} {\bibinfo
  {journal} {Phys. Rev. B}\ }\textbf {\bibinfo {volume} {101}},\ \bibinfo
  {pages} {060506} (\bibinfo {year} {2020})}\BibitemShut {NoStop}%
\bibitem [{\citenamefont {Peña-Alvarez}\ \emph {et~al.}(2020)\citenamefont
  {Peña-Alvarez}, \citenamefont {Li}, \citenamefont {Kelsall}, \citenamefont
  {Binns}, \citenamefont {Dalladay-Simpson}, \citenamefont {Hermann},
  \citenamefont {Howie},\ and\ \citenamefont {Gregoryanz}}]{pena-alvarez2020}%
  \BibitemOpen
  \bibfield  {author} {\bibinfo {author} {\bibfnamefont {M.}~\bibnamefont
  {Peña-Alvarez}}, \bibinfo {author} {\bibfnamefont {B.}~\bibnamefont {Li}},
  \bibinfo {author} {\bibfnamefont {L.~C.}\ \bibnamefont {Kelsall}}, \bibinfo
  {author} {\bibfnamefont {J.}~\bibnamefont {Binns}}, \bibinfo {author}
  {\bibfnamefont {P.}~\bibnamefont {Dalladay-Simpson}}, \bibinfo {author}
  {\bibfnamefont {A.}~\bibnamefont {Hermann}}, \bibinfo {author} {\bibfnamefont
  {R.~T.}\ \bibnamefont {Howie}}, \ and\ \bibinfo {author} {\bibfnamefont
  {E.}~\bibnamefont {Gregoryanz}},\ }\bibfield  {title} {\enquote {\bibinfo
  {title} {Synthesis of superconducting cobalt trihydride},}\ }\href {\doibase
  10.1021/acs.jpclett.0c01807} {\bibfield  {journal} {\bibinfo  {journal} {The
  Journal of Physical Chemistry Letters}\ }\textbf {\bibinfo {volume} {11}},\
  \bibinfo {pages} {6420} (\bibinfo {year} {2020})}\BibitemShut {NoStop}%
\bibitem [{\citenamefont {Peng}\ \emph {et~al.}(2017)\citenamefont {Peng},
  \citenamefont {Sun}, \citenamefont {Pickard}, \citenamefont {Needs},
  \citenamefont {Wu},\ and\ \citenamefont {Ma}}]{peng2017}%
  \BibitemOpen
  \bibfield  {author} {\bibinfo {author} {\bibfnamefont {F.}~\bibnamefont
  {Peng}}, \bibinfo {author} {\bibfnamefont {Y.}~\bibnamefont {Sun}}, \bibinfo
  {author} {\bibfnamefont {C.~J.}\ \bibnamefont {Pickard}}, \bibinfo {author}
  {\bibfnamefont {R.~J.}\ \bibnamefont {Needs}}, \bibinfo {author}
  {\bibfnamefont {Q.}~\bibnamefont {Wu}}, \ and\ \bibinfo {author}
  {\bibfnamefont {Y.}~\bibnamefont {Ma}},\ }\bibfield  {title} {\enquote
  {\bibinfo {title} {Hydrogen clathrate structures in rare earth hydrides at
  high pressures: Possible route to room-temperature superconductivity},}\
  }\href {\doibase 10.1103/PhysRevLett.119.107001} {\bibfield  {journal}
  {\bibinfo  {journal} {Phys. Rev. Lett.}\ }\textbf {\bibinfo {volume} {119}},\
  \bibinfo {pages} {107001} (\bibinfo {year} {2017})}\BibitemShut {NoStop}%
\bibitem [{\citenamefont {Qian}\ \emph {et~al.}(2017)\citenamefont {Qian},
  \citenamefont {Sheng}, \citenamefont {Yan}, \citenamefont {Chen},\ and\
  \citenamefont {Song}}]{qian2017}%
  \BibitemOpen
  \bibfield  {author} {\bibinfo {author} {\bibfnamefont {S.}~\bibnamefont
  {Qian}}, \bibinfo {author} {\bibfnamefont {X.}~\bibnamefont {Sheng}},
  \bibinfo {author} {\bibfnamefont {X.}~\bibnamefont {Yan}}, \bibinfo {author}
  {\bibfnamefont {Y.}~\bibnamefont {Chen}}, \ and\ \bibinfo {author}
  {\bibfnamefont {B.}~\bibnamefont {Song}},\ }\bibfield  {title} {\enquote
  {\bibinfo {title} {Theoretical study of stability and superconductivity of
  ${\mathrm{sch}}_{n}$ ($n=4--8$) at high pressure},}\ }\href {\doibase
  10.1103/PhysRevB.96.094513} {\bibfield  {journal} {\bibinfo  {journal} {Phys.
  Rev. B}\ }\textbf {\bibinfo {volume} {96}},\ \bibinfo {pages} {094513}
  (\bibinfo {year} {2017})}\BibitemShut {NoStop}%
\bibitem [{\citenamefont {Saha}\ \emph {et~al.}(2023)\citenamefont {Saha},
  \citenamefont {Di~Cataldo}, \citenamefont {Giannessi}, \citenamefont
  {Cucciari}, \citenamefont {von~der Linden},\ and\ \citenamefont
  {Boeri}}]{saha2023}%
  \BibitemOpen
  \bibfield  {author} {\bibinfo {author} {\bibfnamefont {S.}~\bibnamefont
  {Saha}}, \bibinfo {author} {\bibfnamefont {S.}~\bibnamefont {Di~Cataldo}},
  \bibinfo {author} {\bibfnamefont {F.}~\bibnamefont {Giannessi}}, \bibinfo
  {author} {\bibfnamefont {A.}~\bibnamefont {Cucciari}}, \bibinfo {author}
  {\bibfnamefont {W.}~\bibnamefont {von~der Linden}}, \ and\ \bibinfo {author}
  {\bibfnamefont {L.}~\bibnamefont {Boeri}},\ }\bibfield  {title} {\enquote
  {\bibinfo {title} {Mapping superconductivity in high-pressure hydrides: The
  superhydra project},}\ }\href {\doibase 10.1103/PhysRevMaterials.7.054806}
  {\bibfield  {journal} {\bibinfo  {journal} {Phys. Rev. Mater.}\ }\textbf
  {\bibinfo {volume} {7}},\ \bibinfo {pages} {054806} (\bibinfo {year}
  {2023})}\BibitemShut {NoStop}%
\bibitem [{\citenamefont {Salke}\ \emph {et~al.}(2019)\citenamefont {Salke},
  \citenamefont {Esfahani}, \citenamefont {Zhang}, \citenamefont {Kruglov},
  \citenamefont {Zhou}, \citenamefont {Wang}, \citenamefont {Greenberg},
  \citenamefont {Prakapenka}, \citenamefont {Liu}, \citenamefont {Oganov},\
  and\ \citenamefont {Lin}}]{salke2019}%
  \BibitemOpen
  \bibfield  {author} {\bibinfo {author} {\bibfnamefont {N.~P.}\ \bibnamefont
  {Salke}}, \bibinfo {author} {\bibfnamefont {M.~M.~D.}\ \bibnamefont
  {Esfahani}}, \bibinfo {author} {\bibfnamefont {Y.}~\bibnamefont {Zhang}},
  \bibinfo {author} {\bibfnamefont {I.~A.}\ \bibnamefont {Kruglov}}, \bibinfo
  {author} {\bibfnamefont {J.}~\bibnamefont {Zhou}}, \bibinfo {author}
  {\bibfnamefont {Y.}~\bibnamefont {Wang}}, \bibinfo {author} {\bibfnamefont
  {E.}~\bibnamefont {Greenberg}}, \bibinfo {author} {\bibfnamefont {V.~B.}\
  \bibnamefont {Prakapenka}}, \bibinfo {author} {\bibfnamefont
  {J.}~\bibnamefont {Liu}}, \bibinfo {author} {\bibfnamefont {A.~R.}\
  \bibnamefont {Oganov}}, \ and\ \bibinfo {author} {\bibfnamefont {J.~F.}\
  \bibnamefont {Lin}},\ }\bibfield  {title} {\enquote {\bibinfo {title}
  {Synthesis of clathrate cerium superhydride ceh9 at 80-100 gpa with atomic
  hydrogen sublattice},}\ }\href {\doibase 10.1038/s41467-019-12326-y}
  {\bibfield  {journal} {\bibinfo  {journal} {Nature Communications}\ }\textbf
  {\bibinfo {volume} {10}},\ \bibinfo {pages} {4453} (\bibinfo {year}
  {2019})}\BibitemShut {NoStop}%
\bibitem [{\citenamefont {Sano}\ \emph {et~al.}(2016)\citenamefont {Sano},
  \citenamefont {Koretsune}, \citenamefont {Tadano}, \citenamefont {Akashi},\
  and\ \citenamefont {Arita}}]{sano2016}%
  \BibitemOpen
  \bibfield  {author} {\bibinfo {author} {\bibfnamefont {W.}~\bibnamefont
  {Sano}}, \bibinfo {author} {\bibfnamefont {T.}~\bibnamefont {Koretsune}},
  \bibinfo {author} {\bibfnamefont {T.}~\bibnamefont {Tadano}}, \bibinfo
  {author} {\bibfnamefont {R.}~\bibnamefont {Akashi}}, \ and\ \bibinfo {author}
  {\bibfnamefont {R.}~\bibnamefont {Arita}},\ }\bibfield  {title} {\enquote
  {\bibinfo {title} {Effect of van hove singularities on
  high-${T}_{\mathrm{c}}$ superconductivity in ${\mathrm{h}}_{3}\mathrm{S}$},}\
  }\href {\doibase 10.1103/PhysRevB.93.094525} {\bibfield  {journal} {\bibinfo
  {journal} {Phys. Rev. B}\ }\textbf {\bibinfo {volume} {93}},\ \bibinfo
  {pages} {094525} (\bibinfo {year} {2016})}\BibitemShut {NoStop}%
\bibitem [{\citenamefont {Scheler}\ \emph {et~al.}(2011)\citenamefont
  {Scheler}, \citenamefont {Degtyareva}, \citenamefont {Marqu\'es},
  \citenamefont {Guillaume}, \citenamefont {Proctor}, \citenamefont {Evans},\
  and\ \citenamefont {Gregoryanz}}]{scheler2011}%
  \BibitemOpen
  \bibfield  {author} {\bibinfo {author} {\bibfnamefont {T.}~\bibnamefont
  {Scheler}}, \bibinfo {author} {\bibfnamefont {O.}~\bibnamefont {Degtyareva}},
  \bibinfo {author} {\bibfnamefont {M.}~\bibnamefont {Marqu\'es}}, \bibinfo
  {author} {\bibfnamefont {C.~L.}\ \bibnamefont {Guillaume}}, \bibinfo {author}
  {\bibfnamefont {J.~E.}\ \bibnamefont {Proctor}}, \bibinfo {author}
  {\bibfnamefont {S.}~\bibnamefont {Evans}}, \ and\ \bibinfo {author}
  {\bibfnamefont {E.}~\bibnamefont {Gregoryanz}},\ }\bibfield  {title}
  {\enquote {\bibinfo {title} {Synthesis and properties of platinum hydride},}\
  }\href {\doibase 10.1103/PhysRevB.83.214106} {\bibfield  {journal} {\bibinfo
  {journal} {Phys. Rev. B}\ }\textbf {\bibinfo {volume} {83}},\ \bibinfo
  {pages} {214106} (\bibinfo {year} {2011})}\BibitemShut {NoStop}%
\bibitem [{\citenamefont {Semenok}\ \emph {et~al.}(2018)\citenamefont
  {Semenok}, \citenamefont {Kvashnin}, \citenamefont {Kruglov},\ and\
  \citenamefont {Oganov}}]{semenok2018}%
  \BibitemOpen
  \bibfield  {author} {\bibinfo {author} {\bibfnamefont {D.~V.}\ \bibnamefont
  {Semenok}}, \bibinfo {author} {\bibfnamefont {A.~G.}\ \bibnamefont
  {Kvashnin}}, \bibinfo {author} {\bibfnamefont {I.~A.}\ \bibnamefont
  {Kruglov}}, \ and\ \bibinfo {author} {\bibfnamefont {A.~R.}\ \bibnamefont
  {Oganov}},\ }\bibfield  {title} {\enquote {\bibinfo {title} {Actinium
  hydrides ach10, ach12, and ach16 as high-temperature conventional
  superconductors},}\ }\href {\doibase 10.1021/acs.jpclett.8b00615} {\bibfield
  {journal} {\bibinfo  {journal} {The Journal of Physical Chemistry Letters}\
  }\textbf {\bibinfo {volume} {9}},\ \bibinfo {pages} {1920} (\bibinfo {year}
  {2018})}\BibitemShut {NoStop}%
\bibitem [{\citenamefont {Semenok}\ \emph
  {et~al.}(2020{\natexlab{b}})\citenamefont {Semenok}, \citenamefont
  {Kvashnin}, \citenamefont {Ivanova}, \citenamefont {Svitlyk}, \citenamefont
  {Fominski}, \citenamefont {Sadakov}, \citenamefont {Sobolevskiy},
  \citenamefont {Pudalov}, \citenamefont {Troyan},\ and\ \citenamefont
  {Oganov}}]{semenok2020}%
  \BibitemOpen
  \bibfield  {author} {\bibinfo {author} {\bibfnamefont {D.~V.}\ \bibnamefont
  {Semenok}}, \bibinfo {author} {\bibfnamefont {A.~G.}\ \bibnamefont
  {Kvashnin}}, \bibinfo {author} {\bibfnamefont {A.~G.}\ \bibnamefont
  {Ivanova}}, \bibinfo {author} {\bibfnamefont {V.}~\bibnamefont {Svitlyk}},
  \bibinfo {author} {\bibfnamefont {V.~Y.}\ \bibnamefont {Fominski}}, \bibinfo
  {author} {\bibfnamefont {A.~V.}\ \bibnamefont {Sadakov}}, \bibinfo {author}
  {\bibfnamefont {O.~A.}\ \bibnamefont {Sobolevskiy}}, \bibinfo {author}
  {\bibfnamefont {V.~M.}\ \bibnamefont {Pudalov}}, \bibinfo {author}
  {\bibfnamefont {I.~A.}\ \bibnamefont {Troyan}}, \ and\ \bibinfo {author}
  {\bibfnamefont {A.~R.}\ \bibnamefont {Oganov}},\ }\bibfield  {title}
  {\enquote {\bibinfo {title} {Superconductivity at 161 k in thorium hydride
  thh10: Synthesis and properties},}\ }\href {\doibase
  https://doi.org/10.1016/j.mattod.2019.10.005} {\bibfield  {journal} {\bibinfo
   {journal} {Materials Today}\ }\textbf {\bibinfo {volume} {33}},\ \bibinfo
  {pages} {36} (\bibinfo {year} {2020}{\natexlab{b}})}\BibitemShut {NoStop}%
\bibitem [{\citenamefont {Shamp}\ and\ \citenamefont
  {Zurek}(2015)}]{shamp2015}%
  \BibitemOpen
  \bibfield  {author} {\bibinfo {author} {\bibfnamefont {A.}~\bibnamefont
  {Shamp}}\ and\ \bibinfo {author} {\bibfnamefont {E.}~\bibnamefont {Zurek}},\
  }\bibfield  {title} {\enquote {\bibinfo {title} {Superconducting
  high-pressure phases composed of hydrogen and iodine},}\ }\href {\doibase
  10.1021/acs.jpclett.5b01839} {\bibfield  {journal} {\bibinfo  {journal} {The
  Journal of Physical Chemistry Letters}\ }\textbf {\bibinfo {volume} {6}},\
  \bibinfo {pages} {4067} (\bibinfo {year} {2015})}\BibitemShut {NoStop}%
\bibitem [{\citenamefont {Shamp}\ \emph {et~al.}(2016)\citenamefont {Shamp},
  \citenamefont {Terpstra}, \citenamefont {Bi}, \citenamefont {Falls},
  \citenamefont {Avery},\ and\ \citenamefont {Zurek}}]{shamp2016}%
  \BibitemOpen
  \bibfield  {author} {\bibinfo {author} {\bibfnamefont {A.}~\bibnamefont
  {Shamp}}, \bibinfo {author} {\bibfnamefont {T.}~\bibnamefont {Terpstra}},
  \bibinfo {author} {\bibfnamefont {T.}~\bibnamefont {Bi}}, \bibinfo {author}
  {\bibfnamefont {Z.}~\bibnamefont {Falls}}, \bibinfo {author} {\bibfnamefont
  {P.}~\bibnamefont {Avery}}, \ and\ \bibinfo {author} {\bibfnamefont
  {E.}~\bibnamefont {Zurek}},\ }\bibfield  {title} {\enquote {\bibinfo {title}
  {Decomposition products of phosphine under pressure: Ph2 stable and
  superconducting?}}\ }\href {\doibase 10.1021/jacs.5b10180} {\bibfield
  {journal} {\bibinfo  {journal} {Journal of the American Chemical Society}\
  }\textbf {\bibinfo {volume} {138}},\ \bibinfo {pages} {1884} (\bibinfo {year}
  {2016})}\BibitemShut {NoStop}%
\bibitem [{\citenamefont {Shanavas}\ \emph {et~al.}(2016)\citenamefont
  {Shanavas}, \citenamefont {Lindsay},\ and\ \citenamefont
  {Parker}}]{shanavas2016}%
  \BibitemOpen
  \bibfield  {author} {\bibinfo {author} {\bibfnamefont {K.~V.}\ \bibnamefont
  {Shanavas}}, \bibinfo {author} {\bibfnamefont {L.}~\bibnamefont {Lindsay}}, \
  and\ \bibinfo {author} {\bibfnamefont {D.~S.}\ \bibnamefont {Parker}},\
  }\bibfield  {title} {\enquote {\bibinfo {title} {Electronic structure and
  electron-phonon coupling in tih2},}\ }\href {\doibase 10.1038/srep28102}
  {\bibfield  {journal} {\bibinfo  {journal} {Scientific Reports}\ }\textbf
  {\bibinfo {volume} {6}},\ \bibinfo {pages} {28102} (\bibinfo {year}
  {2016})}\BibitemShut {NoStop}%
\bibitem [{\citenamefont {Shao}\ \emph {et~al.}(2018)\citenamefont {Shao},
  \citenamefont {Huang}, \citenamefont {Duan}, \citenamefont {Ma},
  \citenamefont {Yu}, \citenamefont {Xie}, \citenamefont {Li}, \citenamefont
  {Tian}, \citenamefont {Liu},\ and\ \citenamefont {Cui}}]{shao2018}%
  \BibitemOpen
  \bibfield  {author} {\bibinfo {author} {\bibfnamefont {Z.}~\bibnamefont
  {Shao}}, \bibinfo {author} {\bibfnamefont {Y.}~\bibnamefont {Huang}},
  \bibinfo {author} {\bibfnamefont {D.}~\bibnamefont {Duan}}, \bibinfo {author}
  {\bibfnamefont {Y.}~\bibnamefont {Ma}}, \bibinfo {author} {\bibfnamefont
  {H.}~\bibnamefont {Yu}}, \bibinfo {author} {\bibfnamefont {H.}~\bibnamefont
  {Xie}}, \bibinfo {author} {\bibfnamefont {D.}~\bibnamefont {Li}}, \bibinfo
  {author} {\bibfnamefont {F.}~\bibnamefont {Tian}}, \bibinfo {author}
  {\bibfnamefont {B.}~\bibnamefont {Liu}}, \ and\ \bibinfo {author}
  {\bibfnamefont {T.}~\bibnamefont {Cui}},\ }\bibfield  {title} {\enquote
  {\bibinfo {title} {Stable structures and superconductivity of an at–h
  system at high pressure},}\ }\href {\doibase 10.1039/C8CP04317E} {\bibfield
  {journal} {\bibinfo  {journal} {Phys. Chem. Chem. Phys.}\ }\textbf {\bibinfo
  {volume} {20}},\ \bibinfo {pages} {24783} (\bibinfo {year}
  {2018})}\BibitemShut {NoStop}%
\bibitem [{\citenamefont {Shao}\ \emph {et~al.}(2019)\citenamefont {Shao},
  \citenamefont {Duan}, \citenamefont {Ma}, \citenamefont {Yu}, \citenamefont
  {Song}, \citenamefont {Xie}, \citenamefont {Li}, \citenamefont {Tian},
  \citenamefont {Liu},\ and\ \citenamefont {Cui}}]{shao2019}%
  \BibitemOpen
  \bibfield  {author} {\bibinfo {author} {\bibfnamefont {Z.}~\bibnamefont
  {Shao}}, \bibinfo {author} {\bibfnamefont {D.}~\bibnamefont {Duan}}, \bibinfo
  {author} {\bibfnamefont {Y.}~\bibnamefont {Ma}}, \bibinfo {author}
  {\bibfnamefont {H.}~\bibnamefont {Yu}}, \bibinfo {author} {\bibfnamefont
  {H.}~\bibnamefont {Song}}, \bibinfo {author} {\bibfnamefont {H.}~\bibnamefont
  {Xie}}, \bibinfo {author} {\bibfnamefont {D.}~\bibnamefont {Li}}, \bibinfo
  {author} {\bibfnamefont {F.}~\bibnamefont {Tian}}, \bibinfo {author}
  {\bibfnamefont {B.}~\bibnamefont {Liu}}, \ and\ \bibinfo {author}
  {\bibfnamefont {T.}~\bibnamefont {Cui}},\ }\bibfield  {title} {\enquote
  {\bibinfo {title} {Unique phase diagram and superconductivity of calcium
  hydrides at high pressures},}\ }\href {\doibase
  10.1021/acs.inorgchem.8b03165} {\bibfield  {journal} {\bibinfo  {journal}
  {Inorganic Chemistry}\ }\textbf {\bibinfo {volume} {58}},\ \bibinfo {pages}
  {2558} (\bibinfo {year} {2019})}\BibitemShut {NoStop}%
\bibitem [{\citenamefont {Shao}\ \emph
  {et~al.}(2021{\natexlab{a}})\citenamefont {Shao}, \citenamefont {Chen},
  \citenamefont {Chen}, \citenamefont {Zhang}, \citenamefont {Huang},\ and\
  \citenamefont {Cui}}]{shao2021}%
  \BibitemOpen
  \bibfield  {author} {\bibinfo {author} {\bibfnamefont {M.}~\bibnamefont
  {Shao}}, \bibinfo {author} {\bibfnamefont {S.}~\bibnamefont {Chen}}, \bibinfo
  {author} {\bibfnamefont {W.}~\bibnamefont {Chen}}, \bibinfo {author}
  {\bibfnamefont {K.}~\bibnamefont {Zhang}}, \bibinfo {author} {\bibfnamefont
  {X.}~\bibnamefont {Huang}}, \ and\ \bibinfo {author} {\bibfnamefont
  {T.}~\bibnamefont {Cui}},\ }\bibfield  {title} {\enquote {\bibinfo {title}
  {Superconducting sch3 and luh3 at megabar pressures},}\ }\href {\doibase
  10.1021/acs.inorgchem.1c01960} {\bibfield  {journal} {\bibinfo  {journal}
  {Inorganic Chemistry}\ }\textbf {\bibinfo {volume} {60}},\ \bibinfo {pages}
  {15330} (\bibinfo {year} {2021}{\natexlab{a}})}\BibitemShut {NoStop}%
\bibitem [{\citenamefont {Shao}\ \emph
  {et~al.}(2021{\natexlab{b}})\citenamefont {Shao}, \citenamefont {Chen},
  \citenamefont {Zhang}, \citenamefont {Huang},\ and\ \citenamefont
  {Cui}}]{shao2021A}%
  \BibitemOpen
  \bibfield  {author} {\bibinfo {author} {\bibfnamefont {M.}~\bibnamefont
  {Shao}}, \bibinfo {author} {\bibfnamefont {W.}~\bibnamefont {Chen}}, \bibinfo
  {author} {\bibfnamefont {K.}~\bibnamefont {Zhang}}, \bibinfo {author}
  {\bibfnamefont {X.}~\bibnamefont {Huang}}, \ and\ \bibinfo {author}
  {\bibfnamefont {T.}~\bibnamefont {Cui}},\ }\bibfield  {title} {\enquote
  {\bibinfo {title} {High-pressure synthesis of superconducting clathratelike
  $\mathrm{Y}{\mathrm{h}}_{4}$},}\ }\href {\doibase
  10.1103/PhysRevB.104.174509} {\bibfield  {journal} {\bibinfo  {journal}
  {Phys. Rev. B}\ }\textbf {\bibinfo {volume} {104}},\ \bibinfo {pages}
  {174509} (\bibinfo {year} {2021}{\natexlab{b}})}\BibitemShut {NoStop}%
\bibitem [{\citenamefont {Shipley}\ \emph {et~al.}(2021)\citenamefont
  {Shipley}, \citenamefont {Hutcheon}, \citenamefont {Needs},\ and\
  \citenamefont {Pickard}}]{shipley2021}%
  \BibitemOpen
  \bibfield  {author} {\bibinfo {author} {\bibfnamefont {A.~M.}\ \bibnamefont
  {Shipley}}, \bibinfo {author} {\bibfnamefont {M.~J.}\ \bibnamefont
  {Hutcheon}}, \bibinfo {author} {\bibfnamefont {R.~J.}\ \bibnamefont {Needs}},
  \ and\ \bibinfo {author} {\bibfnamefont {C.~J.}\ \bibnamefont {Pickard}},\
  }\bibfield  {title} {\enquote {\bibinfo {title} {High-throughput discovery of
  high-temperature conventional superconductors},}\ }\href {\doibase
  10.1103/PhysRevB.104.054501} {\bibfield  {journal} {\bibinfo  {journal}
  {Phys. Rev. B}\ }\textbf {\bibinfo {volume} {104}},\ \bibinfo {pages}
  {054501} (\bibinfo {year} {2021})}\BibitemShut {NoStop}%
\bibitem [{\citenamefont {Snider}\ \emph {et~al.}(2021)\citenamefont {Snider},
  \citenamefont {Dasenbrock-Gammon}, \citenamefont {McBride}, \citenamefont
  {Wang}, \citenamefont {Meyers}, \citenamefont {Lawler}, \citenamefont
  {Zurek}, \citenamefont {Salamat},\ and\ \citenamefont {Dias}}]{snider2021}%
  \BibitemOpen
  \bibfield  {author} {\bibinfo {author} {\bibfnamefont {E.}~\bibnamefont
  {Snider}}, \bibinfo {author} {\bibfnamefont {N.}~\bibnamefont
  {Dasenbrock-Gammon}}, \bibinfo {author} {\bibfnamefont {R.}~\bibnamefont
  {McBride}}, \bibinfo {author} {\bibfnamefont {X.}~\bibnamefont {Wang}},
  \bibinfo {author} {\bibfnamefont {N.}~\bibnamefont {Meyers}}, \bibinfo
  {author} {\bibfnamefont {K.~V.}\ \bibnamefont {Lawler}}, \bibinfo {author}
  {\bibfnamefont {E.}~\bibnamefont {Zurek}}, \bibinfo {author} {\bibfnamefont
  {A.}~\bibnamefont {Salamat}}, \ and\ \bibinfo {author} {\bibfnamefont
  {R.~P.}\ \bibnamefont {Dias}},\ }\bibfield  {title} {\enquote {\bibinfo
  {title} {Synthesis of yttrium superhydride superconductor with a transition
  temperature up to 262 k by catalytic hydrogenation at high pressures},}\
  }\href {\doibase 10.1103/PhysRevLett.126.117003} {\bibfield  {journal}
  {\bibinfo  {journal} {Phys. Rev. Lett.}\ }\textbf {\bibinfo {volume} {126}},\
  \bibinfo {pages} {117003} (\bibinfo {year} {2021})}\BibitemShut {NoStop}%
\bibitem [{\citenamefont {Somayazulu}\ \emph {et~al.}(2019)\citenamefont
  {Somayazulu}, \citenamefont {Ahart}, \citenamefont {Mishra}, \citenamefont
  {Geballe}, \citenamefont {Baldini}, \citenamefont {Meng}, \citenamefont
  {Struzhkin},\ and\ \citenamefont {Hemley}}]{somayazulu2019}%
  \BibitemOpen
  \bibfield  {author} {\bibinfo {author} {\bibfnamefont {M.}~\bibnamefont
  {Somayazulu}}, \bibinfo {author} {\bibfnamefont {M.}~\bibnamefont {Ahart}},
  \bibinfo {author} {\bibfnamefont {A.~K.}\ \bibnamefont {Mishra}}, \bibinfo
  {author} {\bibfnamefont {Z.~M.}\ \bibnamefont {Geballe}}, \bibinfo {author}
  {\bibfnamefont {M.}~\bibnamefont {Baldini}}, \bibinfo {author} {\bibfnamefont
  {Y.}~\bibnamefont {Meng}}, \bibinfo {author} {\bibfnamefont {V.~V.}\
  \bibnamefont {Struzhkin}}, \ and\ \bibinfo {author} {\bibfnamefont {R.~J.}\
  \bibnamefont {Hemley}},\ }\bibfield  {title} {\enquote {\bibinfo {title}
  {Evidence for superconductivity above 260 k in lanthanum superhydride at
  megabar pressures},}\ }\href {\doibase 10.1103/PhysRevLett.122.027001}
  {\bibfield  {journal} {\bibinfo  {journal} {Phys. Rev. Lett.}\ }\textbf
  {\bibinfo {volume} {122}},\ \bibinfo {pages} {027001} (\bibinfo {year}
  {2019})}\BibitemShut {NoStop}%
\bibitem [{\citenamefont {Song}\ \emph {et~al.}(2020)\citenamefont {Song},
  \citenamefont {Duan}, \citenamefont {Cui},\ and\ \citenamefont
  {Kresin}}]{song2020}%
  \BibitemOpen
  \bibfield  {author} {\bibinfo {author} {\bibfnamefont {H.}~\bibnamefont
  {Song}}, \bibinfo {author} {\bibfnamefont {D.}~\bibnamefont {Duan}}, \bibinfo
  {author} {\bibfnamefont {T.}~\bibnamefont {Cui}}, \ and\ \bibinfo {author}
  {\bibfnamefont {V.~Z.}\ \bibnamefont {Kresin}},\ }\bibfield  {title}
  {\enquote {\bibinfo {title} {High-${T}_{c}$ state of lanthanum hydrides},}\
  }\href {\doibase 10.1103/PhysRevB.102.014510} {\bibfield  {journal} {\bibinfo
   {journal} {Phys. Rev. B}\ }\textbf {\bibinfo {volume} {102}},\ \bibinfo
  {pages} {014510} (\bibinfo {year} {2020})}\BibitemShut {NoStop}%
\bibitem [{\citenamefont {Song}\ \emph {et~al.}(2021)\citenamefont {Song},
  \citenamefont {Zhang}, \citenamefont {Cui}, \citenamefont {Pickard},
  \citenamefont {Kresin},\ and\ \citenamefont {Duan}}]{song2021}%
  \BibitemOpen
  \bibfield  {author} {\bibinfo {author} {\bibfnamefont {H.}~\bibnamefont
  {Song}}, \bibinfo {author} {\bibfnamefont {Z.}~\bibnamefont {Zhang}},
  \bibinfo {author} {\bibfnamefont {T.}~\bibnamefont {Cui}}, \bibinfo {author}
  {\bibfnamefont {C.~J.}\ \bibnamefont {Pickard}}, \bibinfo {author}
  {\bibfnamefont {V.~Z.}\ \bibnamefont {Kresin}}, \ and\ \bibinfo {author}
  {\bibfnamefont {D.}~\bibnamefont {Duan}},\ }\bibfield  {title} {\enquote
  {\bibinfo {title} {High tc superconductivity in heavy rare earth hydrides},}\
  }\href {\doibase 10.1088/0256-307X/38/10/107401} {\bibfield  {journal}
  {\bibinfo  {journal} {Chinese Physics Letters}\ }\textbf {\bibinfo {volume}
  {38}},\ \bibinfo {pages} {107401} (\bibinfo {year} {2021})}\BibitemShut
  {NoStop}%
\bibitem [{\citenamefont {Sun}\ \emph {et~al.}(2020)\citenamefont {Sun},
  \citenamefont {Kuang}, \citenamefont {Keen}, \citenamefont {Lu},\ and\
  \citenamefont {Hermann}}]{sun2020}%
  \BibitemOpen
  \bibfield  {author} {\bibinfo {author} {\bibfnamefont {W.}~\bibnamefont
  {Sun}}, \bibinfo {author} {\bibfnamefont {X.}~\bibnamefont {Kuang}}, \bibinfo
  {author} {\bibfnamefont {H.~D.~J.}\ \bibnamefont {Keen}}, \bibinfo {author}
  {\bibfnamefont {C.}~\bibnamefont {Lu}}, \ and\ \bibinfo {author}
  {\bibfnamefont {A.}~\bibnamefont {Hermann}},\ }\bibfield  {title} {\enquote
  {\bibinfo {title} {Second group of high-pressure high-temperature lanthanide
  polyhydride superconductors},}\ }\href {\doibase 10.1103/PhysRevB.102.144524}
  {\bibfield  {journal} {\bibinfo  {journal} {Phys. Rev. B}\ }\textbf {\bibinfo
  {volume} {102}},\ \bibinfo {pages} {144524} (\bibinfo {year}
  {2020})}\BibitemShut {NoStop}%
\bibitem [{\citenamefont {Szczesniak}\ and\ \citenamefont
  {Zemla}(2015)}]{szczesniak2015}%
  \BibitemOpen
  \bibfield  {author} {\bibinfo {author} {\bibfnamefont {D.}~\bibnamefont
  {Szczesniak}}\ and\ \bibinfo {author} {\bibfnamefont {T.~P.}\ \bibnamefont
  {Zemla}},\ }\bibfield  {title} {\enquote {\bibinfo {title} {On the
  high-pressure superconducting phase in platinum hydride},}\ }\href {\doibase
  10.1088/0953-2048/28/8/085018} {\bibfield  {journal} {\bibinfo  {journal}
  {Superconductor Science and Technology}\ }\textbf {\bibinfo {volume} {28}},\
  \bibinfo {pages} {085018} (\bibinfo {year} {2015})}\BibitemShut {NoStop}%
\bibitem [{\citenamefont {Szczesniak}\ and\ \citenamefont
  {Durajski}(2016)}]{szczesniak2016}%
  \BibitemOpen
  \bibfield  {author} {\bibinfo {author} {\bibfnamefont {R.}~\bibnamefont
  {Szczesniak}}\ and\ \bibinfo {author} {\bibfnamefont {A.~P.}\ \bibnamefont
  {Durajski}},\ }\bibfield  {title} {\enquote {\bibinfo {title}
  {Superconductivity well above room temperature in compressed mgh6},}\ }\href
  {\doibase 10.1007/s11467-016-0578-1} {\bibfield  {journal} {\bibinfo
  {journal} {Frontiers of Physics}\ }\textbf {\bibinfo {volume} {11}},\
  \bibinfo {pages} {117406} (\bibinfo {year} {2016})}\BibitemShut {NoStop}%
\bibitem [{\citenamefont {Tanaka}\ \emph {et~al.}(2017)\citenamefont {Tanaka},
  \citenamefont {Tse},\ and\ \citenamefont {Liu}}]{tanaka2017}%
  \BibitemOpen
  \bibfield  {author} {\bibinfo {author} {\bibfnamefont {K.}~\bibnamefont
  {Tanaka}}, \bibinfo {author} {\bibfnamefont {J.~S.}\ \bibnamefont {Tse}}, \
  and\ \bibinfo {author} {\bibfnamefont {H.}~\bibnamefont {Liu}},\ }\bibfield
  {title} {\enquote {\bibinfo {title} {Electron-phonon coupling mechanisms for
  hydrogen-rich metals at high pressure},}\ }\href {\doibase
  10.1103/PhysRevB.96.100502} {\bibfield  {journal} {\bibinfo  {journal} {Phys.
  Rev. B}\ }\textbf {\bibinfo {volume} {96}},\ \bibinfo {pages} {100502}
  (\bibinfo {year} {2017})}\BibitemShut {NoStop}%
\bibitem [{\citenamefont {Tikhonov}\ \emph {et~al.}(2023)\citenamefont
  {Tikhonov}, \citenamefont {Feng}, \citenamefont {Wang},\ and\ \citenamefont
  {Feng}}]{tikhonov2023}%
  \BibitemOpen
  \bibfield  {author} {\bibinfo {author} {\bibfnamefont {E.}~\bibnamefont
  {Tikhonov}}, \bibinfo {author} {\bibfnamefont {J.}~\bibnamefont {Feng}},
  \bibinfo {author} {\bibfnamefont {Y.}~\bibnamefont {Wang}}, \ and\ \bibinfo
  {author} {\bibfnamefont {Q.}~\bibnamefont {Feng}},\ }\bibfield  {title}
  {\enquote {\bibinfo {title} {High-pressure stability and superconductivity of
  vanadium hydrides*},}\ }\href {\doibase
  https://doi.org/10.1016/j.physb.2022.414603} {\bibfield  {journal} {\bibinfo
  {journal} {Physica B: Condensed Matter}\ }\textbf {\bibinfo {volume} {651}},\
  \bibinfo {pages} {414603} (\bibinfo {year} {2023})}\BibitemShut {NoStop}%
\bibitem [{\citenamefont {Troyan}\ \emph {et~al.}(2016)\citenamefont {Troyan},
  \citenamefont {Gavriliuk}, \citenamefont {Rüffer}, \citenamefont {Chumakov},
  \citenamefont {Mironovich}, \citenamefont {Lyubutin}, \citenamefont
  {Perekalin}, \citenamefont {Drozdov},\ and\ \citenamefont
  {Eremets}}]{troyan2016}%
  \BibitemOpen
  \bibfield  {author} {\bibinfo {author} {\bibfnamefont {I.}~\bibnamefont
  {Troyan}}, \bibinfo {author} {\bibfnamefont {A.}~\bibnamefont {Gavriliuk}},
  \bibinfo {author} {\bibfnamefont {R.}~\bibnamefont {Rüffer}}, \bibinfo
  {author} {\bibfnamefont {A.}~\bibnamefont {Chumakov}}, \bibinfo {author}
  {\bibfnamefont {A.}~\bibnamefont {Mironovich}}, \bibinfo {author}
  {\bibfnamefont {I.}~\bibnamefont {Lyubutin}}, \bibinfo {author}
  {\bibfnamefont {D.}~\bibnamefont {Perekalin}}, \bibinfo {author}
  {\bibfnamefont {A.~P.}\ \bibnamefont {Drozdov}}, \ and\ \bibinfo {author}
  {\bibfnamefont {M.~I.}\ \bibnamefont {Eremets}},\ }\bibfield  {title}
  {\enquote {\bibinfo {title} {Observation of superconductivity in hydrogen
  sulfide from nuclear resonant scattering},}\ }\href {\doibase
  10.1126/science.aac8176} {\bibfield  {journal} {\bibinfo  {journal}
  {Science}\ }\textbf {\bibinfo {volume} {351}},\ \bibinfo {pages} {1303}
  (\bibinfo {year} {2016})}\BibitemShut {NoStop}%
\bibitem [{\citenamefont {Tse}\ \emph {et~al.}(2007)\citenamefont {Tse},
  \citenamefont {Yao},\ and\ \citenamefont {Tanaka}}]{tse2007}%
  \BibitemOpen
  \bibfield  {author} {\bibinfo {author} {\bibfnamefont {J.~S.}\ \bibnamefont
  {Tse}}, \bibinfo {author} {\bibfnamefont {Y.}~\bibnamefont {Yao}}, \ and\
  \bibinfo {author} {\bibfnamefont {K.}~\bibnamefont {Tanaka}},\ }\bibfield
  {title} {\enquote {\bibinfo {title} {Novel superconductivity in metallic
  ${\mathrm{snh}}_{4}$ under high pressure},}\ }\href {\doibase
  10.1103/PhysRevLett.98.117004} {\bibfield  {journal} {\bibinfo  {journal}
  {Phys. Rev. Lett.}\ }\textbf {\bibinfo {volume} {98}},\ \bibinfo {pages}
  {117004} (\bibinfo {year} {2007})}\BibitemShut {NoStop}%
\bibitem [{\citenamefont {Tsuppayakorn-aek}\ \emph {et~al.}(2020)\citenamefont
  {Tsuppayakorn-aek}, \citenamefont {Pinsook}, \citenamefont {Luo},
  \citenamefont {Ahuja},\ and\ \citenamefont
  {Bovornratanaraks}}]{tsuppayakorn-aek2020}%
  \BibitemOpen
  \bibfield  {author} {\bibinfo {author} {\bibfnamefont {P.}~\bibnamefont
  {Tsuppayakorn-aek}}, \bibinfo {author} {\bibfnamefont {U.}~\bibnamefont
  {Pinsook}}, \bibinfo {author} {\bibfnamefont {W.}~\bibnamefont {Luo}},
  \bibinfo {author} {\bibfnamefont {R.}~\bibnamefont {Ahuja}}, \ and\ \bibinfo
  {author} {\bibfnamefont {T.}~\bibnamefont {Bovornratanaraks}},\ }\bibfield
  {title} {\enquote {\bibinfo {title} {Superconductivity of superhydride ceh10
  under high pressure},}\ }\href {\doibase 10.1088/2053-1591/ababc2} {\bibfield
   {journal} {\bibinfo  {journal} {Materials Research Express}\ }\textbf
  {\bibinfo {volume} {7}},\ \bibinfo {pages} {086001} (\bibinfo {year}
  {2020})}\BibitemShut {NoStop}%
\bibitem [{\citenamefont {Tsuppayakorn-aek}\ \emph {et~al.}(2021)\citenamefont
  {Tsuppayakorn-aek}, \citenamefont {Phaisangittisakul}, \citenamefont
  {Ahuja},\ and\ \citenamefont {Bovornratanaraks}}]{tsuppayakorn-aek2021}%
  \BibitemOpen
  \bibfield  {author} {\bibinfo {author} {\bibfnamefont {P.}~\bibnamefont
  {Tsuppayakorn-aek}}, \bibinfo {author} {\bibfnamefont {N.}~\bibnamefont
  {Phaisangittisakul}}, \bibinfo {author} {\bibfnamefont {R.}~\bibnamefont
  {Ahuja}}, \ and\ \bibinfo {author} {\bibfnamefont {T.}~\bibnamefont
  {Bovornratanaraks}},\ }\bibfield  {title} {\enquote {\bibinfo {title}
  {High-temperature superconductor of sodalite-like clathrate hafnium
  hexahydride},}\ }\href {\doibase 10.1038/s41598-021-95112-5} {\bibfield
  {journal} {\bibinfo  {journal} {Scientific Reports}\ }\textbf {\bibinfo
  {volume} {11}},\ \bibinfo {pages} {16403} (\bibinfo {year}
  {2021})}\BibitemShut {NoStop}%
\bibitem [{\citenamefont {Tsuppayakorn-aek}\ \emph {et~al.}(2023)\citenamefont
  {Tsuppayakorn-aek}, \citenamefont {Majumdar}, \citenamefont {Ahuja},
  \citenamefont {Bovornratanaraks},\ and\ \citenamefont
  {Luo}}]{tsuppayakorn-aek2023}%
  \BibitemOpen
  \bibfield  {author} {\bibinfo {author} {\bibfnamefont {P.}~\bibnamefont
  {Tsuppayakorn-aek}}, \bibinfo {author} {\bibfnamefont {A.}~\bibnamefont
  {Majumdar}}, \bibinfo {author} {\bibfnamefont {R.}~\bibnamefont {Ahuja}},
  \bibinfo {author} {\bibfnamefont {T.}~\bibnamefont {Bovornratanaraks}}, \
  and\ \bibinfo {author} {\bibfnamefont {W.}~\bibnamefont {Luo}},\ }\bibfield
  {title} {\enquote {\bibinfo {title} {Superconducting state of the van der
  waals layered pdh2 structure at high pressure},}\ }\href {\doibase
  https://doi.org/10.1016/j.ijhydene.2022.12.312} {\bibfield  {journal}
  {\bibinfo  {journal} {International Journal of Hydrogen Energy}\ }\textbf
  {\bibinfo {volume} {48}},\ \bibinfo {pages} {16769} (\bibinfo {year}
  {2023})}\BibitemShut {NoStop}%
\bibitem [{\citenamefont {Wang}\ \emph {et~al.}(2012)\citenamefont {Wang},
  \citenamefont {Tse}, \citenamefont {Tanaka}, \citenamefont {Iitaka},\ and\
  \citenamefont {Ma}}]{wang2012}%
  \BibitemOpen
  \bibfield  {author} {\bibinfo {author} {\bibfnamefont {H.}~\bibnamefont
  {Wang}}, \bibinfo {author} {\bibfnamefont {J.~S.}\ \bibnamefont {Tse}},
  \bibinfo {author} {\bibfnamefont {K.}~\bibnamefont {Tanaka}}, \bibinfo
  {author} {\bibfnamefont {T.}~\bibnamefont {Iitaka}}, \ and\ \bibinfo {author}
  {\bibfnamefont {Y.}~\bibnamefont {Ma}},\ }\bibfield  {title} {\enquote
  {\bibinfo {title} {Superconductive sodalite-like clathrate calcium hydride at
  high pressures},}\ }\href {\doibase 10.1073/pnas.1118168109} {\bibfield
  {journal} {\bibinfo  {journal} {Proceedings of the National Academy of
  Sciences}\ }\textbf {\bibinfo {volume} {109}},\ \bibinfo {pages} {6463}
  (\bibinfo {year} {2012})}\BibitemShut {NoStop}%
\bibitem [{\citenamefont {Wang}\ \emph {et~al.}(2018)\citenamefont {Wang},
  \citenamefont {Duan}, \citenamefont {Yu}, \citenamefont {Xie}, \citenamefont
  {Huang}, \citenamefont {Ma}, \citenamefont {Tian}, \citenamefont {Li},
  \citenamefont {Liu},\ and\ \citenamefont {Cui}}]{wang2017}%
  \BibitemOpen
  \bibfield  {author} {\bibinfo {author} {\bibfnamefont {L.}~\bibnamefont
  {Wang}}, \bibinfo {author} {\bibfnamefont {D.}~\bibnamefont {Duan}}, \bibinfo
  {author} {\bibfnamefont {H.}~\bibnamefont {Yu}}, \bibinfo {author}
  {\bibfnamefont {H.}~\bibnamefont {Xie}}, \bibinfo {author} {\bibfnamefont
  {X.}~\bibnamefont {Huang}}, \bibinfo {author} {\bibfnamefont
  {Y.}~\bibnamefont {Ma}}, \bibinfo {author} {\bibfnamefont {F.}~\bibnamefont
  {Tian}}, \bibinfo {author} {\bibfnamefont {D.}~\bibnamefont {Li}}, \bibinfo
  {author} {\bibfnamefont {B.}~\bibnamefont {Liu}}, \ and\ \bibinfo {author}
  {\bibfnamefont {T.}~\bibnamefont {Cui}},\ }\bibfield  {title} {\enquote
  {\bibinfo {title} {High-pressure formation of cobalt polyhydrides: A
  first-principle study},}\ }\href {\doibase 10.1021/acs.inorgchem.7b02371}
  {\bibfield  {journal} {\bibinfo  {journal} {Inorganic Chemistry}\ }\textbf
  {\bibinfo {volume} {57}},\ \bibinfo {pages} {181} (\bibinfo {year}
  {2018})}\BibitemShut {NoStop}%
\bibitem [{\citenamefont {Wang}\ \emph {et~al.}(2019)\citenamefont {Wang},
  \citenamefont {Zhang}, \citenamefont {Chen}, \citenamefont {Wu},
  \citenamefont {Zang},\ and\ \citenamefont {Lu}}]{wang2019}%
  \BibitemOpen
  \bibfield  {author} {\bibinfo {author} {\bibfnamefont {D.}~\bibnamefont
  {Wang}}, \bibinfo {author} {\bibfnamefont {H.}~\bibnamefont {Zhang}},
  \bibinfo {author} {\bibfnamefont {H.-L.}\ \bibnamefont {Chen}}, \bibinfo
  {author} {\bibfnamefont {J.}~\bibnamefont {Wu}}, \bibinfo {author}
  {\bibfnamefont {Q.-J.}\ \bibnamefont {Zang}}, \ and\ \bibinfo {author}
  {\bibfnamefont {W.-C.}\ \bibnamefont {Lu}},\ }\bibfield  {title} {\enquote
  {\bibinfo {title} {Theoretical study on uh4, uh8 and uh10 at high
  pressure},}\ }\href {\doibase https://doi.org/10.1016/j.physleta.2018.11.048}
  {\bibfield  {journal} {\bibinfo  {journal} {Physics Letters A}\ }\textbf
  {\bibinfo {volume} {383}},\ \bibinfo {pages} {774} (\bibinfo {year}
  {2019})}\BibitemShut {NoStop}%
\bibitem [{\citenamefont {Wang}\ \emph
  {et~al.}(2021{\natexlab{a}})\citenamefont {Wang}, \citenamefont {Zheng},
  \citenamefont {Gu}, \citenamefont {Tan}, \citenamefont {Zhao}, \citenamefont
  {Liu}, \citenamefont {Sun}, \citenamefont {Liu},\ and\ \citenamefont
  {Zhang}}]{wang2021}%
  \BibitemOpen
  \bibfield  {author} {\bibinfo {author} {\bibfnamefont {X.-h.}\ \bibnamefont
  {Wang}}, \bibinfo {author} {\bibfnamefont {F.-w.}\ \bibnamefont {Zheng}},
  \bibinfo {author} {\bibfnamefont {Z.-w.}\ \bibnamefont {Gu}}, \bibinfo
  {author} {\bibfnamefont {F.-l.}\ \bibnamefont {Tan}}, \bibinfo {author}
  {\bibfnamefont {J.-h.}\ \bibnamefont {Zhao}}, \bibinfo {author}
  {\bibfnamefont {C.-l.}\ \bibnamefont {Liu}}, \bibinfo {author} {\bibfnamefont
  {C.-w.}\ \bibnamefont {Sun}}, \bibinfo {author} {\bibfnamefont
  {J.}~\bibnamefont {Liu}}, \ and\ \bibinfo {author} {\bibfnamefont
  {P.}~\bibnamefont {Zhang}},\ }\bibfield  {title} {\enquote {\bibinfo {title}
  {Hydrogen clathrate structures in uranium hydrides at high pressures},}\
  }\href {\doibase 10.1021/acsomega.0c05794} {\bibfield  {journal} {\bibinfo
  {journal} {ACS Omega}\ }\textbf {\bibinfo {volume} {6}},\ \bibinfo {pages}
  {3946} (\bibinfo {year} {2021}{\natexlab{a}})}\BibitemShut {NoStop}%
\bibitem [{\citenamefont {Wang}\ \emph
  {et~al.}(2021{\natexlab{b}})\citenamefont {Wang}, \citenamefont {Shan},
  \citenamefont {Chen}, \citenamefont {Sun}, \citenamefont {Yang},
  \citenamefont {Ma}, \citenamefont {Wang}, \citenamefont {Yu}, \citenamefont
  {Zhang}, \citenamefont {Chen}, \citenamefont {Cheng}, \citenamefont {Dong},
  \citenamefont {Chen},\ and\ \citenamefont {Zhao}}]{wang2021A}%
  \BibitemOpen
  \bibfield  {author} {\bibinfo {author} {\bibfnamefont {N.~N.}\ \bibnamefont
  {Wang}}, \bibinfo {author} {\bibfnamefont {P.~F.}\ \bibnamefont {Shan}},
  \bibinfo {author} {\bibfnamefont {K.~Y.}\ \bibnamefont {Chen}}, \bibinfo
  {author} {\bibfnamefont {J.~P.}\ \bibnamefont {Sun}}, \bibinfo {author}
  {\bibfnamefont {P.~T.}\ \bibnamefont {Yang}}, \bibinfo {author}
  {\bibfnamefont {X.~L.}\ \bibnamefont {Ma}}, \bibinfo {author} {\bibfnamefont
  {B.~S.}\ \bibnamefont {Wang}}, \bibinfo {author} {\bibfnamefont {X.~H.}\
  \bibnamefont {Yu}}, \bibinfo {author} {\bibfnamefont {S.}~\bibnamefont
  {Zhang}}, \bibinfo {author} {\bibfnamefont {G.~F.}\ \bibnamefont {Chen}},
  \bibinfo {author} {\bibfnamefont {J.-G.}\ \bibnamefont {Cheng}}, \bibinfo
  {author} {\bibfnamefont {X.~L.}\ \bibnamefont {Dong}}, \bibinfo {author}
  {\bibfnamefont {X.~H.}\ \bibnamefont {Chen}}, \ and\ \bibinfo {author}
  {\bibfnamefont {Z.~X.}\ \bibnamefont {Zhao}},\ }\bibfield  {title} {\enquote
  {\bibinfo {title} {A low-tc superconducting modification of th4h15
  synthesized under high pressure},}\ }\href {\doibase
  10.1088/1361-6668/abdcc2} {\bibfield  {journal} {\bibinfo  {journal}
  {Superconductor Science and Technology}\ }\textbf {\bibinfo {volume} {34}},\
  \bibinfo {pages} {034006} (\bibinfo {year} {2021}{\natexlab{b}})}\BibitemShut
  {NoStop}%
\bibitem [{\citenamefont {Wang}\ \emph {et~al.}(2022)\citenamefont {Wang},
  \citenamefont {Wang}, \citenamefont {Sun}, \citenamefont {Ma}, \citenamefont
  {Wang}, \citenamefont {Zou}, \citenamefont {Liu}, \citenamefont {Zhou},\ and\
  \citenamefont {Wang}}]{wang2022}%
  \BibitemOpen
  \bibfield  {author} {\bibinfo {author} {\bibfnamefont {Y.}~\bibnamefont
  {Wang}}, \bibinfo {author} {\bibfnamefont {K.}~\bibnamefont {Wang}}, \bibinfo
  {author} {\bibfnamefont {Y.}~\bibnamefont {Sun}}, \bibinfo {author}
  {\bibfnamefont {L.}~\bibnamefont {Ma}}, \bibinfo {author} {\bibfnamefont
  {Y.}~\bibnamefont {Wang}}, \bibinfo {author} {\bibfnamefont {B.}~\bibnamefont
  {Zou}}, \bibinfo {author} {\bibfnamefont {G.}~\bibnamefont {Liu}}, \bibinfo
  {author} {\bibfnamefont {M.}~\bibnamefont {Zhou}}, \ and\ \bibinfo {author}
  {\bibfnamefont {H.}~\bibnamefont {Wang}},\ }\bibfield  {title} {\enquote
  {\bibinfo {title} {Synthesis and superconductivity in yttrium superhydrides
  under high pressure},}\ }\href {\doibase 10.1088/1674-1056/ac872e} {\bibfield
   {journal} {\bibinfo  {journal} {Chinese Physics B}\ }\textbf {\bibinfo
  {volume} {31}},\ \bibinfo {pages} {106201} (\bibinfo {year}
  {2022})}\BibitemShut {NoStop}%
\bibitem [{\citenamefont {Wang}\ \emph
  {et~al.}(2023{\natexlab{a}})\citenamefont {Wang}, \citenamefont {Zhang},
  \citenamefont {Li}, \citenamefont {Wang}, \citenamefont {Liu}, \citenamefont
  {Ma}, \citenamefont {Xu}, \citenamefont {Liu},\ and\ \citenamefont
  {Ma}}]{wang2023}%
  \BibitemOpen
  \bibfield  {author} {\bibinfo {author} {\bibfnamefont {Q.}~\bibnamefont
  {Wang}}, \bibinfo {author} {\bibfnamefont {S.}~\bibnamefont {Zhang}},
  \bibinfo {author} {\bibfnamefont {H.}~\bibnamefont {Li}}, \bibinfo {author}
  {\bibfnamefont {H.}~\bibnamefont {Wang}}, \bibinfo {author} {\bibfnamefont
  {G.}~\bibnamefont {Liu}}, \bibinfo {author} {\bibfnamefont {J.}~\bibnamefont
  {Ma}}, \bibinfo {author} {\bibfnamefont {H.}~\bibnamefont {Xu}}, \bibinfo
  {author} {\bibfnamefont {H.}~\bibnamefont {Liu}}, \ and\ \bibinfo {author}
  {\bibfnamefont {Y.}~\bibnamefont {Ma}},\ }\bibfield  {title} {\enquote
  {\bibinfo {title} {Coexistence of superconductivity and electride states in
  ca2h with an antifluorite-type motif under compression},}\ }\href {\doibase
  10.1039/D3TA04418A} {\bibfield  {journal} {\bibinfo  {journal} {J. Mater.
  Chem. A}\ }\textbf {\bibinfo {volume} {11}},\ \bibinfo {pages} {21345}
  (\bibinfo {year} {2023}{\natexlab{a}})}\BibitemShut {NoStop}%
\bibitem [{\citenamefont {Wang}\ \emph
  {et~al.}(2023{\natexlab{b}})\citenamefont {Wang}, \citenamefont {Zhang},
  \citenamefont {Zhang}, \citenamefont {Li}, \citenamefont {Ju}, \citenamefont
  {Sun}, \citenamefont {Dou},\ and\ \citenamefont {Jin}}]{wang2023A}%
  \BibitemOpen
  \bibfield  {author} {\bibinfo {author} {\bibfnamefont {Y.-Q.}\ \bibnamefont
  {Wang}}, \bibinfo {author} {\bibfnamefont {C.-Z.}\ \bibnamefont {Zhang}},
  \bibinfo {author} {\bibfnamefont {J.-Q.}\ \bibnamefont {Zhang}}, \bibinfo
  {author} {\bibfnamefont {S.}~\bibnamefont {Li}}, \bibinfo {author}
  {\bibfnamefont {M.}~\bibnamefont {Ju}}, \bibinfo {author} {\bibfnamefont
  {W.-G.}\ \bibnamefont {Sun}}, \bibinfo {author} {\bibfnamefont {X.-L.}\
  \bibnamefont {Dou}}, \ and\ \bibinfo {author} {\bibfnamefont {Y.-Y.}\
  \bibnamefont {Jin}},\ }\bibfield  {title} {\enquote {\bibinfo {title} {A
  ten-fold coordinated high-pressure structure in hafnium dihydrogen with
  increasing superconducting transition temperature induced by enhancive
  pressure},}\ }\href {\doibase 10.1088/1674-1056/acc934} {\bibfield  {journal}
  {\bibinfo  {journal} {Chinese Physics B}\ }\textbf {\bibinfo {volume} {32}},\
  \bibinfo {pages} {097402} (\bibinfo {year} {2023}{\natexlab{b}})}\BibitemShut
  {NoStop}%
\bibitem [{\citenamefont {Wei}\ \emph {et~al.}(2016)\citenamefont {Wei},
  \citenamefont {Yuan}, \citenamefont {Khan}, \citenamefont {Ji}, \citenamefont
  {Gu},\ and\ \citenamefont {Wei}}]{wei2016}%
  \BibitemOpen
  \bibfield  {author} {\bibinfo {author} {\bibfnamefont {Y.-K.}\ \bibnamefont
  {Wei}}, \bibinfo {author} {\bibfnamefont {J.-N.}\ \bibnamefont {Yuan}},
  \bibinfo {author} {\bibfnamefont {F.~I.}\ \bibnamefont {Khan}}, \bibinfo
  {author} {\bibfnamefont {G.-F.}\ \bibnamefont {Ji}}, \bibinfo {author}
  {\bibfnamefont {Z.-W.}\ \bibnamefont {Gu}}, \ and\ \bibinfo {author}
  {\bibfnamefont {D.-Q.}\ \bibnamefont {Wei}},\ }\bibfield  {title} {\enquote
  {\bibinfo {title} {Pressure induced superconductivity and electronic
  structure properties of scandium hydrides using first principles
  calculations},}\ }\href {\doibase 10.1039/C6RA11862C} {\bibfield  {journal}
  {\bibinfo  {journal} {RSC Adv.}\ }\textbf {\bibinfo {volume} {6}},\ \bibinfo
  {pages} {81534} (\bibinfo {year} {2016})}\BibitemShut {NoStop}%
\bibitem [{\citenamefont {Wu}\ \emph {et~al.}(2018)\citenamefont {Wu},
  \citenamefont {Zhao}, \citenamefont {Chen}, \citenamefont {Wang},
  \citenamefont {Chen}, \citenamefont {Guo}, \citenamefont {Zang},\ and\
  \citenamefont {Lu}}]{wu2018}%
  \BibitemOpen
  \bibfield  {author} {\bibinfo {author} {\bibfnamefont {J.}~\bibnamefont
  {Wu}}, \bibinfo {author} {\bibfnamefont {L.-Z.}\ \bibnamefont {Zhao}},
  \bibinfo {author} {\bibfnamefont {H.-L.}\ \bibnamefont {Chen}}, \bibinfo
  {author} {\bibfnamefont {D.}~\bibnamefont {Wang}}, \bibinfo {author}
  {\bibfnamefont {J.-Y.}\ \bibnamefont {Chen}}, \bibinfo {author}
  {\bibfnamefont {X.}~\bibnamefont {Guo}}, \bibinfo {author} {\bibfnamefont
  {Q.-J.}\ \bibnamefont {Zang}}, \ and\ \bibinfo {author} {\bibfnamefont
  {W.-C.}\ \bibnamefont {Lu}},\ }\bibfield  {title} {\enquote {\bibinfo {title}
  {Structures and superconducting properties of ultra-hydrogen-rich selenium
  hydride h6se},}\ }\href {\doibase https://doi.org/10.1002/pssb.201800224}
  {\bibfield  {journal} {\bibinfo  {journal} {Physica Status Solidi (b)}\
  }\textbf {\bibinfo {volume} {255}},\ \bibinfo {pages} {1800224} (\bibinfo
  {year} {2018})}\BibitemShut {NoStop}%
\bibitem [{\citenamefont {Xiao}\ \emph
  {et~al.}(2019{\natexlab{a}})\citenamefont {Xiao}, \citenamefont {Duan},
  \citenamefont {Xie}, \citenamefont {Shao}, \citenamefont {Li}, \citenamefont
  {Tian}, \citenamefont {Song}, \citenamefont {Yu}, \citenamefont {Bao},\ and\
  \citenamefont {Cui}}]{xiao2019}%
  \BibitemOpen
  \bibfield  {author} {\bibinfo {author} {\bibfnamefont {X.}~\bibnamefont
  {Xiao}}, \bibinfo {author} {\bibfnamefont {D.}~\bibnamefont {Duan}}, \bibinfo
  {author} {\bibfnamefont {H.}~\bibnamefont {Xie}}, \bibinfo {author}
  {\bibfnamefont {Z.}~\bibnamefont {Shao}}, \bibinfo {author} {\bibfnamefont
  {D.}~\bibnamefont {Li}}, \bibinfo {author} {\bibfnamefont {F.}~\bibnamefont
  {Tian}}, \bibinfo {author} {\bibfnamefont {H.}~\bibnamefont {Song}}, \bibinfo
  {author} {\bibfnamefont {H.}~\bibnamefont {Yu}}, \bibinfo {author}
  {\bibfnamefont {K.}~\bibnamefont {Bao}}, \ and\ \bibinfo {author}
  {\bibfnamefont {T.}~\bibnamefont {Cui}},\ }\bibfield  {title} {\enquote
  {\bibinfo {title} {Structure and superconductivity of protactinium hydrides
  under high pressure},}\ }\href {\doibase 10.1088/1361-648X/ab1d03} {\bibfield
   {journal} {\bibinfo  {journal} {Journal of Physics: Condensed Matter}\
  }\textbf {\bibinfo {volume} {31}},\ \bibinfo {pages} {315403} (\bibinfo
  {year} {2019}{\natexlab{a}})}\BibitemShut {NoStop}%
\bibitem [{\citenamefont {Xiao}\ \emph
  {et~al.}(2019{\natexlab{b}})\citenamefont {Xiao}, \citenamefont {Duan},
  \citenamefont {Xie}, \citenamefont {Shao}, \citenamefont {Li}, \citenamefont
  {Tian}, \citenamefont {Song}, \citenamefont {Yu}, \citenamefont {Bao},\ and\
  \citenamefont {Cui}}]{xiao2019A}%
  \BibitemOpen
  \bibfield  {author} {\bibinfo {author} {\bibfnamefont {X.}~\bibnamefont
  {Xiao}}, \bibinfo {author} {\bibfnamefont {D.}~\bibnamefont {Duan}}, \bibinfo
  {author} {\bibfnamefont {H.}~\bibnamefont {Xie}}, \bibinfo {author}
  {\bibfnamefont {Z.}~\bibnamefont {Shao}}, \bibinfo {author} {\bibfnamefont
  {D.}~\bibnamefont {Li}}, \bibinfo {author} {\bibfnamefont {F.}~\bibnamefont
  {Tian}}, \bibinfo {author} {\bibfnamefont {H.}~\bibnamefont {Song}}, \bibinfo
  {author} {\bibfnamefont {H.}~\bibnamefont {Yu}}, \bibinfo {author}
  {\bibfnamefont {K.}~\bibnamefont {Bao}}, \ and\ \bibinfo {author}
  {\bibfnamefont {T.}~\bibnamefont {Cui}},\ }\bibfield  {title} {\enquote
  {\bibinfo {title} {Structure and superconductivity of protactinium hydrides
  under high pressure},}\ }\href {\doibase 10.1088/1361-648X/ab1d03} {\bibfield
   {journal} {\bibinfo  {journal} {Journal of Physics Condensed Matter}\
  }\textbf {\bibinfo {volume} {31}},\ \bibinfo {pages} {315403} (\bibinfo
  {year} {2019}{\natexlab{b}})}\BibitemShut {NoStop}%
\bibitem [{\citenamefont {Xie}\ \emph {et~al.}(2014)\citenamefont {Xie},
  \citenamefont {Li}, \citenamefont {Oganov},\ and\ \citenamefont
  {Wang}}]{xie2014}%
  \BibitemOpen
  \bibfield  {author} {\bibinfo {author} {\bibfnamefont {Y.}~\bibnamefont
  {Xie}}, \bibinfo {author} {\bibfnamefont {Q.}~\bibnamefont {Li}}, \bibinfo
  {author} {\bibfnamefont {A.~R.}\ \bibnamefont {Oganov}}, \ and\ \bibinfo
  {author} {\bibfnamefont {H.}~\bibnamefont {Wang}},\ }\bibfield  {title}
  {\enquote {\bibinfo {title} {{Superconductivity of lithium-doped hydrogen
  under high pressure}},}\ }\href {\doibase 10.1107/S2053229613028337}
  {\bibfield  {journal} {\bibinfo  {journal} {Acta Crystallographica Section
  C}\ }\textbf {\bibinfo {volume} {70}},\ \bibinfo {pages} {104} (\bibinfo
  {year} {2014})}\BibitemShut {NoStop}%
\bibitem [{\citenamefont {Xie}\ \emph {et~al.}(2020{\natexlab{a}})\citenamefont
  {Xie}, \citenamefont {Yao}, \citenamefont {Feng}, \citenamefont {Duan},
  \citenamefont {Song}, \citenamefont {Zhang}, \citenamefont {Jiang},
  \citenamefont {Redfern}, \citenamefont {Kresin}, \citenamefont {Pickard},\
  and\ \citenamefont {Cui}}]{xie2020}%
  \BibitemOpen
  \bibfield  {author} {\bibinfo {author} {\bibfnamefont {H.}~\bibnamefont
  {Xie}}, \bibinfo {author} {\bibfnamefont {Y.}~\bibnamefont {Yao}}, \bibinfo
  {author} {\bibfnamefont {X.}~\bibnamefont {Feng}}, \bibinfo {author}
  {\bibfnamefont {D.}~\bibnamefont {Duan}}, \bibinfo {author} {\bibfnamefont
  {H.}~\bibnamefont {Song}}, \bibinfo {author} {\bibfnamefont {Z.}~\bibnamefont
  {Zhang}}, \bibinfo {author} {\bibfnamefont {S.}~\bibnamefont {Jiang}},
  \bibinfo {author} {\bibfnamefont {S.~A.~T.}\ \bibnamefont {Redfern}},
  \bibinfo {author} {\bibfnamefont {V.~Z.}\ \bibnamefont {Kresin}}, \bibinfo
  {author} {\bibfnamefont {C.~J.}\ \bibnamefont {Pickard}}, \ and\ \bibinfo
  {author} {\bibfnamefont {T.}~\bibnamefont {Cui}},\ }\bibfield  {title}
  {\enquote {\bibinfo {title} {Hydrogen pentagraphenelike structure stabilized
  by hafnium: A high-temperature conventional superconductor},}\ }\href
  {\doibase 10.1103/PhysRevLett.125.217001} {\bibfield  {journal} {\bibinfo
  {journal} {Phys. Rev. Lett.}\ }\textbf {\bibinfo {volume} {125}},\ \bibinfo
  {pages} {217001} (\bibinfo {year} {2020}{\natexlab{a}})}\BibitemShut
  {NoStop}%
\bibitem [{\citenamefont {Xie}\ \emph {et~al.}(2020{\natexlab{b}})\citenamefont
  {Xie}, \citenamefont {Zhang}, \citenamefont {Duan}, \citenamefont {Huang},
  \citenamefont {Huang}, \citenamefont {Song}, \citenamefont {Feng},
  \citenamefont {Yao}, \citenamefont {Pickard},\ and\ \citenamefont
  {Cui}}]{xie2020A}%
  \BibitemOpen
  \bibfield  {author} {\bibinfo {author} {\bibfnamefont {H.}~\bibnamefont
  {Xie}}, \bibinfo {author} {\bibfnamefont {W.}~\bibnamefont {Zhang}}, \bibinfo
  {author} {\bibfnamefont {D.}~\bibnamefont {Duan}}, \bibinfo {author}
  {\bibfnamefont {X.}~\bibnamefont {Huang}}, \bibinfo {author} {\bibfnamefont
  {Y.}~\bibnamefont {Huang}}, \bibinfo {author} {\bibfnamefont
  {H.}~\bibnamefont {Song}}, \bibinfo {author} {\bibfnamefont {X.}~\bibnamefont
  {Feng}}, \bibinfo {author} {\bibfnamefont {Y.}~\bibnamefont {Yao}}, \bibinfo
  {author} {\bibfnamefont {C.~J.}\ \bibnamefont {Pickard}}, \ and\ \bibinfo
  {author} {\bibfnamefont {T.}~\bibnamefont {Cui}},\ }\bibfield  {title}
  {\enquote {\bibinfo {title} {Superconducting zirconium polyhydrides at
  moderate pressures},}\ }\href {\doibase 10.1021/acs.jpclett.9b03632}
  {\bibfield  {journal} {\bibinfo  {journal} {The Journal of Physical Chemistry
  Letters}\ }\textbf {\bibinfo {volume} {11}},\ \bibinfo {pages} {646}
  (\bibinfo {year} {2020}{\natexlab{b}})}\BibitemShut {NoStop}%
\bibitem [{\citenamefont {Yan}\ \emph {et~al.}(2015)\citenamefont {Yan},
  \citenamefont {Chen}, \citenamefont {Kuang},\ and\ \citenamefont
  {Xiang}}]{yan2015}%
  \BibitemOpen
  \bibfield  {author} {\bibinfo {author} {\bibfnamefont {X.}~\bibnamefont
  {Yan}}, \bibinfo {author} {\bibfnamefont {Y.}~\bibnamefont {Chen}}, \bibinfo
  {author} {\bibfnamefont {X.}~\bibnamefont {Kuang}}, \ and\ \bibinfo {author}
  {\bibfnamefont {S.}~\bibnamefont {Xiang}},\ }\bibfield  {title} {\enquote
  {\bibinfo {title} {{Structure, stability, and superconductivity of new Xe–H
  compounds under high pressure}},}\ }\href {\doibase 10.1063/1.4931931}
  {\bibfield  {journal} {\bibinfo  {journal} {The Journal of Chemical Physics}\
  }\textbf {\bibinfo {volume} {143}},\ \bibinfo {pages} {124310} (\bibinfo
  {year} {2015})}\BibitemShut {NoStop}%
\bibitem [{\citenamefont {Yan}\ \emph {et~al.}(2022)\citenamefont {Yan},
  \citenamefont {Ding}, \citenamefont {Zhang}, \citenamefont {Bergara},
  \citenamefont {Liu}, \citenamefont {Wang}, \citenamefont {Zhou},\ and\
  \citenamefont {Yang}}]{yan2022}%
  \BibitemOpen
  \bibfield  {author} {\bibinfo {author} {\bibfnamefont {X.}~\bibnamefont
  {Yan}}, \bibinfo {author} {\bibfnamefont {S.}~\bibnamefont {Ding}}, \bibinfo
  {author} {\bibfnamefont {X.}~\bibnamefont {Zhang}}, \bibinfo {author}
  {\bibfnamefont {A.}~\bibnamefont {Bergara}}, \bibinfo {author} {\bibfnamefont
  {Y.}~\bibnamefont {Liu}}, \bibinfo {author} {\bibfnamefont {Y.}~\bibnamefont
  {Wang}}, \bibinfo {author} {\bibfnamefont {X.-F.}\ \bibnamefont {Zhou}}, \
  and\ \bibinfo {author} {\bibfnamefont {G.}~\bibnamefont {Yang}},\ }\bibfield
  {title} {\enquote {\bibinfo {title} {Enhanced superconductivity in
  ${\mathrm{cuh}}_{2}$ monolayers},}\ }\href {\doibase
  10.1103/PhysRevB.106.014514} {\bibfield  {journal} {\bibinfo  {journal}
  {Phys. Rev. B}\ }\textbf {\bibinfo {volume} {106}},\ \bibinfo {pages}
  {014514} (\bibinfo {year} {2022})}\BibitemShut {NoStop}%
\bibitem [{\citenamefont {Yan}\ \emph {et~al.}(2023)\citenamefont {Yan},
  \citenamefont {Zhang}, \citenamefont {Chen},\ and\ \citenamefont
  {Kuang}}]{yan2023}%
  \BibitemOpen
  \bibfield  {author} {\bibinfo {author} {\bibfnamefont {X.~Z.}\ \bibnamefont
  {Yan}}, \bibinfo {author} {\bibfnamefont {Z.~L.}\ \bibnamefont {Zhang}},
  \bibinfo {author} {\bibfnamefont {Y.~M.}\ \bibnamefont {Chen}}, \ and\
  \bibinfo {author} {\bibfnamefont {F.~G.}\ \bibnamefont {Kuang}},\ }\bibfield
  {title} {\enquote {\bibinfo {title} {Prediction of superconductivity in
  clathrate er hydrides under high pressure},}\ }\href {\doibase
  10.3390/cryst13050792} {\bibfield  {journal} {\bibinfo  {journal} {Crystals}\
  }\textbf {\bibinfo {volume} {13}},\ \bibinfo {pages} {792} (\bibinfo {year}
  {2023})}\BibitemShut {NoStop}%
\bibitem [{\citenamefont {Yang}\ \emph {et~al.}(2019)\citenamefont {Yang},
  \citenamefont {Lu}, \citenamefont {Li}, \citenamefont {Xue}, \citenamefont
  {Zang}, \citenamefont {Ho},\ and\ \citenamefont {Wang}}]{yang2019}%
  \BibitemOpen
  \bibfield  {author} {\bibinfo {author} {\bibfnamefont {W.-H.}\ \bibnamefont
  {Yang}}, \bibinfo {author} {\bibfnamefont {W.-C.}\ \bibnamefont {Lu}},
  \bibinfo {author} {\bibfnamefont {S.-D.}\ \bibnamefont {Li}}, \bibinfo
  {author} {\bibfnamefont {X.-Y.}\ \bibnamefont {Xue}}, \bibinfo {author}
  {\bibfnamefont {Q.-J.}\ \bibnamefont {Zang}}, \bibinfo {author}
  {\bibfnamefont {K.~M.}\ \bibnamefont {Ho}}, \ and\ \bibinfo {author}
  {\bibfnamefont {C.~Z.}\ \bibnamefont {Wang}},\ }\bibfield  {title} {\enquote
  {\bibinfo {title} {Novel superconducting structures of bh2 under high
  pressure},}\ }\href {\doibase 10.1039/C9CP00310J} {\bibfield  {journal}
  {\bibinfo  {journal} {Phys. Chem. Chem. Phys.}\ }\textbf {\bibinfo {volume}
  {21}},\ \bibinfo {pages} {5466} (\bibinfo {year} {2019})}\BibitemShut
  {NoStop}%
\bibitem [{\citenamefont {Yang}\ \emph {et~al.}(2023)\citenamefont {Yang},
  \citenamefont {Jiang},\ and\ \citenamefont {Zhao}}]{yang2023}%
  \BibitemOpen
  \bibfield  {author} {\bibinfo {author} {\bibfnamefont {Q.}~\bibnamefont
  {Yang}}, \bibinfo {author} {\bibfnamefont {X.}~\bibnamefont {Jiang}}, \ and\
  \bibinfo {author} {\bibfnamefont {J.}~\bibnamefont {Zhao}},\ }\bibfield
  {title} {\enquote {\bibinfo {title} {Coexistence of zero-dimensional
  electride state and superconductivity in alh2 monolayer},}\ }\href {\doibase
  10.1088/0256-307X/40/10/107401} {\bibfield  {journal} {\bibinfo  {journal}
  {Chinese Physics Letters}\ }\textbf {\bibinfo {volume} {40}},\ \bibinfo
  {pages} {107401} (\bibinfo {year} {2023})}\BibitemShut {NoStop}%
\bibitem [{\citenamefont {Yao}\ \emph {et~al.}(2007)\citenamefont {Yao},
  \citenamefont {Tse}, \citenamefont {Ma},\ and\ \citenamefont
  {Tanaka}}]{yao2007}%
  \BibitemOpen
  \bibfield  {author} {\bibinfo {author} {\bibfnamefont {Y.}~\bibnamefont
  {Yao}}, \bibinfo {author} {\bibfnamefont {J.~S.}\ \bibnamefont {Tse}},
  \bibinfo {author} {\bibfnamefont {Y.}~\bibnamefont {Ma}}, \ and\ \bibinfo
  {author} {\bibfnamefont {K.}~\bibnamefont {Tanaka}},\ }\bibfield  {title}
  {\enquote {\bibinfo {title} {Superconductivity in high-pressure sih4},}\
  }\href {\doibase 10.1209/0295-5075/78/37003} {\bibfield  {journal} {\bibinfo
  {journal} {Europhysics Letters}\ }\textbf {\bibinfo {volume} {78}},\ \bibinfo
  {pages} {37003} (\bibinfo {year} {2007})}\BibitemShut {NoStop}%
\bibitem [{\citenamefont {Yao}\ \emph {et~al.}(2024)\citenamefont {Yao},
  \citenamefont {Wang}, \citenamefont {Jeon}, \citenamefont {Liu},
  \citenamefont {Bok}, \citenamefont {Bang}, \citenamefont {Jia},\ and\
  \citenamefont {Cho}}]{yao2023}%
  \BibitemOpen
  \bibfield  {author} {\bibinfo {author} {\bibfnamefont {S.}~\bibnamefont
  {Yao}}, \bibinfo {author} {\bibfnamefont {C.}~\bibnamefont {Wang}}, \bibinfo
  {author} {\bibfnamefont {H.}~\bibnamefont {Jeon}}, \bibinfo {author}
  {\bibfnamefont {L.}~\bibnamefont {Liu}}, \bibinfo {author} {\bibfnamefont
  {J.~M.}\ \bibnamefont {Bok}}, \bibinfo {author} {\bibfnamefont
  {Y.}~\bibnamefont {Bang}}, \bibinfo {author} {\bibfnamefont {Y.}~\bibnamefont
  {Jia}}, \ and\ \bibinfo {author} {\bibfnamefont {J.-H.}\ \bibnamefont
  {Cho}},\ }\bibfield  {title} {\enquote {\bibinfo {title} {High-pressure
  stability and superconductivity of clathrate thorium hydrides},}\ }\href
  {\doibase https://doi.org/10.1002/pssb.202300451} {\bibfield  {journal}
  {\bibinfo  {journal} {Physica Status Solidi (b)}\ }\textbf {\bibinfo {volume}
  {261}},\ \bibinfo {pages} {2300451} (\bibinfo {year} {2024})}\BibitemShut
  {NoStop}%
\bibitem [{\citenamefont {Ye}\ \emph {et~al.}(2018)\citenamefont {Ye},
  \citenamefont {Zarifi}, \citenamefont {Zurek}, \citenamefont {Hoffmann},\
  and\ \citenamefont {Ashcroft}}]{ye2018}%
  \BibitemOpen
  \bibfield  {author} {\bibinfo {author} {\bibfnamefont {X.}~\bibnamefont
  {Ye}}, \bibinfo {author} {\bibfnamefont {N.}~\bibnamefont {Zarifi}}, \bibinfo
  {author} {\bibfnamefont {E.}~\bibnamefont {Zurek}}, \bibinfo {author}
  {\bibfnamefont {R.}~\bibnamefont {Hoffmann}}, \ and\ \bibinfo {author}
  {\bibfnamefont {N.~W.}\ \bibnamefont {Ashcroft}},\ }\bibfield  {title}
  {\enquote {\bibinfo {title} {High hydrides of scandium under pressure:
  Potential superconductors},}\ }\href {\doibase 10.1021/acs.jpcc.7b12124}
  {\bibfield  {journal} {\bibinfo  {journal} {The Journal of Physical Chemistry
  C}\ }\textbf {\bibinfo {volume} {122}},\ \bibinfo {pages} {6298} (\bibinfo
  {year} {2018})}\BibitemShut {NoStop}%
\bibitem [{\citenamefont {He}\ and\ \citenamefont {Shi}(2023)}]{yong2023}%
  \BibitemOpen
  \bibfield  {author} {\bibinfo {author} {\bibfnamefont {Y.}~\bibnamefont
  {He}}\ and\ \bibinfo {author} {\bibfnamefont {J.-j.}\ \bibnamefont {Shi}},\
  }\bibfield  {title} {\enquote {\bibinfo {title} {Few-hydrogen high-tc
  superconductivity in (be4)2h nanosuperlattice with promising ductility under
  ambient pressure},}\ }\href {\doibase 10.1021/acs.nanolett.3c02213}
  {\bibfield  {journal} {\bibinfo  {journal} {Nano Letters}\ }\textbf {\bibinfo
  {volume} {23}},\ \bibinfo {pages} {8126} (\bibinfo {year}
  {2023})}\BibitemShut {NoStop}%
\bibitem [{\citenamefont {Yu}\ \emph {et~al.}(2014)\citenamefont {Yu},
  \citenamefont {Zeng}, \citenamefont {Oganov}, \citenamefont {Hu},
  \citenamefont {Frapper},\ and\ \citenamefont {Zhang}}]{yu2014}%
  \BibitemOpen
  \bibfield  {author} {\bibinfo {author} {\bibfnamefont {S.}~\bibnamefont
  {Yu}}, \bibinfo {author} {\bibfnamefont {Q.}~\bibnamefont {Zeng}}, \bibinfo
  {author} {\bibfnamefont {A.~R.}\ \bibnamefont {Oganov}}, \bibinfo {author}
  {\bibfnamefont {C.}~\bibnamefont {Hu}}, \bibinfo {author} {\bibfnamefont
  {G.}~\bibnamefont {Frapper}}, \ and\ \bibinfo {author} {\bibfnamefont
  {L.}~\bibnamefont {Zhang}},\ }\bibfield  {title} {\enquote {\bibinfo {title}
  {{Exploration of stable compounds, crystal structures, and superconductivity
  in the Be-H system}},}\ }\href {\doibase 10.1063/1.4898145} {\bibfield
  {journal} {\bibinfo  {journal} {AIP Advances}\ }\textbf {\bibinfo {volume}
  {4}},\ \bibinfo {pages} {107118} (\bibinfo {year} {2014})}\BibitemShut
  {NoStop}%
\bibitem [{\citenamefont {Yu}\ \emph {et~al.}(2015)\citenamefont {Yu},
  \citenamefont {Jia}, \citenamefont {Frapper}, \citenamefont {Li},
  \citenamefont {Oganov}, \citenamefont {Zeng},\ and\ \citenamefont
  {Zhang}}]{yu2015}%
  \BibitemOpen
  \bibfield  {author} {\bibinfo {author} {\bibfnamefont {S.}~\bibnamefont
  {Yu}}, \bibinfo {author} {\bibfnamefont {X.}~\bibnamefont {Jia}}, \bibinfo
  {author} {\bibfnamefont {G.}~\bibnamefont {Frapper}}, \bibinfo {author}
  {\bibfnamefont {D.}~\bibnamefont {Li}}, \bibinfo {author} {\bibfnamefont
  {A.~R.}\ \bibnamefont {Oganov}}, \bibinfo {author} {\bibfnamefont
  {Q.}~\bibnamefont {Zeng}}, \ and\ \bibinfo {author} {\bibfnamefont
  {L.}~\bibnamefont {Zhang}},\ }\bibfield  {title} {\enquote {\bibinfo {title}
  {Pressure-driven formation and stabilization of superconductive chromium
  hydrides},}\ }\href {\doibase 10.1038/srep17764} {\bibfield  {journal}
  {\bibinfo  {journal} {Scientific Reports}\ }\textbf {\bibinfo {volume} {5}},\
  \bibinfo {pages} {17764} (\bibinfo {year} {2015})}\BibitemShut {NoStop}%
\bibitem [{\citenamefont {Yuan}\ \emph {et~al.}(2019)\citenamefont {Yuan},
  \citenamefont {Li}, \citenamefont {Fang}, \citenamefont {Liu}, \citenamefont
  {Pei}, \citenamefont {Li}, \citenamefont {Zheng}, \citenamefont {Yang},\ and\
  \citenamefont {Wang}}]{yuan2019}%
  \BibitemOpen
  \bibfield  {author} {\bibinfo {author} {\bibfnamefont {Y.}~\bibnamefont
  {Yuan}}, \bibinfo {author} {\bibfnamefont {Y.}~\bibnamefont {Li}}, \bibinfo
  {author} {\bibfnamefont {G.}~\bibnamefont {Fang}}, \bibinfo {author}
  {\bibfnamefont {G.}~\bibnamefont {Liu}}, \bibinfo {author} {\bibfnamefont
  {C.}~\bibnamefont {Pei}}, \bibinfo {author} {\bibfnamefont {X.}~\bibnamefont
  {Li}}, \bibinfo {author} {\bibfnamefont {H.}~\bibnamefont {Zheng}}, \bibinfo
  {author} {\bibfnamefont {K.}~\bibnamefont {Yang}}, \ and\ \bibinfo {author}
  {\bibfnamefont {L.}~\bibnamefont {Wang}},\ }\bibfield  {title} {\enquote
  {\bibinfo {title} {{Stoichiometric evolutions of PH3 under high pressure:
  implication for high-Tc superconducting hydrides}},}\ }\href {\doibase
  10.1093/nsr/nwz010} {\bibfield  {journal} {\bibinfo  {journal} {National
  Science Review}\ }\textbf {\bibinfo {volume} {6}},\ \bibinfo {pages} {524}
  (\bibinfo {year} {2019})}\BibitemShut {NoStop}%
\bibitem [{\citenamefont {Zeng}\ \emph {et~al.}(2017)\citenamefont {Zeng},
  \citenamefont {Yu}, \citenamefont {Li}, \citenamefont {Oganov},\ and\
  \citenamefont {Frapper}}]{zeng2017}%
  \BibitemOpen
  \bibfield  {author} {\bibinfo {author} {\bibfnamefont {Q.}~\bibnamefont
  {Zeng}}, \bibinfo {author} {\bibfnamefont {S.}~\bibnamefont {Yu}}, \bibinfo
  {author} {\bibfnamefont {D.}~\bibnamefont {Li}}, \bibinfo {author}
  {\bibfnamefont {A.~R.}\ \bibnamefont {Oganov}}, \ and\ \bibinfo {author}
  {\bibfnamefont {G.}~\bibnamefont {Frapper}},\ }\bibfield  {title} {\enquote
  {\bibinfo {title} {Emergence of novel hydrogen chlorides under high
  pressure},}\ }\href {\doibase 10.1039/C6CP08708F} {\bibfield  {journal}
  {\bibinfo  {journal} {Phys. Chem. Chem. Phys.}\ }\textbf {\bibinfo {volume}
  {19}},\ \bibinfo {pages} {8236} (\bibinfo {year} {2017})}\BibitemShut
  {NoStop}%
\bibitem [{\citenamefont {Zhang}\ \emph {et~al.}(2010)\citenamefont {Zhang},
  \citenamefont {Chen}, \citenamefont {Li}, \citenamefont {Struzhkin},
  \citenamefont {Hemley}, \citenamefont {Mao}, \citenamefont {Zhang},\ and\
  \citenamefont {Lin}}]{zhang2010}%
  \BibitemOpen
  \bibfield  {author} {\bibinfo {author} {\bibfnamefont {C.}~\bibnamefont
  {Zhang}}, \bibinfo {author} {\bibfnamefont {X.~J.}\ \bibnamefont {Chen}},
  \bibinfo {author} {\bibfnamefont {Y.~L.}\ \bibnamefont {Li}}, \bibinfo
  {author} {\bibfnamefont {V.~V.}\ \bibnamefont {Struzhkin}}, \bibinfo {author}
  {\bibfnamefont {R.~J.}\ \bibnamefont {Hemley}}, \bibinfo {author}
  {\bibfnamefont {H.~K.}\ \bibnamefont {Mao}}, \bibinfo {author} {\bibfnamefont
  {R.~Q.}\ \bibnamefont {Zhang}}, \ and\ \bibinfo {author} {\bibfnamefont
  {H.~Q.}\ \bibnamefont {Lin}},\ }\bibfield  {title} {\enquote {\bibinfo
  {title} {Superconductivity in hydrogen-rich material: Geh4},}\ }\href
  {\doibase 10.1007/s10948-010-0675-2} {\bibfield  {journal} {\bibinfo
  {journal} {Journal of Superconductivity and Novel Magnetism}\ }\textbf
  {\bibinfo {volume} {23}},\ \bibinfo {pages} {717} (\bibinfo {year}
  {2010})}\BibitemShut {NoStop}%
\bibitem [{\citenamefont {Zhang}\ \emph {et~al.}(2011)\citenamefont {Zhang},
  \citenamefont {Chen},\ and\ \citenamefont {Lin}}]{zhang2011}%
  \BibitemOpen
  \bibfield  {author} {\bibinfo {author} {\bibfnamefont {C.}~\bibnamefont
  {Zhang}}, \bibinfo {author} {\bibfnamefont {X.-J.}\ \bibnamefont {Chen}}, \
  and\ \bibinfo {author} {\bibfnamefont {H.-Q.}\ \bibnamefont {Lin}},\
  }\bibfield  {title} {\enquote {\bibinfo {title} {Phase transitions and
  electron–phonon coupling in platinum hydride},}\ }\href {\doibase
  10.1088/0953-8984/24/3/035701} {\bibfield  {journal} {\bibinfo  {journal}
  {Journal of Physics: Condensed Matter}\ }\textbf {\bibinfo {volume} {24}},\
  \bibinfo {pages} {035701} (\bibinfo {year} {2011})}\BibitemShut {NoStop}%
\bibitem [{\citenamefont {Zhang}\ \emph
  {et~al.}(2015{\natexlab{a}})\citenamefont {Zhang}, \citenamefont {Jin},
  \citenamefont {Lv}, \citenamefont {Zhuang}, \citenamefont {Lv}, \citenamefont
  {Liu}, \citenamefont {Bao}, \citenamefont {Li}, \citenamefont {Liu},\ and\
  \citenamefont {Cui}}]{zhang2015}%
  \BibitemOpen
  \bibfield  {author} {\bibinfo {author} {\bibfnamefont {H.}~\bibnamefont
  {Zhang}}, \bibinfo {author} {\bibfnamefont {X.}~\bibnamefont {Jin}}, \bibinfo
  {author} {\bibfnamefont {Y.}~\bibnamefont {Lv}}, \bibinfo {author}
  {\bibfnamefont {Q.}~\bibnamefont {Zhuang}}, \bibinfo {author} {\bibfnamefont
  {Q.}~\bibnamefont {Lv}}, \bibinfo {author} {\bibfnamefont {Y.}~\bibnamefont
  {Liu}}, \bibinfo {author} {\bibfnamefont {K.}~\bibnamefont {Bao}}, \bibinfo
  {author} {\bibfnamefont {D.}~\bibnamefont {Li}}, \bibinfo {author}
  {\bibfnamefont {B.}~\bibnamefont {Liu}}, \ and\ \bibinfo {author}
  {\bibfnamefont {T.}~\bibnamefont {Cui}},\ }\bibfield  {title} {\enquote
  {\bibinfo {title} {Investigation of stable germane structures under
  high-pressure},}\ }\href {\doibase 10.1039/C5CP03807C} {\bibfield  {journal}
  {\bibinfo  {journal} {Phys. Chem. Chem. Phys.}\ }\textbf {\bibinfo {volume}
  {17}},\ \bibinfo {pages} {27630} (\bibinfo {year}
  {2015}{\natexlab{a}})}\BibitemShut {NoStop}%
\bibitem [{\citenamefont {Zhang}\ \emph
  {et~al.}(2015{\natexlab{b}})\citenamefont {Zhang}, \citenamefont {Jin},
  \citenamefont {Lv}, \citenamefont {Zhuang}, \citenamefont {Liu},
  \citenamefont {Lv}, \citenamefont {Li}, \citenamefont {Bao}, \citenamefont
  {Liu},\ and\ \citenamefont {Cui}}]{zhang2015A}%
  \BibitemOpen
  \bibfield  {author} {\bibinfo {author} {\bibfnamefont {H.}~\bibnamefont
  {Zhang}}, \bibinfo {author} {\bibfnamefont {X.}~\bibnamefont {Jin}}, \bibinfo
  {author} {\bibfnamefont {Y.}~\bibnamefont {Lv}}, \bibinfo {author}
  {\bibfnamefont {Q.}~\bibnamefont {Zhuang}}, \bibinfo {author} {\bibfnamefont
  {Y.}~\bibnamefont {Liu}}, \bibinfo {author} {\bibfnamefont {Q.}~\bibnamefont
  {Lv}}, \bibinfo {author} {\bibfnamefont {D.}~\bibnamefont {Li}}, \bibinfo
  {author} {\bibfnamefont {K.}~\bibnamefont {Bao}}, \bibinfo {author}
  {\bibfnamefont {B.}~\bibnamefont {Liu}}, \ and\ \bibinfo {author}
  {\bibfnamefont {T.}~\bibnamefont {Cui}},\ }\bibfield  {title} {\enquote
  {\bibinfo {title} {A novel stable hydrogen-rich snh8 under high pressure},}\
  }\href {\doibase 10.1039/C5RA20428C} {\bibfield  {journal} {\bibinfo
  {journal} {RSC Adv.}\ }\textbf {\bibinfo {volume} {5}},\ \bibinfo {pages}
  {107637} (\bibinfo {year} {2015}{\natexlab{b}})}\BibitemShut {NoStop}%
\bibitem [{\citenamefont {Zhang}\ \emph
  {et~al.}(2015{\natexlab{c}})\citenamefont {Zhang}, \citenamefont {Jin},
  \citenamefont {Lv}, \citenamefont {Zhuang}, \citenamefont {Liu},
  \citenamefont {Lv}, \citenamefont {Bao}, \citenamefont {Li}, \citenamefont
  {Liu},\ and\ \citenamefont {Cui}}]{zhang2015B}%
  \BibitemOpen
  \bibfield  {author} {\bibinfo {author} {\bibfnamefont {H.}~\bibnamefont
  {Zhang}}, \bibinfo {author} {\bibfnamefont {X.}~\bibnamefont {Jin}}, \bibinfo
  {author} {\bibfnamefont {Y.}~\bibnamefont {Lv}}, \bibinfo {author}
  {\bibfnamefont {Q.}~\bibnamefont {Zhuang}}, \bibinfo {author} {\bibfnamefont
  {Y.}~\bibnamefont {Liu}}, \bibinfo {author} {\bibfnamefont {Q.}~\bibnamefont
  {Lv}}, \bibinfo {author} {\bibfnamefont {K.}~\bibnamefont {Bao}}, \bibinfo
  {author} {\bibfnamefont {D.}~\bibnamefont {Li}}, \bibinfo {author}
  {\bibfnamefont {B.}~\bibnamefont {Liu}}, \ and\ \bibinfo {author}
  {\bibfnamefont {T.}~\bibnamefont {Cui}},\ }\bibfield  {title} {\enquote
  {\bibinfo {title} {High-temperature superconductivity in compressed solid
  silane},}\ }\href {\doibase 10.1038/srep08845} {\bibfield  {journal}
  {\bibinfo  {journal} {Scientific Reports}\ }\textbf {\bibinfo {volume} {5}},\
  \bibinfo {pages} {8845} (\bibinfo {year} {2015}{\natexlab{c}})}\BibitemShut
  {NoStop}%
\bibitem [{\citenamefont {Zhang}\ \emph
  {et~al.}(2015{\natexlab{d}})\citenamefont {Zhang}, \citenamefont {Wang},
  \citenamefont {Zhang}, \citenamefont {Liu}, \citenamefont {Zhong},
  \citenamefont {Song}, \citenamefont {Yang}, \citenamefont {Zhang},\ and\
  \citenamefont {Ma}}]{zhang2015C}%
  \BibitemOpen
  \bibfield  {author} {\bibinfo {author} {\bibfnamefont {S.}~\bibnamefont
  {Zhang}}, \bibinfo {author} {\bibfnamefont {Y.}~\bibnamefont {Wang}},
  \bibinfo {author} {\bibfnamefont {J.}~\bibnamefont {Zhang}}, \bibinfo
  {author} {\bibfnamefont {H.}~\bibnamefont {Liu}}, \bibinfo {author}
  {\bibfnamefont {X.}~\bibnamefont {Zhong}}, \bibinfo {author} {\bibfnamefont
  {H.~F.}\ \bibnamefont {Song}}, \bibinfo {author} {\bibfnamefont
  {G.}~\bibnamefont {Yang}}, \bibinfo {author} {\bibfnamefont {L.}~\bibnamefont
  {Zhang}}, \ and\ \bibinfo {author} {\bibfnamefont {Y.}~\bibnamefont {Ma}},\
  }\bibfield  {title} {\enquote {\bibinfo {title} {Phase diagram and
  high-temperature superconductivity of compressed selenium hydrides},}\ }\href
  {\doibase 10.1038/srep15433} {\bibfield  {journal} {\bibinfo  {journal}
  {Scientific Reports}\ }\textbf {\bibinfo {volume} {5}},\ \bibinfo {pages}
  {15433} (\bibinfo {year} {2015}{\natexlab{d}})}\BibitemShut {NoStop}%
\bibitem [{\citenamefont {Zhang}\ \emph {et~al.}(2016)\citenamefont {Zhang},
  \citenamefont {Jin}, \citenamefont {Lv}, \citenamefont {Zhuang},
  \citenamefont {Li}, \citenamefont {Bao}, \citenamefont {Li}, \citenamefont
  {Liu},\ and\ \citenamefont {Cui}}]{zhang2016}%
  \BibitemOpen
  \bibfield  {author} {\bibinfo {author} {\bibfnamefont {H.}~\bibnamefont
  {Zhang}}, \bibinfo {author} {\bibfnamefont {X.}~\bibnamefont {Jin}}, \bibinfo
  {author} {\bibfnamefont {Y.}~\bibnamefont {Lv}}, \bibinfo {author}
  {\bibfnamefont {Q.}~\bibnamefont {Zhuang}}, \bibinfo {author} {\bibfnamefont
  {Y.}~\bibnamefont {Li}}, \bibinfo {author} {\bibfnamefont {K.}~\bibnamefont
  {Bao}}, \bibinfo {author} {\bibfnamefont {D.}~\bibnamefont {Li}}, \bibinfo
  {author} {\bibfnamefont {B.}~\bibnamefont {Liu}}, \ and\ \bibinfo {author}
  {\bibfnamefont {T.}~\bibnamefont {Cui}},\ }\bibfield  {title} {\enquote
  {\bibinfo {title} {Pressure-induced phase transition of snh4: a new layered
  structure},}\ }\href {\doibase 10.1039/C5RA27037E} {\bibfield  {journal}
  {\bibinfo  {journal} {RSC Adv.}\ }\textbf {\bibinfo {volume} {6}},\ \bibinfo
  {pages} {10456} (\bibinfo {year} {2016})}\BibitemShut {NoStop}%
\bibitem [{\citenamefont {Zhang}\ \emph {et~al.}(2020)\citenamefont {Zhang},
  \citenamefont {McMahon}, \citenamefont {Oganov}, \citenamefont {Li},
  \citenamefont {Dong}, \citenamefont {Dong},\ and\ \citenamefont
  {Wang}}]{zhang2020}%
  \BibitemOpen
  \bibfield  {author} {\bibinfo {author} {\bibfnamefont {J.}~\bibnamefont
  {Zhang}}, \bibinfo {author} {\bibfnamefont {J.~M.}\ \bibnamefont {McMahon}},
  \bibinfo {author} {\bibfnamefont {A.~R.}\ \bibnamefont {Oganov}}, \bibinfo
  {author} {\bibfnamefont {X.}~\bibnamefont {Li}}, \bibinfo {author}
  {\bibfnamefont {X.}~\bibnamefont {Dong}}, \bibinfo {author} {\bibfnamefont
  {H.}~\bibnamefont {Dong}}, \ and\ \bibinfo {author} {\bibfnamefont
  {S.}~\bibnamefont {Wang}},\ }\bibfield  {title} {\enquote {\bibinfo {title}
  {High-temperature superconductivity in the ti-h system at high pressures},}\
  }\href {\doibase 10.1103/PhysRevB.101.134108} {\bibfield  {journal} {\bibinfo
   {journal} {Phys. Rev. B}\ }\textbf {\bibinfo {volume} {101}},\ \bibinfo
  {pages} {134108} (\bibinfo {year} {2020})}\BibitemShut {NoStop}%
\bibitem [{\citenamefont {Zhang}\ \emph
  {et~al.}(2022{\natexlab{a}})\citenamefont {Zhang}, \citenamefont {He},
  \citenamefont {Li}, \citenamefont {Zhang}, \citenamefont {Min}, \citenamefont
  {Zhang}, \citenamefont {Lu}, \citenamefont {Zhao}, \citenamefont {Shi},
  \citenamefont {Peng}, \citenamefont {Wang}, \citenamefont {Feng},
  \citenamefont {Yu}, \citenamefont {Wang}, \citenamefont {Prakapenka},
  \citenamefont {Chariton}, \citenamefont {Liu},\ and\ \citenamefont
  {Jin}}]{zhang2022}%
  \BibitemOpen
  \bibfield  {author} {\bibinfo {author} {\bibfnamefont {C.}~\bibnamefont
  {Zhang}}, \bibinfo {author} {\bibfnamefont {X.}~\bibnamefont {He}}, \bibinfo
  {author} {\bibfnamefont {Z.}~\bibnamefont {Li}}, \bibinfo {author}
  {\bibfnamefont {S.}~\bibnamefont {Zhang}}, \bibinfo {author} {\bibfnamefont
  {B.}~\bibnamefont {Min}}, \bibinfo {author} {\bibfnamefont {J.}~\bibnamefont
  {Zhang}}, \bibinfo {author} {\bibfnamefont {K.}~\bibnamefont {Lu}}, \bibinfo
  {author} {\bibfnamefont {J.}~\bibnamefont {Zhao}}, \bibinfo {author}
  {\bibfnamefont {L.}~\bibnamefont {Shi}}, \bibinfo {author} {\bibfnamefont
  {Y.}~\bibnamefont {Peng}}, \bibinfo {author} {\bibfnamefont {X.}~\bibnamefont
  {Wang}}, \bibinfo {author} {\bibfnamefont {S.}~\bibnamefont {Feng}}, \bibinfo
  {author} {\bibfnamefont {R.}~\bibnamefont {Yu}}, \bibinfo {author}
  {\bibfnamefont {L.}~\bibnamefont {Wang}}, \bibinfo {author} {\bibfnamefont
  {V.}~\bibnamefont {Prakapenka}}, \bibinfo {author} {\bibfnamefont
  {S.}~\bibnamefont {Chariton}}, \bibinfo {author} {\bibfnamefont
  {H.}~\bibnamefont {Liu}}, \ and\ \bibinfo {author} {\bibfnamefont
  {C.}~\bibnamefont {Jin}},\ }\bibfield  {title} {\enquote {\bibinfo {title}
  {Superconductivity above 80 k in polyhydrides of hafnium},}\ }\href
  {\doibase https://doi.org/10.1016/j.mtphys.2022.100826} {\bibfield  {journal}
  {\bibinfo  {journal} {Materials Today Physics}\ }\textbf {\bibinfo {volume}
  {27}},\ \bibinfo {pages} {100826} (\bibinfo {year}
  {2022}{\natexlab{a}})}\BibitemShut {NoStop}%
\bibitem [{\citenamefont {Zhang}\ \emph
  {et~al.}(2022{\natexlab{b}})\citenamefont {Zhang}, \citenamefont {Wang},
  \citenamefont {Tang}, \citenamefont {Duan}, \citenamefont {Wang},
  \citenamefont {Li}, \citenamefont {Ju}, \citenamefont {Sun}, \citenamefont
  {Jin},\ and\ \citenamefont {Zhang}}]{zhang2022A}%
  \BibitemOpen
  \bibfield  {author} {\bibinfo {author} {\bibfnamefont {J.}~\bibnamefont
  {Zhang}}, \bibinfo {author} {\bibfnamefont {Y.}~\bibnamefont {Wang}},
  \bibinfo {author} {\bibfnamefont {L.}~\bibnamefont {Tang}}, \bibinfo {author}
  {\bibfnamefont {J.}~\bibnamefont {Duan}}, \bibinfo {author} {\bibfnamefont
  {J.}~\bibnamefont {Wang}}, \bibinfo {author} {\bibfnamefont {S.}~\bibnamefont
  {Li}}, \bibinfo {author} {\bibfnamefont {M.}~\bibnamefont {Ju}}, \bibinfo
  {author} {\bibfnamefont {W.}~\bibnamefont {Sun}}, \bibinfo {author}
  {\bibfnamefont {Y.}~\bibnamefont {Jin}}, \ and\ \bibinfo {author}
  {\bibfnamefont {C.}~\bibnamefont {Zhang}},\ }\bibfield  {title} {\enquote
  {\bibinfo {title} {Exploring high pressure structural transformations,
  electronic properties and superconducting properties of mh2 (m = nb,
  ta)},}\ }\href {\doibase https://doi.org/10.1016/j.arabjc.2022.104347}
  {\bibfield  {journal} {\bibinfo  {journal} {Arabian Journal of Chemistry}\
  }\textbf {\bibinfo {volume} {15}},\ \bibinfo {pages} {104347} (\bibinfo
  {year} {2022}{\natexlab{b}})}\BibitemShut {NoStop}%
\bibitem [{\citenamefont {Zhang}\ \emph
  {et~al.}(2022{\natexlab{c}})\citenamefont {Zhang}, \citenamefont {He},
  \citenamefont {Li}, \citenamefont {Zhang}, \citenamefont {Feng},
  \citenamefont {Wang}, \citenamefont {Yu},\ and\ \citenamefont
  {Jin}}]{zhang2022B}%
  \BibitemOpen
  \bibfield  {author} {\bibinfo {author} {\bibfnamefont {C.}~\bibnamefont
  {Zhang}}, \bibinfo {author} {\bibfnamefont {X.}~\bibnamefont {He}}, \bibinfo
  {author} {\bibfnamefont {Z.}~\bibnamefont {Li}}, \bibinfo {author}
  {\bibfnamefont {S.}~\bibnamefont {Zhang}}, \bibinfo {author} {\bibfnamefont
  {S.}~\bibnamefont {Feng}}, \bibinfo {author} {\bibfnamefont {X.}~\bibnamefont
  {Wang}}, \bibinfo {author} {\bibfnamefont {R.}~\bibnamefont {Yu}}, \ and\
  \bibinfo {author} {\bibfnamefont {C.}~\bibnamefont {Jin}},\ }\bibfield
  {title} {\enquote {\bibinfo {title} {Superconductivity in zirconium
  polyhydrides with tc above 70 k},}\ }\href {\doibase
  https://doi.org/10.1016/j.scib.2022.03.001} {\bibfield  {journal} {\bibinfo
  {journal} {Science Bulletin}\ }\textbf {\bibinfo {volume} {67}},\ \bibinfo
  {pages} {907} (\bibinfo {year} {2022}{\natexlab{c}})}\BibitemShut {NoStop}%
\bibitem [{\citenamefont {Zheng}\ \emph {et~al.}(2018)\citenamefont {Zheng},
  \citenamefont {Zhang}, \citenamefont {Sun}, \citenamefont {Zhang},
  \citenamefont {Lin}, \citenamefont {Yang},\ and\ \citenamefont
  {Bergara}}]{zheng2018}%
  \BibitemOpen
  \bibfield  {author} {\bibinfo {author} {\bibfnamefont {S.}~\bibnamefont
  {Zheng}}, \bibinfo {author} {\bibfnamefont {S.}~\bibnamefont {Zhang}},
  \bibinfo {author} {\bibfnamefont {Y.}~\bibnamefont {Sun}}, \bibinfo {author}
  {\bibfnamefont {J.}~\bibnamefont {Zhang}}, \bibinfo {author} {\bibfnamefont
  {J.}~\bibnamefont {Lin}}, \bibinfo {author} {\bibfnamefont {G.}~\bibnamefont
  {Yang}}, \ and\ \bibinfo {author} {\bibfnamefont {A.}~\bibnamefont
  {Bergara}},\ }\bibfield  {title} {\enquote {\bibinfo {title} {Structural and
  superconducting properties of tungsten hydrides under high pressure},}\
  }\href {\doibase 10.3389/fphy.2018.00101} {\bibfield  {journal} {\bibinfo
  {journal} {Frontiers in Physics}\ }\textbf {\bibinfo {volume} {6}},\ \bibinfo
  {pages} {101} (\bibinfo {year} {2018})}\BibitemShut {NoStop}%
\bibitem [{\citenamefont {Zhong}\ \emph {et~al.}(2012)\citenamefont {Zhong},
  \citenamefont {Zhang}, \citenamefont {Chen}, \citenamefont {Li},
  \citenamefont {Zhang},\ and\ \citenamefont {Lin}}]{zhong2012}%
  \BibitemOpen
  \bibfield  {author} {\bibinfo {author} {\bibfnamefont {G.}~\bibnamefont
  {Zhong}}, \bibinfo {author} {\bibfnamefont {C.}~\bibnamefont {Zhang}},
  \bibinfo {author} {\bibfnamefont {X.}~\bibnamefont {Chen}}, \bibinfo {author}
  {\bibfnamefont {Y.}~\bibnamefont {Li}}, \bibinfo {author} {\bibfnamefont
  {R.}~\bibnamefont {Zhang}}, \ and\ \bibinfo {author} {\bibfnamefont
  {H.}~\bibnamefont {Lin}},\ }\bibfield  {title} {\enquote {\bibinfo {title}
  {Structural, electronic, dynamical, and superconducting properties in dense
  geh4(h2)2},}\ }\href {\doibase 10.1021/jp211051r} {\bibfield  {journal}
  {\bibinfo  {journal} {The Journal of Physical Chemistry C}\ }\textbf
  {\bibinfo {volume} {116}},\ \bibinfo {pages} {5225} (\bibinfo {year}
  {2012})}\BibitemShut {NoStop}%
\bibitem [{\citenamefont {Zhong}\ \emph {et~al.}(2016)\citenamefont {Zhong},
  \citenamefont {Wang}, \citenamefont {Zhang}, \citenamefont {Liu},
  \citenamefont {Zhang}, \citenamefont {Song}, \citenamefont {Yang},
  \citenamefont {Zhang},\ and\ \citenamefont {Ma}}]{zhong2016}%
  \BibitemOpen
  \bibfield  {author} {\bibinfo {author} {\bibfnamefont {X.}~\bibnamefont
  {Zhong}}, \bibinfo {author} {\bibfnamefont {H.}~\bibnamefont {Wang}},
  \bibinfo {author} {\bibfnamefont {J.}~\bibnamefont {Zhang}}, \bibinfo
  {author} {\bibfnamefont {H.}~\bibnamefont {Liu}}, \bibinfo {author}
  {\bibfnamefont {S.}~\bibnamefont {Zhang}}, \bibinfo {author} {\bibfnamefont
  {H.-F.}\ \bibnamefont {Song}}, \bibinfo {author} {\bibfnamefont
  {G.}~\bibnamefont {Yang}}, \bibinfo {author} {\bibfnamefont {L.}~\bibnamefont
  {Zhang}}, \ and\ \bibinfo {author} {\bibfnamefont {Y.}~\bibnamefont {Ma}},\
  }\bibfield  {title} {\enquote {\bibinfo {title} {Tellurium hydrides at high
  pressures: High-temperature superconductors},}\ }\href {\doibase
  10.1103/PhysRevLett.116.057002} {\bibfield  {journal} {\bibinfo  {journal}
  {Phys. Rev. Lett.}\ }\textbf {\bibinfo {volume} {116}},\ \bibinfo {pages}
  {057002} (\bibinfo {year} {2016})}\BibitemShut {NoStop}%
\bibitem [{\citenamefont {Zhong}\ \emph {et~al.}(2022)\citenamefont {Zhong},
  \citenamefont {Sun}, \citenamefont {Iitaka}, \citenamefont {Xu},
  \citenamefont {Liu}, \citenamefont {Hemley}, \citenamefont {Chen},\ and\
  \citenamefont {Ma}}]{zhong2022}%
  \BibitemOpen
  \bibfield  {author} {\bibinfo {author} {\bibfnamefont {X.}~\bibnamefont
  {Zhong}}, \bibinfo {author} {\bibfnamefont {Y.}~\bibnamefont {Sun}}, \bibinfo
  {author} {\bibfnamefont {T.}~\bibnamefont {Iitaka}}, \bibinfo {author}
  {\bibfnamefont {M.}~\bibnamefont {Xu}}, \bibinfo {author} {\bibfnamefont
  {H.}~\bibnamefont {Liu}}, \bibinfo {author} {\bibfnamefont {R.~J.}\
  \bibnamefont {Hemley}}, \bibinfo {author} {\bibfnamefont {C.}~\bibnamefont
  {Chen}}, \ and\ \bibinfo {author} {\bibfnamefont {Y.}~\bibnamefont {Ma}},\
  }\bibfield  {title} {\enquote {\bibinfo {title} {Prediction of
  above-room-temperature superconductivity in lanthanide/actinide extreme
  superhydrides},}\ }\href {\doibase 10.1021/jacs.2c05834} {\bibfield
  {journal} {\bibinfo  {journal} {Journal of the American Chemical Society}\
  }\textbf {\bibinfo {volume} {144}},\ \bibinfo {pages} {13394} (\bibinfo
  {year} {2022})}\BibitemShut {NoStop}%
\bibitem [{\citenamefont {Zhou}\ \emph {et~al.}(2011)\citenamefont {Zhou},
  \citenamefont {Oganov}, \citenamefont {Dong}, \citenamefont {Zhang},
  \citenamefont {Tian},\ and\ \citenamefont {Wang}}]{zhou2011}%
  \BibitemOpen
  \bibfield  {author} {\bibinfo {author} {\bibfnamefont {X.-F.}\ \bibnamefont
  {Zhou}}, \bibinfo {author} {\bibfnamefont {A.~R.}\ \bibnamefont {Oganov}},
  \bibinfo {author} {\bibfnamefont {X.}~\bibnamefont {Dong}}, \bibinfo {author}
  {\bibfnamefont {L.}~\bibnamefont {Zhang}}, \bibinfo {author} {\bibfnamefont
  {Y.}~\bibnamefont {Tian}}, \ and\ \bibinfo {author} {\bibfnamefont {H.-T.}\
  \bibnamefont {Wang}},\ }\bibfield  {title} {\enquote {\bibinfo {title}
  {Superconducting high-pressure phase of platinum hydride from first
  principles},}\ }\href {\doibase 10.1103/PhysRevB.84.054543} {\bibfield
  {journal} {\bibinfo  {journal} {Phys. Rev. B}\ }\textbf {\bibinfo {volume}
  {84}},\ \bibinfo {pages} {054543} (\bibinfo {year} {2011})}\BibitemShut
  {NoStop}%
\bibitem [{\citenamefont {Zhou}\ \emph {et~al.}(2012)\citenamefont {Zhou},
  \citenamefont {Jin}, \citenamefont {Meng}, \citenamefont {Bao}, \citenamefont
  {Ma}, \citenamefont {Liu},\ and\ \citenamefont {Cui}}]{zhou2012}%
  \BibitemOpen
  \bibfield  {author} {\bibinfo {author} {\bibfnamefont {D.}~\bibnamefont
  {Zhou}}, \bibinfo {author} {\bibfnamefont {X.}~\bibnamefont {Jin}}, \bibinfo
  {author} {\bibfnamefont {X.}~\bibnamefont {Meng}}, \bibinfo {author}
  {\bibfnamefont {G.}~\bibnamefont {Bao}}, \bibinfo {author} {\bibfnamefont
  {Y.}~\bibnamefont {Ma}}, \bibinfo {author} {\bibfnamefont {B.}~\bibnamefont
  {Liu}}, \ and\ \bibinfo {author} {\bibfnamefont {T.}~\bibnamefont {Cui}},\
  }\bibfield  {title} {\enquote {\bibinfo {title} {Ab initio study revealing a
  layered structure in hydrogen-rich kh${}_{6}$ under high pressure},}\ }\href
  {\doibase 10.1103/PhysRevB.86.014118} {\bibfield  {journal} {\bibinfo
  {journal} {Phys. Rev. B}\ }\textbf {\bibinfo {volume} {86}},\ \bibinfo
  {pages} {014118} (\bibinfo {year} {2012})}\BibitemShut {NoStop}%
\bibitem [{\citenamefont {Zhou}\ \emph
  {et~al.}(2020{\natexlab{a}})\citenamefont {Zhou}, \citenamefont {Semenok},
  \citenamefont {Duan}, \citenamefont {Xie}, \citenamefont {Chen},
  \citenamefont {Huang}, \citenamefont {Li}, \citenamefont {Liu}, \citenamefont
  {Oganov},\ and\ \citenamefont {Cui}}]{zhou2020}%
  \BibitemOpen
  \bibfield  {author} {\bibinfo {author} {\bibfnamefont {D.}~\bibnamefont
  {Zhou}}, \bibinfo {author} {\bibfnamefont {D.~V.}\ \bibnamefont {Semenok}},
  \bibinfo {author} {\bibfnamefont {D.}~\bibnamefont {Duan}}, \bibinfo {author}
  {\bibfnamefont {H.}~\bibnamefont {Xie}}, \bibinfo {author} {\bibfnamefont
  {W.}~\bibnamefont {Chen}}, \bibinfo {author} {\bibfnamefont {X.}~\bibnamefont
  {Huang}}, \bibinfo {author} {\bibfnamefont {X.}~\bibnamefont {Li}}, \bibinfo
  {author} {\bibfnamefont {B.}~\bibnamefont {Liu}}, \bibinfo {author}
  {\bibfnamefont {A.~R.}\ \bibnamefont {Oganov}}, \ and\ \bibinfo {author}
  {\bibfnamefont {T.}~\bibnamefont {Cui}},\ }\bibfield  {title} {\enquote
  {\bibinfo {title} {Superconducting praseodymium superhydrides},}\ }\href
  {\doibase 10.1126/sciadv.aax6849} {\bibfield  {journal} {\bibinfo  {journal}
  {Science Advances}\ }\textbf {\bibinfo {volume} {6}},\ \bibinfo {pages}
  {eaax6849} (\bibinfo {year} {2020}{\natexlab{a}})}\BibitemShut {NoStop}%
\bibitem [{\citenamefont {Zhou}\ \emph
  {et~al.}(2020{\natexlab{b}})\citenamefont {Zhou}, \citenamefont {Semenok},
  \citenamefont {Xie}, \citenamefont {Huang}, \citenamefont {Duan},
  \citenamefont {Aperis}, \citenamefont {Oppeneer}, \citenamefont {Galasso},
  \citenamefont {Kartsev}, \citenamefont {Kvashnin}, \citenamefont {Oganov},\
  and\ \citenamefont {Cui}}]{zhou2020A}%
  \BibitemOpen
  \bibfield  {author} {\bibinfo {author} {\bibfnamefont {D.}~\bibnamefont
  {Zhou}}, \bibinfo {author} {\bibfnamefont {D.~V.}\ \bibnamefont {Semenok}},
  \bibinfo {author} {\bibfnamefont {H.}~\bibnamefont {Xie}}, \bibinfo {author}
  {\bibfnamefont {X.}~\bibnamefont {Huang}}, \bibinfo {author} {\bibfnamefont
  {D.}~\bibnamefont {Duan}}, \bibinfo {author} {\bibfnamefont {A.}~\bibnamefont
  {Aperis}}, \bibinfo {author} {\bibfnamefont {P.~M.}\ \bibnamefont
  {Oppeneer}}, \bibinfo {author} {\bibfnamefont {M.}~\bibnamefont {Galasso}},
  \bibinfo {author} {\bibfnamefont {A.~I.}\ \bibnamefont {Kartsev}}, \bibinfo
  {author} {\bibfnamefont {A.~G.}\ \bibnamefont {Kvashnin}}, \bibinfo {author}
  {\bibfnamefont {A.~R.}\ \bibnamefont {Oganov}}, \ and\ \bibinfo {author}
  {\bibfnamefont {T.}~\bibnamefont {Cui}},\ }\bibfield  {title} {\enquote
  {\bibinfo {title} {High-pressure synthesis of magnetic neodymium
  polyhydrides},}\ }\href {\doibase 10.1021/jacs.9b10439} {\bibfield  {journal}
  {\bibinfo  {journal} {Journal of the American Chemical Society}\ }\textbf
  {\bibinfo {volume} {142}},\ \bibinfo {pages} {2803} (\bibinfo {year}
  {2020}{\natexlab{b}})}\BibitemShut {NoStop}%
\bibitem [{\citenamefont {Zhuang}\ \emph
  {et~al.}(2017{\natexlab{a}})\citenamefont {Zhuang}, \citenamefont {Jin},
  \citenamefont {Cui}, \citenamefont {Ma}, \citenamefont {Lv}, \citenamefont
  {Li}, \citenamefont {Zhang}, \citenamefont {Meng},\ and\ \citenamefont
  {Bao}}]{zhuang2017}%
  \BibitemOpen
  \bibfield  {author} {\bibinfo {author} {\bibfnamefont {Q.}~\bibnamefont
  {Zhuang}}, \bibinfo {author} {\bibfnamefont {X.}~\bibnamefont {Jin}},
  \bibinfo {author} {\bibfnamefont {T.}~\bibnamefont {Cui}}, \bibinfo {author}
  {\bibfnamefont {Y.}~\bibnamefont {Ma}}, \bibinfo {author} {\bibfnamefont
  {Q.}~\bibnamefont {Lv}}, \bibinfo {author} {\bibfnamefont {Y.}~\bibnamefont
  {Li}}, \bibinfo {author} {\bibfnamefont {H.}~\bibnamefont {Zhang}}, \bibinfo
  {author} {\bibfnamefont {X.}~\bibnamefont {Meng}}, \ and\ \bibinfo {author}
  {\bibfnamefont {K.}~\bibnamefont {Bao}},\ }\bibfield  {title} {\enquote
  {\bibinfo {title} {Pressure-stabilized superconductive ionic tantalum
  hydrides},}\ }\href {\doibase 10.1021/acs.inorgchem.6b02822} {\bibfield
  {journal} {\bibinfo  {journal} {Inorganic Chemistry}\ }\textbf {\bibinfo
  {volume} {56}},\ \bibinfo {pages} {3901} (\bibinfo {year}
  {2017}{\natexlab{a}})}\BibitemShut {NoStop}%
\bibitem [{\citenamefont {Zhuang}\ \emph
  {et~al.}(2017{\natexlab{b}})\citenamefont {Zhuang}, \citenamefont {Jin},
  \citenamefont {Lv}, \citenamefont {Li}, \citenamefont {Shao}, \citenamefont
  {Liu}, \citenamefont {Li}, \citenamefont {Zhang}, \citenamefont {Meng},
  \citenamefont {Bao},\ and\ \citenamefont {Cui}}]{zhuang2017A}%
  \BibitemOpen
  \bibfield  {author} {\bibinfo {author} {\bibfnamefont {Q.}~\bibnamefont
  {Zhuang}}, \bibinfo {author} {\bibfnamefont {X.}~\bibnamefont {Jin}},
  \bibinfo {author} {\bibfnamefont {Q.}~\bibnamefont {Lv}}, \bibinfo {author}
  {\bibfnamefont {Y.}~\bibnamefont {Li}}, \bibinfo {author} {\bibfnamefont
  {Z.}~\bibnamefont {Shao}}, \bibinfo {author} {\bibfnamefont {Z.}~\bibnamefont
  {Liu}}, \bibinfo {author} {\bibfnamefont {X.}~\bibnamefont {Li}}, \bibinfo
  {author} {\bibfnamefont {H.}~\bibnamefont {Zhang}}, \bibinfo {author}
  {\bibfnamefont {X.}~\bibnamefont {Meng}}, \bibinfo {author} {\bibfnamefont
  {K.}~\bibnamefont {Bao}}, \ and\ \bibinfo {author} {\bibfnamefont
  {T.}~\bibnamefont {Cui}},\ }\bibfield  {title} {\enquote {\bibinfo {title}
  {Investigation of superconductivity in compressed vanadium hydrides},}\
  }\href {\doibase 10.1039/C7CP03435K} {\bibfield  {journal} {\bibinfo
  {journal} {Phys. Chem. Chem. Phys.}\ }\textbf {\bibinfo {volume} {19}},\
  \bibinfo {pages} {26280} (\bibinfo {year} {2017}{\natexlab{b}})}\BibitemShut
  {NoStop}%
\bibitem [{\citenamefont {Zhuang}\ \emph {et~al.}(2018)\citenamefont {Zhuang},
  \citenamefont {Jin}, \citenamefont {Cui}, \citenamefont {Zhang},
  \citenamefont {Li}, \citenamefont {Li}, \citenamefont {Bao},\ and\
  \citenamefont {Liu}}]{zhuang2018}%
  \BibitemOpen
  \bibfield  {author} {\bibinfo {author} {\bibfnamefont {Q.}~\bibnamefont
  {Zhuang}}, \bibinfo {author} {\bibfnamefont {X.}~\bibnamefont {Jin}},
  \bibinfo {author} {\bibfnamefont {T.}~\bibnamefont {Cui}}, \bibinfo {author}
  {\bibfnamefont {D.}~\bibnamefont {Zhang}}, \bibinfo {author} {\bibfnamefont
  {Y.}~\bibnamefont {Li}}, \bibinfo {author} {\bibfnamefont {X.}~\bibnamefont
  {Li}}, \bibinfo {author} {\bibfnamefont {K.}~\bibnamefont {Bao}}, \ and\
  \bibinfo {author} {\bibfnamefont {B.}~\bibnamefont {Liu}},\ }\bibfield
  {title} {\enquote {\bibinfo {title} {Effect of electrons scattered by optical
  phonons on superconductivity in $m{\mathrm{h}}_{3}$ ($m=\mathrm{S}$, ti, v,
  se)},}\ }\href {\doibase 10.1103/PhysRevB.98.024514} {\bibfield  {journal}
  {\bibinfo  {journal} {Phys. Rev. B}\ }\textbf {\bibinfo {volume} {98}},\
  \bibinfo {pages} {024514} (\bibinfo {year} {2018})}\BibitemShut {NoStop}%
\bibitem [{\citenamefont {Gavriliuk}\ \emph {et~al.}(2023)\citenamefont
  {Gavriliuk}, \citenamefont {Troyan}, \citenamefont {Struzhkin}, \citenamefont
  {Trunov}, \citenamefont {Aksenov}, \citenamefont {Mironovich}, \citenamefont
  {Ivanova},\ and\ \citenamefont {Lyubutin}}]{gavriliuk2023}%
  \BibitemOpen
  \bibfield  {author} {\bibinfo {author} {\bibfnamefont {A.~G.}\ \bibnamefont
  {Gavriliuk}}, \bibinfo {author} {\bibfnamefont {I.~A.}\ \bibnamefont
  {Troyan}}, \bibinfo {author} {\bibfnamefont {V.~V.}\ \bibnamefont
  {Struzhkin}}, \bibinfo {author} {\bibfnamefont {D.~N.}\ \bibnamefont
  {Trunov}}, \bibinfo {author} {\bibfnamefont {S.~N.}\ \bibnamefont {Aksenov}},
  \bibinfo {author} {\bibfnamefont {A.~A.}\ \bibnamefont {Mironovich}},
  \bibinfo {author} {\bibfnamefont {A.~G.}\ \bibnamefont {Ivanova}}, \ and\
  \bibinfo {author} {\bibfnamefont {I.~S.}\ \bibnamefont {Lyubutin}},\
  }\bibfield  {title} {\enquote {\bibinfo {title} {Synthesis and
  superconducting properties of some phases of iron polyhydrides at high
  pressures},}\ }\href {\doibase https://doi.org/10.1134/S0021364023603159}
  {\bibfield  {journal} {\bibinfo  {journal} {JETP Letters}\ }\textbf {\bibinfo
  {volume} {118}},\ \bibinfo {pages} {742} (\bibinfo {year}
  {2023})}\BibitemShut {NoStop}%
\bibitem [{\citenamefont {Kunreuther}(2002)}]{Kunreuther_2002}%
  \BibitemOpen
  \bibfield  {author} {\bibinfo {author} {\bibfnamefont {H.}~\bibnamefont
  {Kunreuther}},\ }\bibfield  {title} {\enquote {\bibinfo {title} {Risk
  analysis and risk management in an uncertain world1},}\ }\href {\doibase
  https://doi.org/10.1111/0272-4332.00057} {\bibfield  {journal} {\bibinfo
  {journal} {Risk Analysis}\ }\textbf {\bibinfo {volume} {22}},\ \bibinfo
  {pages} {655} (\bibinfo {year} {2002})}\BibitemShut {NoStop}%
\bibitem [{\citenamefont {Tamminen}\ \emph {et~al.}(2013)\citenamefont
  {Tamminen}, \citenamefont {Juutilainen},\ and\ \citenamefont
  {Röning}}]{TAMMINEN20134577}%
  \BibitemOpen
  \bibfield  {author} {\bibinfo {author} {\bibfnamefont {S.}~\bibnamefont
  {Tamminen}}, \bibinfo {author} {\bibfnamefont {I.}~\bibnamefont
  {Juutilainen}}, \ and\ \bibinfo {author} {\bibfnamefont {J.}~\bibnamefont
  {Röning}},\ }\bibfield  {title} {\enquote {\bibinfo {title} {Exceedance
  probability estimation for a quality test consisting of multiple
  measurements},}\ }\href {\doibase https://doi.org/10.1016/j.eswa.2013.01.056}
  {\bibfield  {journal} {\bibinfo  {journal} {Expert Systems with
  Applications}\ }\textbf {\bibinfo {volume} {40}},\ \bibinfo {pages} {4577}
  (\bibinfo {year} {2013})}\BibitemShut {NoStop}%
\bibitem [{\citenamefont {Guo}\ \emph {et~al.}(2024)\citenamefont {Guo},
  \citenamefont {Semenok}, \citenamefont {Shutov}, \citenamefont {Zhou},
  \citenamefont {Chen}, \citenamefont {Wang}, \citenamefont {Zhang},
  \citenamefont {Wu}, \citenamefont {Luther}, \citenamefont {Helm},
  \citenamefont {Huang},\ and\ \citenamefont {Cui}}]{guo2024}%
  \BibitemOpen
  \bibfield  {author} {\bibinfo {author} {\bibfnamefont {J.}~\bibnamefont
  {Guo}}, \bibinfo {author} {\bibfnamefont {D.}~\bibnamefont {Semenok}},
  \bibinfo {author} {\bibfnamefont {G.}~\bibnamefont {Shutov}}, \bibinfo
  {author} {\bibfnamefont {D.}~\bibnamefont {Zhou}}, \bibinfo {author}
  {\bibfnamefont {S.}~\bibnamefont {Chen}}, \bibinfo {author} {\bibfnamefont
  {Y.}~\bibnamefont {Wang}}, \bibinfo {author} {\bibfnamefont {K.}~\bibnamefont
  {Zhang}}, \bibinfo {author} {\bibfnamefont {X.}~\bibnamefont {Wu}}, \bibinfo
  {author} {\bibfnamefont {S.}~\bibnamefont {Luther}}, \bibinfo {author}
  {\bibfnamefont {T.}~\bibnamefont {Helm}}, \bibinfo {author} {\bibfnamefont
  {X.}~\bibnamefont {Huang}}, \ and\ \bibinfo {author} {\bibfnamefont
  {T.}~\bibnamefont {Cui}},\ }\bibfield  {title} {\enquote {\bibinfo {title}
  {{Unusual metallic state in superconducting A15-type La4H23}},}\ }\href
  {\doibase 10.1093/nsr/nwae149} {\bibfield  {journal} {\bibinfo  {journal}
  {Natl. Sci. Rev.}\ ,\ \bibinfo {pages} {nwae149}} (\bibinfo {year}
  {2024})}\BibitemShut {NoStop}%
\bibitem [{\citenamefont {Cross}\ \emph {et~al.}(2024)\citenamefont {Cross},
  \citenamefont {Buhot}, \citenamefont {Brooks}, \citenamefont {Thomas},
  \citenamefont {Kleppe}, \citenamefont {Lord},\ and\ \citenamefont
  {Friedemann}}]{cross2024}%
  \BibitemOpen
  \bibfield  {author} {\bibinfo {author} {\bibfnamefont {S.}~\bibnamefont
  {Cross}}, \bibinfo {author} {\bibfnamefont {J.}~\bibnamefont {Buhot}},
  \bibinfo {author} {\bibfnamefont {A.}~\bibnamefont {Brooks}}, \bibinfo
  {author} {\bibfnamefont {W.}~\bibnamefont {Thomas}}, \bibinfo {author}
  {\bibfnamefont {A.}~\bibnamefont {Kleppe}}, \bibinfo {author} {\bibfnamefont
  {O.}~\bibnamefont {Lord}}, \ and\ \bibinfo {author} {\bibfnamefont
  {S.}~\bibnamefont {Friedemann}},\ }\bibfield  {title} {\enquote {\bibinfo
  {title} {High-temperature superconductivity in
  ${\mathrm{la}}_{4}{\mathrm{h}}_{23}$ below 100 gpa},}\ }\href {\doibase
  10.1103/PhysRevB.109.L020503} {\bibfield  {journal} {\bibinfo  {journal}
  {Phys. Rev. B}\ }\textbf {\bibinfo {volume} {109}},\ \bibinfo {pages}
  {L020503} (\bibinfo {year} {2024})}\BibitemShut {NoStop}%
\bibitem [{\citenamefont {He}\ \emph {et~al.}(2024)\citenamefont {He},
  \citenamefont {Zhang}, \citenamefont {Li}, \citenamefont {Lu}, \citenamefont
  {Zhang}, \citenamefont {Min}, \citenamefont {Zhang}, \citenamefont {Shi},
  \citenamefont {Feng}, \citenamefont {Liu}, \citenamefont {Song},
  \citenamefont {Wang}, \citenamefont {Peng}, \citenamefont {Wang},
  \citenamefont {Prakapenka}, \citenamefont {Chariton}, \citenamefont {Liu},\
  and\ \citenamefont {Jin}}]{he2024}%
  \BibitemOpen
  \bibfield  {author} {\bibinfo {author} {\bibfnamefont {X.}~\bibnamefont
  {He}}, \bibinfo {author} {\bibfnamefont {C.}~\bibnamefont {Zhang}}, \bibinfo
  {author} {\bibfnamefont {Z.}~\bibnamefont {Li}}, \bibinfo {author}
  {\bibfnamefont {K.}~\bibnamefont {Lu}}, \bibinfo {author} {\bibfnamefont
  {S.}~\bibnamefont {Zhang}}, \bibinfo {author} {\bibfnamefont
  {B.}~\bibnamefont {Min}}, \bibinfo {author} {\bibfnamefont {J.}~\bibnamefont
  {Zhang}}, \bibinfo {author} {\bibfnamefont {L.}~\bibnamefont {Shi}}, \bibinfo
  {author} {\bibfnamefont {S.}~\bibnamefont {Feng}}, \bibinfo {author}
  {\bibfnamefont {Q.}~\bibnamefont {Liu}}, \bibinfo {author} {\bibfnamefont
  {J.}~\bibnamefont {Song}}, \bibinfo {author} {\bibfnamefont {X.}~\bibnamefont
  {Wang}}, \bibinfo {author} {\bibfnamefont {Y.}~\bibnamefont {Peng}}, \bibinfo
  {author} {\bibfnamefont {L.}~\bibnamefont {Wang}}, \bibinfo {author}
  {\bibfnamefont {V.}~\bibnamefont {Prakapenka}}, \bibinfo {author}
  {\bibfnamefont {S.}~\bibnamefont {Chariton}}, \bibinfo {author}
  {\bibfnamefont {H.}~\bibnamefont {Liu}}, \ and\ \bibinfo {author}
  {\bibfnamefont {C.}~\bibnamefont {Jin}},\ }\bibfield  {title} {\enquote
  {\bibinfo {title} {Superconductivity discovered in niobium polyhydride at
  high pressures},}\ }\href {\doibase
  https://doi.org/10.1016/j.mtphys.2023.101298} {\bibfield  {journal} {\bibinfo
   {journal} {Mater. Today Phys.}\ }\textbf {\bibinfo {volume} {40}},\ \bibinfo
  {pages} {101298} (\bibinfo {year} {2024})}\BibitemShut {NoStop}%
\end{thebibliography}%
%
\end{document}